\documentclass[fleqn,usenatbib]{mnras}

\usepackage{newtxtext,newtxmath}

\usepackage[T1]{fontenc}
\usepackage{ae,aecompl}

\usepackage{graphicx}	
\usepackage{amsmath}	
\usepackage{amssymb}	
\usepackage{csvsimple}	
\usepackage{pdflscape}	
\usepackage{afterpage}
\usepackage{multirow}
\usepackage{subfigure}
\usepackage{xcolor}

\newcommand\ColdMassa{1.1}		
\newcommand\ColdMassaP{0.4}
\newcommand\ColdMassaN{0.3}
\newcommand\ColdMassaBest{0.9}
\newcommand\WarmMassa{2.2}		
\newcommand\WarmMassaP{3.5}
\newcommand\WarmMassaN{2.1}
\newcommand\WarmMassaBest{7.9 $\times 10^{-6}$}
\newcommand\PPMAPMassa{0.34}
\newcommand\PPMAPMassErra{0.14}
\newcommand\PPMAPMassaWide{0.50}
\newcommand\PPMAPMassErraWide{0.22}

\newcommand\ColdMassb{0.12}		
\newcommand\ColdMassbP{0.02}

\newcommand\ColdMassbBest{0.12}
\newcommand\PPMAPMassb{0.29}
\newcommand\PPMAPMassErrb{0.08}

\newcommand\ColdMassc{0.3}		
\newcommand\ColdMasscP{0.4}
\newcommand\ColdMasscN{0.2}
\newcommand\ColdMasscBest{0.03}
\newcommand\PPMAPMassc{0.51}
\newcommand\PPMAPMassErrc{0.13}
\newcommand\PPMAPMasscWide{0.64}
\newcommand\PPMAPMassErrcWide{0.18}

\newcommand\ColdTempa{26.7}		
\newcommand\ColdTempaP{1.7}
\newcommand\ColdTempaN{1.3}
\newcommand\ColdTempaBest{27.3}
\newcommand\WarmTempa{92.4}		
\newcommand\WarmTempaP{87.6}
\newcommand\WarmTempaN{12.7}
\newcommand\WarmTempaBest{180}

\newcommand\ColdTempb{26.0}		
\newcommand\ColdTempbP{0.7}

\newcommand\ColdTempbBest{25.9}

\newcommand\ColdTempc{30.7}		
\newcommand\ColdTempcP{11.7}
\newcommand\ColdTempcN{4.0}
\newcommand\ColdTempcBest{45.7}

\newcommand\numStudied{71 }		
\newcommand\percOne{41 } 		
\newcommand\numOne{29 }			
\newcommand\numNewDetections{27 }	

\newcommand\numPWNd{4 }			
\newcommand\numShell{23 }		
\newcommand\numCentral{8 }		

\newcommand\numTypeOnea{1 }		
\newcommand\numCoreCollapse{11 } 
\newcommand\numTypeUnknown{13 }	

\newcommand\numLessOne{2 }		
\newcommand\numOnetTen{12 }		
\newcommand\numTentTwenty{1 }	
\newcommand\numTwentyPlus{4 }	
\newcommand\numAgeUnknown{9 }	

\graphicspath{{SubplotCol/}{SubplotCol/Grids/}{SEDPlots/}{PPMAP/}{PPMAP/ppmap_snrs_11nov/}}

\title[Dusty Supernova remnants]{A Catalogue of Galactic Supernova Remnants in the far-infrared: revealing ejecta dust in pulsar wind nebulae}

\author[H. Chawner et al.]{
H. Chawner$^{1}$\thanks{E-mail: ChawnerHS@cardiff.ac.uk},
K. Marsh$^{1}$,
M. Matsuura$^{1}$,
H.L. Gomez$^{1}$,
P. Cigan$^{1}$,
I. De Looze$^{2}$,
\newauthor{
M. J. Barlow$^{2}$,
L. Dunne$^{1}$
A. Noriega-Crespo$^{3}$,
and J. Rho$^{4,5}$}
\\
$^{1}$School of Physics and Astronomy, Cardiff University, Queens Buildings, The Parade, Cardiff, CF24 3AA, UK\\
$^{2}$Department of Physics and Astronomy, University College London, Gower Street, London WC1E 6BT, UK\\
$^{3}$Space Telescope Science Institute, 3700 San Martin Drive, Baltimore, MD 21218, USA\\
$^{4}$SETI Institute, 189 N. Bernardo Ave, Suite 100, Mountain View, CA 94043, USA\\
$^{5}$SOFIA Science Center, NASA Ames Research Center, MS 232, Moffett Field, CA 94035, USA
}

\date{Accepted XXX. Received YYY; in original form ZZZ}

\pubyear{2017}

\begin{document}
\label{firstpage}
\pagerange{\pageref{firstpage}--\pageref{lastpage}}
\maketitle

\begin{abstract}
We search for far-infrared (FIR) counterparts of known supernova remnants (SNRs) in the Galactic plane ($10^{\circ}< \mid l \mid <60^{\circ}$) at 70\,--\,500\,$\mu$m using the \textit{Herschel} Infrared Galactic Plane Survey (Hi-GAL). Of \numStudied sources studied, we find that \numOne (\percOne\,\%) SNRs have a clear FIR detection of dust emission associated with the SNR. Dust from \numCentral of these is in the central region, and \numPWNd indicate pulsar wind nebulae (PWNe) heated ejecta dust. A further \numShell have dust emission in the outer shell structures which is potentially related to swept up material. Many Galactic SNe have dust signatures but we are biased towards detecting ejecta dust in young remnants and those with a heating source (shock or PWN). We estimate the dust temperature and mass contained within three PWNe, G11.2$-$0.3, G21.5$-$0.9, and G29.7$-$0.3 using modified blackbody fits. To more rigorously analyse the dust properties at various temperatures and dust emissivity index $\beta$, we use point process mapping (PPMAP). We find significant quantities of cool dust (at 20-40\,K) with dust masses of M$_d$\,=\,\PPMAPMassa\,$\pm$\,\PPMAPMassErra\,M$_\odot$, M$_d$\,=\,\PPMAPMassb\,$\pm$\,\PPMAPMassErrb\,M$_\odot$, and M$_d$\,=\,\PPMAPMassc\,$\pm$\,\PPMAPMassErrc\,M$_\odot$ for G11.2$-$0.3, G21.5$-$0.9, and G29.7$-$0.3 respectively. We derive the dust emissivity index for the PWN ejecta dust in G21.5$-$0.3 to be $\beta\,=\,1.4 \pm 0.5$ compared to dust in the surrounding medium where $\beta\,=\,1.8 \pm 0.1$.

\end{abstract}




\section{Introduction} \label{Intro}
Historically, evolved stars have been considered the main source of dust in galaxies, especially Asymptotic Giant Branch stars (low- and intermediate-mass stars in the final stages of evolution) \citep[e.g.][]{Dwek1998}. However, the injection rate from evolved stars falls short by up to an order of magnitude if they are to explain the mass of dust which we observe in the interstellar medium (ISM) of galaxies \citep[e.g.][]{Morgan2003, Matsuura2009}. This is especially problematic in dusty high redshift galaxies for which the lifetime of such stars is close to, or longer than, the dust production timescale \citep{Morgan2003, Dwek2007, Michalowski2010, Gall2011, Mancini2015, Rowlands2014, Michalowski2015}. There is a longstanding debate as to whether dust formed in supernovae (SNe) may survive to make up this shortfall. SNe provide ideal conditions for dust formation as there is an abundance of heavy elements in the ejecta and the temperature drops quickly with expansion. Models suggest that core-collapse SNe can quickly produce substantial amounts of heavy materials \citep{Todini2001, Nozawa2003}.

One unresolved question is how much of the ejecta dust will survive the harsh environment of a SNR. Simulations of dust destruction  in the ISM and SNe suggest that sputtering is responsible for a large amount of dust destruction \citep[e.g.][]{Jones1997, Bocchio2014, Micelotta2016} and destruction rates of up to $3.7 \times 10^{-2} M_\odot yr^{-1}$ have been estimated for SNRs in the Magellanic Clouds \citep{Temim2015}. Nevertheless, the mass of dust which survives destructive processes is dependent on a number of factors including the grain size \citep{Nozawa2007}, the SN type (e.g. \citealp{Kozasa2009, Nozawa2010, Biscaro2016}), and the clumpiness of the ejecta \citep[e.g.][]{Biscaro2016}. Observations of dust in a range of SNRs are therefore crucial in confirming whether dust can survive destructive processes and, if so, how much dust is injected into the ISM.

Along with line and continuum emission from gas, infrared dust emission is thought to be one of the key cooling processes of supernova remnants (SNRs) \citep{Dwek1987, Draine1981, Ostriker1973}. Infrared emission contains important information about the dust within a SNR and can help us to determine whether ejecta dust can survive to be injected into the ISM. However, confusion with the ISM makes SNRs extremely difficult to detect in the Galactic plane and was a problem for surveys using the Infrared Astronomical Satellite (IRAS) which observed at 12\,--\,100\,$\mu$m \citep{Saken1992, Arendt1989}. Infrared surveys of SNRs in the Milky Way were completed by \citeauthor{Reach2006} (2006, hereafter R06) and \citeauthor{Goncalves2011} (2011, hereafter PG11) using \textit{Spitzer} IRAC (3.6\,--\,8\,$\mu$m) and MIPS (24 \& 70\,$\mu$m) data respectively. However, shorter infrared wavelengths may miss any cool dust component that might exist in SNRs and therefore we may underestimate the amount of dust formed after the SN explosion \citep[e.g.][]{Barlow2010}.

With the advent of \textit{Herschel Space Observatory} \citep{Pilbratt2010}, massive quantities of colder (\textless 50 K) SN ejecta dust has been detected in a handful of core collapse remnants. \textit{Herschel} detected filaments of supernova ejecta dust in the Crab Nebula at temperatures of 27\,--\,35 K with a mass of up to 0.47 $M_\odot$ \citep{Gomez2012b, Owen2015}, an order of magnitude larger than that estimated using \textit{Spitzer} data up to 70\,$\mu$m \citep{Temim2012}. The filamentary dust is likely heated by non-thermal, synchrotron radiation from the pulsar wind nebula (PWN) \citep[e.g.][]{Davidson1985, MacAlpine2008}.

Similarly to the Crab Nebula, the PWN G54.1$+$0.3 was discovered to have a shell of SN-condensed dust. Analysis of the 15\,--\,870\,$\mu$m emission from \textit{Herschel},  \textit{Spitzer}  and APEX suggests a minimum total dust mass of 0.07\,--\,1.1 M$_\odot$ \citep{Temim2017, Rho2018}, thought to be heated by the PWN or nearby cluster stars.

Dust in the O-rich SNR Cassiopeia A was discovered using IRAS and the Infrared Space Observatory finding a warm dust mass of $10^{-4} - 10^{-2} M_\odot$ \citep[e.g.][]{Braun1987, Dwek1987, Arendt1999, Douvion2001}. Orders of magnitude more  dust at colder temperatures was then detected by \cite{Dunne2003,Dunne2009}, with later \textit{Herschel} and  \textit{Spitzer}  data showing dust in two regions: a warm dust component in the outer, reverse shock region, and a central region of cold (\textless 25 K), unshocked ejecta dust \citep{Barlow2010}.  The total dust mass is as high as 0.3 $-$ 1.0 $M_\odot$ \citep{Dunne2003,Rho2008,Barlow2010,DeLooze2017}. However, a \textit{Herschel} and \textit{Spitzer} study of the more evolved O-rich remnant, G292.0$+$1.8, found that the IR emission is dominated by pre-existing dust in the circumstellar medium (CSM) \citep{Ghavamian2016}. There is immense debate whether reverse shocks destroy newly formed ejecta dust \citep[e.g.][]{Lau2015, Micelotta2016, Biscaro2016}

In contrast to CCSNe, the only \textit{Herschel} observations of Type Ia SNRs to date (with a sample size of 2) found that any dust emission seen is due to ISM/CSM dust, suggesting that they do not produce significant amounts of dust in their ejecta \citep{Gomez2012a}. Nevertheless, \cite{Lau2015} detected 0.02~M$_\odot$ of warm {$\sim$100~K) dust within the 10$^4$~yr old Sgr~A SNR, suggesting the dust has survived the reverse shock.

Outside of the Milky Way, observations of SN 1987A in the Large Magellanic Cloud with \textit{Herschel} and the Atacama Large Millimetre Array found $\sim0.5 M_\odot$ of cold (<25 K) ejecta dust \citep{Matsuura2011, Matsuura2015, Indebetouw2014, Dwek2015}.


The above discussion suggests that core-collapse SNe can produce significant quantities of dust that would help explain high-redshift dust masses. However, this is based on SN dust formation only being verified in a limited number of sources. 
Therefore a more complete survey is required to establish the importance of SN as dust producers, and to investigate how the dust content varies across SNR types, evolutionary stages, and environment. In this paper we present a survey of SNRs detected at FIR wavelengths ($\geq$70\,$\mu$m) with \textit{Herschel} to complement previous SNR dust surveys by R06 and PG11. In Section \ref{FIRSurvey} we introduce the survey, the detected sources, and discuss our selection bias. In Section \ref{DustMasses} we derive dust masses of SNRs with signatures related to ejecta dust emission and in Section \ref{ppmap} we use a more advanced technique to further study dust properties in these sources. Section \ref{conclusions} lists our conclusions.

\section{Survey for Far Infrared Supernova Remnant Emission} \label{FIRSurvey}

To make a catalogue of FIR SNRs we use data obtained by the Photodetector Array Camera and Spectrometer (PACS, \citealp{Poglitsch2010}) and the Spectral and Photometric Imaging Receiver (SPIRE, \citealp{Griffin2010}) on board the \textit{Herschel} Space Observatory (hereafter \textit{Herschel}) \citep{Pilbratt2010} during the \textit{Herschel} Infrared Galactic Plane Survey (Hi-GAL; \citealp{Molinari2011}). Hi-GAL mapped a $2^{\circ}$ latitude strip of the Galactic Plane using two PACS and three SPIRE wavebands centred on 70, 160, 250, 350 and 500\,$\mu$m. In order to provide a comparison with R06, we study \numStudied remnants within the region where $10^{\circ} < \mid l \mid <60^{\circ}$ and $\mid b \mid \leq 1$, which is covered by Hi-GAL I. There are a total of 127 SNRs in this area based on the radio catalogue from \citet{Green2014}.
\textit{Herschel} has a diffraction limited angular resolution of 5.0\,--\,35.9$''$, 30 times better than the IRAS resolution of $\sim\,6-8^\prime$ \citep{Saken1992}.
Also, the PACS 70\,$\mu$m maps have an angular resolution of $6.4''$, an improvement over that of the MIPS 70\,$\mu$m maps of $18''$ \citep{Carey2009}. Higher angular resolution is important to resolve SNR dust features from the foreground/background or surrounding ISM. At all wavelengths, the map noise is dominated by Galactic Plane cirrus confusion \citep{Molinari2013}.

Table~\ref{tab:DetectionComparison} lists the \numStudied Galactic SNRs from HiGal studied in this survey. Each remnant was first inspected as a false colour image combining the 70, 160, and 250\,$\mu$m \textit{Herschel} wavebands, which are regridded and convolved to the resolution of the 250\,$\mu$m band (Figure\,A1). A circle was overlaid to show the SNR radio size (from the \citet{Green2014} catalog) and X-ray location (the radio location from Green's catalogue is used where X-ray is not available). Various colour scales were applied in order to reveal any FIR structures potentially related to the SNR. We also assessed \textit{Herschel} images in individual bands if any potential SNR dust emission was detected in the initial inspection.
The level of FIR detection was then determined by comparing the structure in the \textit{Herschel} images with that at MIR wavelengths using \textit{Spitzer} (R06 and PG11), X-ray and/or radio wavelengths where possible. See Figure\,A2 for the IRAC and MIPS images of our sample. Where available, VLA 20\,cm images from the Multi-Array Galactic Plane Imaging Survey (MAGPIS) \citep{Helfand2006} \footnote{the MAGPIS database is available at https://third.ucllnl.org/gps/} were used (Figure\,A2) for sources within the range $\mid b \mid<0.8^{\circ}$, $5^{\circ}<l<48.5^{\circ}$; 1420\,MHz images from the Canadian Galactic Plane Survey (CGPS) were used for sources in the range $52^{\circ}<l<192^{\circ}$ \citep{Taylor2003}; and 0.843\,GHz images from the MOST Supernova remnant catalogue for sources within the range $245^{\circ}<l<48.5^{\circ}$ \citep{Whiteoak1996}.

Detection levels were assigned based on the classification scheme adopted by R06. That is: 1 = detection (FIR emission which is clearly correlated with radio, MIR, or X-ray structures and can be distinguished from ISM), 2 = possible detection (FIR emission in the region of the SNR, potentially related to radio, MIR, or X-ray structures but confused with ISM), 3 = unlikely detection (detection of FIR emission but probably unrelated to the SNR), and 4 = not detected in the FIR.
In Section~\ref{StudyComparison} we summarise our findings based on Figures\,A1 and A2 and in Section~\ref{CatalogueResults} we provide individual notes on each of the level 1 and level 2 detected sources.

\begin{landscape}
	\begin{table} 
		\caption{Supernova remnants in the Hi-GAL I Survey ($\mid l \mid <60^{\circ}$)} 
		\csvreader[tabular= l c c c c c c c c c c c c,
				table head=\hline SNR & Name & \parbox{1cm}{\centering Size \textsuperscript{a} \\ (arcmin)} & \parbox{1.5cm}{\centering PWN \textsuperscript{b} \\ (FIR)} & \parbox{1cm}{\centering Age \\ (kyr)} & SN Type & Ref \textsuperscript{c} & GLIMPSE \textsuperscript{d} & MIPSGAL \textsuperscript{e} & Hi-GAL \textsuperscript{f} & Dust features \textsuperscript{g} & Comparison \textsuperscript{h} \\\hline\hline] 
		{Detections/SNRDetections.csv}{SNR=\snr, Name=\name, Size=\size, Morphology=\morph, PWN=\pwn, Age=\age, Type=\type, Distance=\dist, References=\refs, GLIMPSE=\irac, MIPSGAL=\mips, HiGal=\hersc, Region=\region, Comparison=\comp} 
				{\snr & \name & \size & \pwn & \age & \type & \refs & \irac & \mips & \hersc & \region & \comp} 
		\\[1.5pt]
		\label{tab:DetectionComparison}  
	\end{table}
\end{landscape}
\begin{landscape}
	\begin{table}
		\contcaption{Supernova remnants in the Hi-GAL I Survey ($\mid l \mid <60^{\circ}$)}
		\begin{filecontents*}{Detections/SNRDetections2.csv}
SNR,Name,Size,Distance,Morphology,PWN,Age,Type,References,GLIMPSE,MIPSGAL,HiGal,Region,Comparison,,Number,Level 1,Level 2,Level 3,Level 4
G39.2$-$0.3,3C396,8x6,6.5,C,Y (N),7,,26; 27,1,1,1,Western shell,R,,42,16,,,
G41.1$-$0.3,3C397  ,4.5x2.5,8 $-$ 10,S,,1.35 $-$ 1.75,Type Ia,27; 28,1,1,3,,R,,43,,,19,
G43.3$-$0.2,W49B  ,4x3,10,S,,1 $-$ 4,Bipolar Ib/Ic,29; 30,1,1,1,Fe/radio loop,R,,44,17,,,
,,,,,,,,,,,,H$_2$ filament,,,,,,,
G54.1$+$0.3,  ,12?,6.2,C,Y (Y),2,Core collapse,31,3,3,1,PWN southern rim,,,45,18,,,
G54.4$-$0.3,(HC40),40,\textgreater 3,S,,95,,32,1,1,1,Outer filaments,,,46,19,,,
G55.0$+$0.3,  ,20x15?,14,S,,$\leq$1900,,33,2,2,2,,,,47,,8,,
G296.8$-$0.3,1156-62,20x14,9.6,,,2 $-$ 10,,34,3,1,3,,,,48,,,20,
G304.6$+$0.1,Kes 17 ,8,\textgreater 9.7,S,,28 $-$ 64,Core collapse,35,1,1,1,Western shell,,,49,20,,,
G310.6$-$0.3,Kes 20B ,8,,S,,,,,2,3,3,,,,50,,,21,
G310.8$-$0.4,Kes 20A ,12,13.7,S,,,,,1,1,1,,,,51,,9,,
G311.5$-$0.3,  ,5,12.5,S,,23,,36,1,1,1,Shell,,,52,21,,,
G332.0$+$0.2,  ,12,,S,,,,,4,3,3,,,,53,,,22,
G332.4$-$0.4,RCW 103 ,10,3.1,S,,2,Type IIP,37; 38,1,1,1,Southern arc,X,,54,22,,,
G332.4$+$0.1,MSH 16-51,15,\textgreater 6.6,S,,3,,39,2,2,3,,,,55,,,23,
,Kes 32,,,,,,,,,,,,,,,,,,
G336.7$+$0.5,,14x10,,S,,,,,4,1,3,,R,,56,,,24,
G337.2$-$0.7,,6,2 $-$ 9.3,S,,0.75 $-$ 3.5,Type Ia,40; 41,4,1,2,,X,,57,,10,,
G340.4$+$0.4,  ,10x7,\textgreater 14,S,,3.1,,42,4,3,3,,,,58,,,25,
G340.6$+$0.3,  ,6,15,S,,2.6,,42,2,1,1,Shell,R,,59,23,,,
G341.2$+$0.9,  ,22x16,6.9,C,Y (N),32.6,Core collapse,43,4,3,3,,,,60,,,26,
G341.9$-$0.3,  ,7,17.3,S,,5,,42,4,3,3,,,,61,,,27,
G342.0$-$0.2,  ,12x9,,S,,11.7,,42,3,3,3,,,,62,,,28,
G344.7$-$0.1,  ,8,6.3 $-$ 8,C,,3,Type Ia,44; 45,1,1,1,North-central region,R,,63,24,,,
G345.7$-$0.2,  ,6,,S,,,,,4,1,2,,,,64,,11,,
G346.6$-$0.2,  ,8,11,S,,47,Type Ib/Ic/II,36,1,2,2,,,,65,,12,,
G347.3$-$0.5,RX J1713.7-3946 ,65x55,1.3,S,,1.59 $-$ 2.1,Type Ib/c,46,3,3,3,,,,66,,,29,
G348.5$-$0.0,,10?,\textless 6.3,S,,,,,1,1,1,Arc,R,,67,25,,,
G348.5$+$0.1  ,CTB37A,15?,6.3 $-$ 9.5,S,,32 $-$ 42,,36,1,1,1,Northern arc,R,,68,26,,,
,,,,,,,,,,,,Breakout region,,,,,,,
G348.7$+$0.3  ,CTB 37B ,17?,13.2,S,,5,Core collapse,47,3,1,1,North-eastern shell,R,,69,27,,,
G349.2$-$0.1,  ,9x6,,S,,,,,3,3,4,,,,70,,,,2
G349.7$+$0.2,  ,2.5x2,11.5,S,,1.8,,48,1,1,1,Eastern shell,R,,71,28,,,
\end{filecontents*}
\csvreader[tabular= l c c c c c c c c c c c c,
			table head=\hline SNR & Name(s) & \parbox{1cm}{\centering Size \textsuperscript{a} \\ (arcmin)} & \parbox{1.5cm}{\centering PWN \textsuperscript{b} \\ (FIR)} & \parbox{1cm}{\centering Age \\ (kyr)} & SN Type & Ref \textsuperscript{c} & GLIMPSE \textsuperscript{d} & MIPSGAL \textsuperscript{e} & HiGAL \textsuperscript{f} & Dust features \textsuperscript{g} & Comparison \textsuperscript{h} \\\hline\hline,table foot = \\\hline]
			{Detections/SNRDetections2.csv}{SNR=\snr, Name=\name, Size=\size, Morphology=\type, PWN=\pwn, Age=\age, Type=\type, Distance=\dist, References=\refs, GLIMPSE=\irac, MIPSGAL=\mips, HiGal=\hersc, Region=\region, Comparison=\comp}
			{\snr & \name & \size & \pwn & \age & \type & \refs & \irac & \mips & \hersc & \region & \comp} 
		\\[1.5pt]
		\textsuperscript{a} Radio size from Green's catalogue.  \textsuperscript{b} `Y' indicates that a source contains an associated PWN, `?' indicates an unconfirmed PWN candidate. FIR detection of the PWN is indicated in brackets.

		Level of detection for each SNR is listed for the following surveys: \textsuperscript{d} \textit{Spitzer} GLIMPSE (IRAC, \citealp{Reach2006}), \textsuperscript{e} \textit{Spitzer} MIPSGAL (MIPS, \citealp{Goncalves2011}), \textsuperscript{f} \textit{Herschel} Hi-GAL (PACS \& SPIRE, this work). \textsuperscript{g} Location of FIR detected dust features. \textsuperscript{h} Waveband of previous detection to which FIR structure is compared: O = optical, R = radio, X = X-ray.


		Detection level of remnants: 1 = likely detection, 2 = possible detection, 3 = unlikely detection but confused, 4 = not detected, - = unstudied.

		\textsuperscript{b} References for SNR age and SN type:
		\textsuperscript{1}\citealp{Borkowski2016},
		\textsuperscript{2}\citealp{Yamauchi2014a},
		\textsuperscript{3}\citealp{Reynolds2006},
		\textsuperscript{4}\citealp{Klochkov2016},
		\textsuperscript{5}\citealp{Helfand2003a},
		\textsuperscript{6}\citealp{Leahy2014},
		\textsuperscript{7}\citealp{Voisin2016},
		\textsuperscript{8}\citealp{Dubner1999},
		\textsuperscript{9}\citealp{Petriella2013},
		\textsuperscript{10}\citealp{Bietenholz2008},
		\textsuperscript{11}\citealp{Bocchino2005},
		\textsuperscript{12}\citealp{Tian2008},
		\textsuperscript{13}\citealp{Kumar2014},
		\textsuperscript{14}\citealp{Chevalier2005},
		\textsuperscript{15}\citealp{Reich1984},
		\textsuperscript{16}\citealp{Misanovic2010}
		\textsuperscript{17}\citealp{Gaensler1999},
		\textsuperscript{18}\citealp{Leahy2008a},
		\textsuperscript{19}\citealp{Morton2007},
		\textsuperscript{20}\citealp{Chen2004},
		\textsuperscript{21}\citealp{Zhou2011},
		\textsuperscript{22}\citealp{Zhou2016},
		\textsuperscript{23}\citealp{Sato2016},
		\textsuperscript{24}\citealp{Wolszczan1991},
		\textsuperscript{25}\citealp{Zhu2013},
		\textsuperscript{26}\citealp{Harrus1999},
		\textsuperscript{27}\citealp{Leahy2016},
		\textsuperscript{28}\citealp{Yamaguchi2015},
		\textsuperscript{29}\citealp{Pye1984, Smith1985, Hwang2000},
		\textsuperscript{30}\citealp{Lopez2013},
		\textsuperscript{31}\citealp{Bocchino2010},
		\textsuperscript{32}\citealp{Park2013},
		\textsuperscript{33}\citealp{Matthews1998},
		\textsuperscript{34}\citealp{Gaensler1998},
		\textsuperscript{35}\citealp{Combi2010},
		\textsuperscript{36}\citealp{Pannuti2014},
		\textsuperscript{37}\citealp{Nugent1984, Carter1997},
		\textsuperscript{38}\citealp{Frank2015},
		\textsuperscript{39}\citealp{Vink2004},
		\textsuperscript{40}\citealp{Rakowski2006},
		\textsuperscript{41}\citealp{Takata2016},
		\textsuperscript{42}\citealp{Caswell1983},
		\textsuperscript{43}\citealp{Frail1994},
		\textsuperscript{44}\citealp{Giacani2011},
		\textsuperscript{45}\citealp{Yamaguchi2012},
		\textsuperscript{46}\citealp{Tsuji2014},
		\textsuperscript{47}\citealp{HESSCollaboration2008b},
		\textsuperscript{48}\citealp{Tian2014}.

	  \label{tab:DetectionComparisonContinued}
	\end{table}
\end{landscape}

\begin{table}
	\begin{tabular}{lcc}
	\hline
	\multicolumn{2}{c}{Detection Type} & Number Detected \\\hline\hline
	SNR Region $\natural$ & Shell / outer shock region & \numShell \\
	& Inner ejecta region & \numCentral \\
	& Confirmed PWN $\dag$ & \numPWNd \\
	&&\\\hline
	Age (kyr) & $\leq$ 1 & \numLessOne \\
	& 1 \textless Age $\leq$ 10 & \numOnetTen \\
	& 10 \textless Age $\leq$ 20 & \numTentTwenty \\
	& \textgreater 20 & \numTwentyPlus \\
	& Unknown & \numAgeUnknown \\
	&&\\\hline
	SN Type & Type Ia & \numTypeOnea \\
	& Core collapse & \numCoreCollapse \\
	& Unknown & \numTypeUnknown \\ \hline

	\end{tabular}
	\caption{Summary of the level 1 detected sample in this work. $\natural$The total number for this classification is larger than the number of level 1 detected sources as dust in some SNRs in detected in both outer and inner regions. \dag SNRs for which there is evidence that the detected central region is associated with the confirmed PWN.}
	\label{tab:DetectionSummary}
\end{table}

\begin{figure}
	\includegraphics[width=1.0\linewidth, trim = 0cm 0.4cm 0cm 0cm, clip]{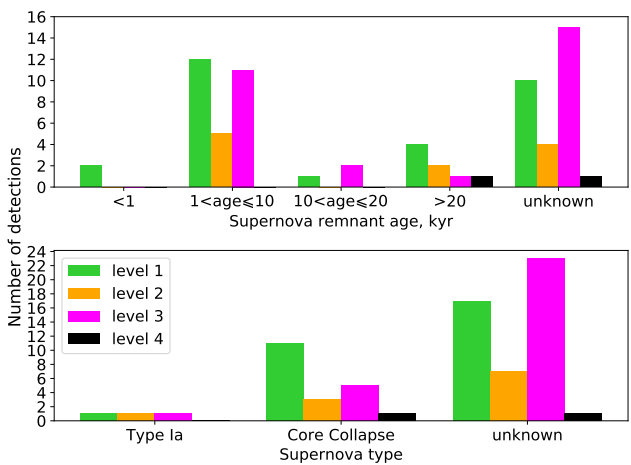}
	\caption{Source types detected in the sample. \textit{Top}: Number of detections of sources of different ages. \textit{Bottom}: Number of detections of different SN types.}
	\label{fig:DetectionSummary}
\end{figure}

\begin{figure}
	\includegraphics[width=1.0\linewidth, trim = 0.3cm 0cm 1cm 1cm, clip]{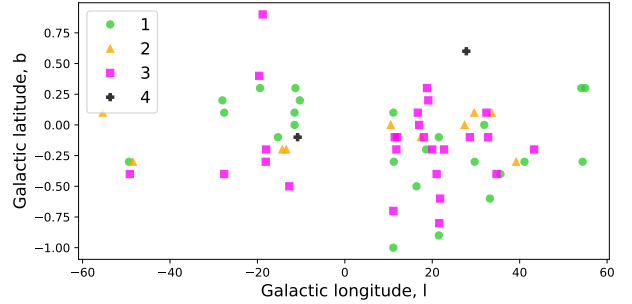}
	\caption{Location of sources from our sample within the Galactic Plane. We do not see any bias in the location of detected sources.}
	\label{fig:SourceLocations}
\end{figure}

Of the \numOne SNRs detected in this survey, 13 are in common with R06 and 21 with PG11. R06 detected 18 of 95 SNRs from GLIMPSE and PG11 detected 39 of 121 SNRs from the MIPSGAL Survey; their classifications of our sample are listed in Table~\ref{tab:DetectionComparison}. We detect 1 additional SNR, G11.1$-$1.0, which was not in the PG11 or R06 samples.

\begin{figure*}
	\centering
	\includegraphics[width=1.0\linewidth]{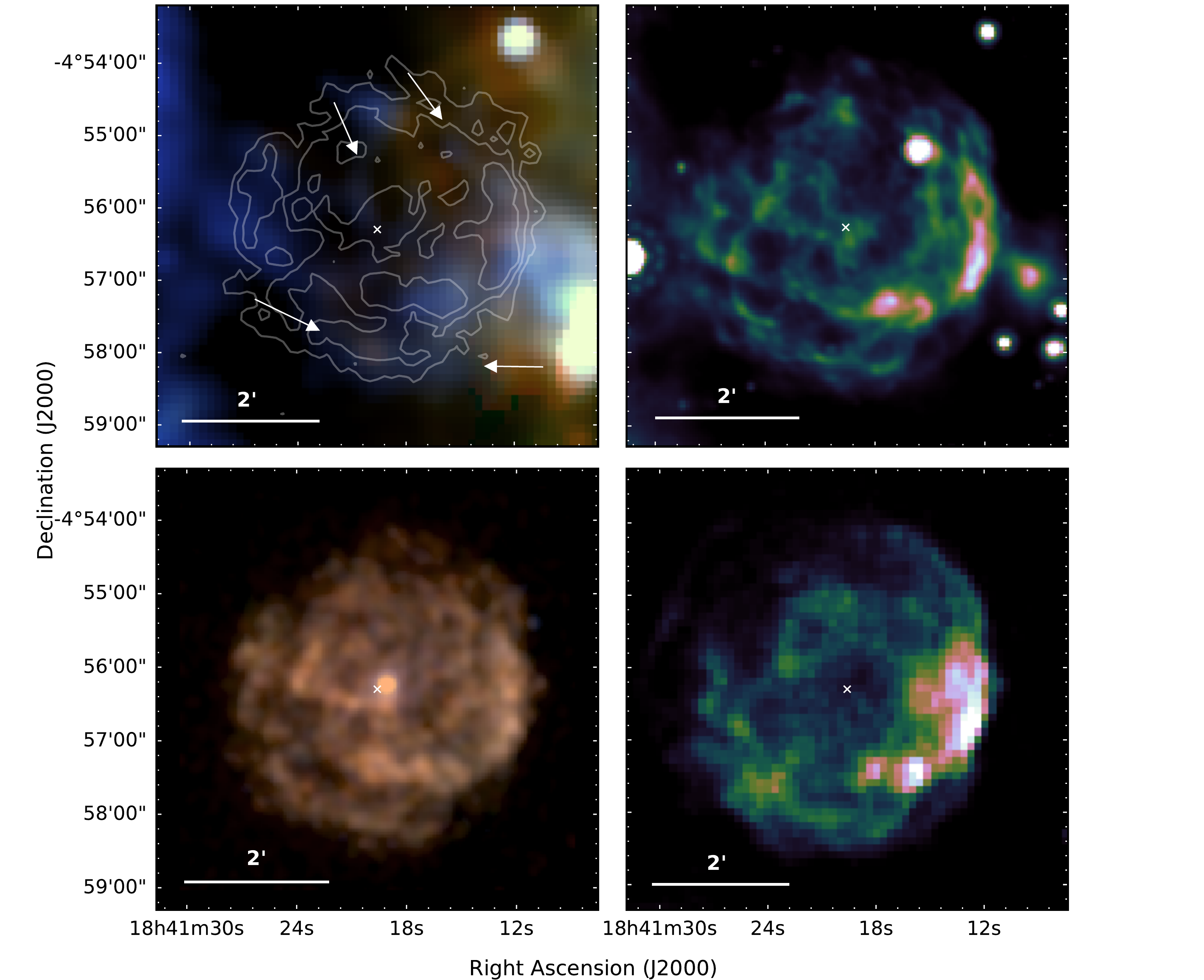}
	\caption{G27.4+0.0 - \textit{Top-left}: \textit{Herschel} colour image with X-ray contours overlaid, colours are red = 250\,$\mu$m, green = 160\,$\mu$m, and blue = 70\,$\mu$m. The same \textit{Herschel} colour combinations are used in Figures  \ref{fig:G11.1-1.0Image} - \ref{fig:G349.7+0.2Image}. The white arrows indicate emission at 70\,$\mu$m which may be associated with SNR filaments detected at other wavelengths. However, extensive dust emission to the west makes the region very confused and FIR emission cannot be conclusively distinguished from the local ISM. There is no dust emission detected at other \textit{Herschel} wavelengths which corresponds to the X-ray contours.
	\textit{Top-right}: \textit{Spitzer} MIPS 24\,$\mu$m image.
	\textit{Bottom-left}: \textit{Chandra} three colour image, colours are red = $0.8 - 1.7$ keV, green = $1.7 - 2.6$ keV, and blue = $2.6 - 7.0$ keV. The pulsar PSR J1841-0456 can be seen at the centre of the SNR.
	\textit{Bottom-right}: VLA 20\,cm radio image.
	The white cross shows the X-ray coordinates of the SNR centre. }
	\label{fig:G27.4+0.0Image}
\end{figure*}

\subsection{Summary of the Sample} \label{StudyComparison}
Table~\ref{tab:DetectionSummary} and Figure~\ref{fig:DetectionSummary} give a summary of the types of SNRs detected in this study. Of our new FIR detections, we observe dust emission from the shell / outer shock region of \numShell SNRs, and within the inner ejecta region (interior to the reverse shock) of \numCentral sources.
We detect \numTypeOnea Type Ia (G344.7$-$0.1) and \numCoreCollapse core collapse SNe; the emission from the Type Ia SNR is thought to arise from a shocked cloud in front of the SNR rather than the ejecta (see Section \ref{CatalogueResults} for more details). Of the SNRs in our survey with SN type classification, 87\% are core collapse with only three Type Ia SNe.
This adds \numNewDetections new sources to the current sample of 3 Galactic objects (4 including LMC) with confirmed cool dust (\textless 50\,K).
Figure~\ref{fig:DetectionSummary} demonstrates that we are biased towards detecting young SNRs as those aged $\leqslant$5\,kyr make up 61\% of the sources in our survey which have estimated ages. We are least likely to detect sources which do not have an estimated age. These sources make up a large proportion of the fainter sources in our sample, making up 53 per cent of sources with a 1\,GHz flux below 5\,Jy.
Furthermore, the majority of these sources are not very well studied and there are few images available to compare morphology.

Figure~\ref{fig:SourceLocations} compares the location of the sources on the sky with their assigned detection level in this work to check if there is any bias due to location (e.g. due to higher levels of confusion expected towards the Galactic centre).  We see little evidence for any such bias.

We find that five PG11 detections are only a level 2 with \textit{Herschel} and a further four are level 3 in this study. Some of these differences are due to limitations of observing at FIR wavelengths.
In many cases the FIR emission is too confused to distinguish between the ISM and any SNR related emission which may be at a similar temperature. Furthermore, the \textit{Spitzer} data ($\leq 24 \mu$m) has higher angular resolution ($< 2''$ at 3.6\,--\,8.0\,$\mu$m and $6''$ at 24\,$\mu$m) than that of \textit{Herschel} and so may be better at resolving dust structures which emit at both MIR and FIR wavelengths. An example of this issue is G27.4$+$0.0 (Figure \ref{fig:G27.4+0.0Image}) from which PG11 detected clear structure at 24\,$\mu$m, similar to the X-ray structure (Figure 22 in PG11).
We detect some emission at 70\,$\mu$m which may be associated with the SNR. However, extensive interstellar dust emission to the west of the SNR makes any related dust emission difficult to distinguish from the local ISM in \textit{Herschel} wavebands.
We note that synchrotron radiation may also contribute at the long wavelengths but since the majority of our detected sources are brightest at the shortest \textit{Herschel} wavelengths, the synchrotron contamination is minimal. We also do not expect the \textit{Herschel} flux to be dominated by line emission; studies of line intensity in two SNRs found a negligible contribution in the \textit{Herschel} wavebands \citep{Gomez2012b, DeLooze2017}.


We detect FIR emission from 4 out of the 9 confirmed PWN sources in our sample: G11.2$-$0.3, G21.5$-$0.9, G29.7$-$0.3 and G54.1$+$0.3. The discovery of cold dust in G54.1$+$0.3 has previously been reported based on FIR and MIR observations \citep{Temim2017, Rho2018}.  We do not detect dust features related to the 4 `unconfirmed' PWN candidates in our sample (G12.0$-$0.1, G18.6$-$0.2, G22.7$-$0.2, and G27.8$+$0.6). The 4 detected PWNe all have ages less than 2.5 kyr which could indicate a lack of dust in older PWNe due to destruction by the reverse shock at later times. However, this is a limited sample making it difficult to come to general conclusions, especially since the reverse shock in G11.2$-$0.3 has already passed the ejecta material \citep{Borkowski2016}.
Both G29.7$-$0.3 and G54.1$+$0.3 were classified as level 3 detections by R06 but have been classified here as level-1 detections. The \textit{Herschel} dust emission is clear in the images due to the dust temperature being above the typical ISM dust, likely due to heating by the central PWN. There is no IRAC detection of the PWNe associated with G11$-$0.2 or G21.5$-$0.9. It may be that this source of heating does not increase the dust temperature by an adequate amount for strong emission in the IRAC wavebands.

\begin{figure*}
	\includegraphics[width=1.0\linewidth, trim = 0cm 5.7cm 0cm 6cm, clip]{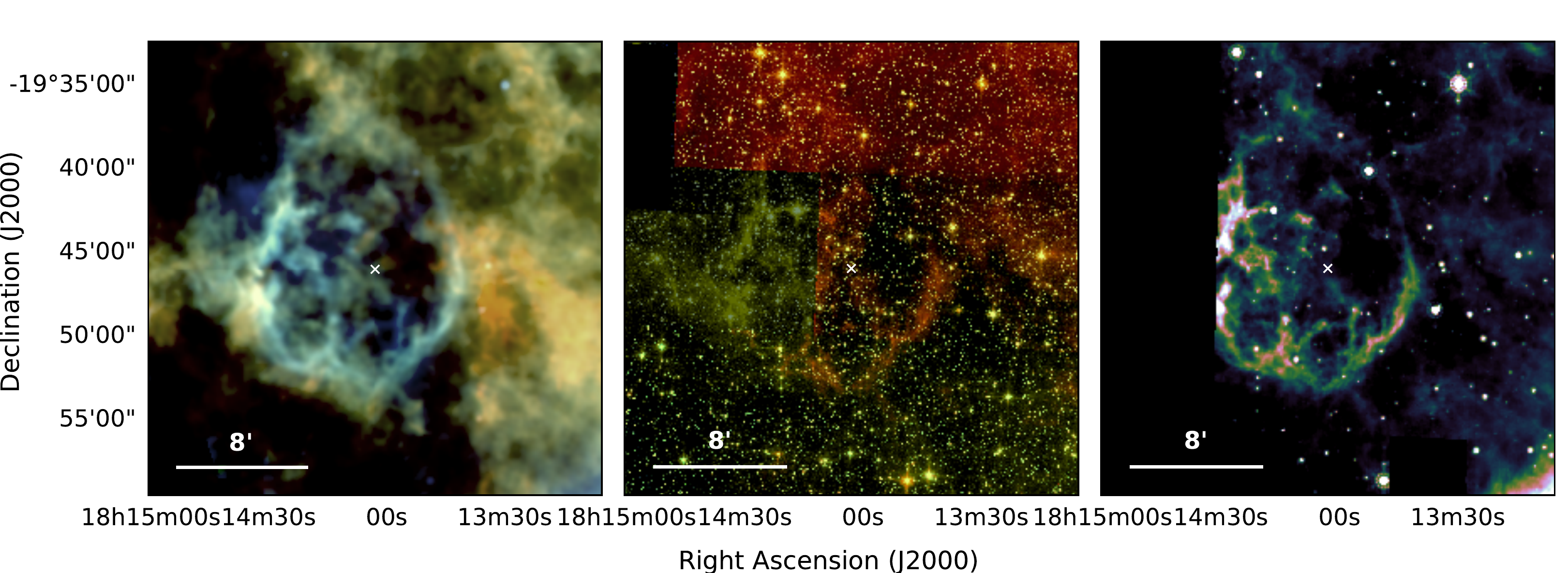}
	\caption{G11.1$-$1.0 - \textit{Left}: \textit{Herschel} colour image.
	\textit{Middle}:  \textit{Spitzer}  IRAC four colour image, colours are red = 8.0 ~$\mu$m, yellow = 4.5 ~$\mu$m, green = 5.8 ~$\mu$m, and blue = 8.0 ~$\mu$m.
	\textit{Right}:  \textit{Spitzer}  MIPS 24\,$\mu$m image.
	In both images, filaments of dust are seen at the outer edges of the shocks.
	The white cross shows the radio coordinates of the SNR centre from \citet{Green2014}.}
	\label{fig:G11.1-1.0Image}
\end{figure*}

Clearly there are limitations to detecting SNRs in the FIR and we have quite complex selection effects. We easily detect dust in SNRs where the dust is at a different temperature to the local ISM or where there is little contaminating foreground/background dust.
This biases us towards younger SNRs, or those with a source of heating, such as a PWN or shock heating.
A further selection effect arises from the availability of radio and X-ray data as these images are used to visually compare FIR structures and determine if any FIR structures correlation with the X-ray and radio structures associated with the SNR. If radio and X-ray data are unavailable it can be difficult to clearly distinguish SN and ISM material.

\subsection{Results for Individual Remnants}\label{CatalogueResults}
Here we provide notes on individual sources for which we detect dust within the SNR (shell or inner ejecta) in the \textit{Herschel} three colour images. Detected remnants (level 1 in Table~\ref{tab:DetectionComparison}) are in bold and possible detections (level 2 in Table~\ref{tab:DetectionComparison}) in italics. \textit{Herschel} FIR and \textit{Spitzer} NIR / MIR images for the entire sample are shown in Figures\,A1 and A2.
\bigskip

\begin{figure*}
	\centering
	\includegraphics[width=1.0\linewidth, trim = 0cm 3.2cm 0cm 3.5cm, clip]{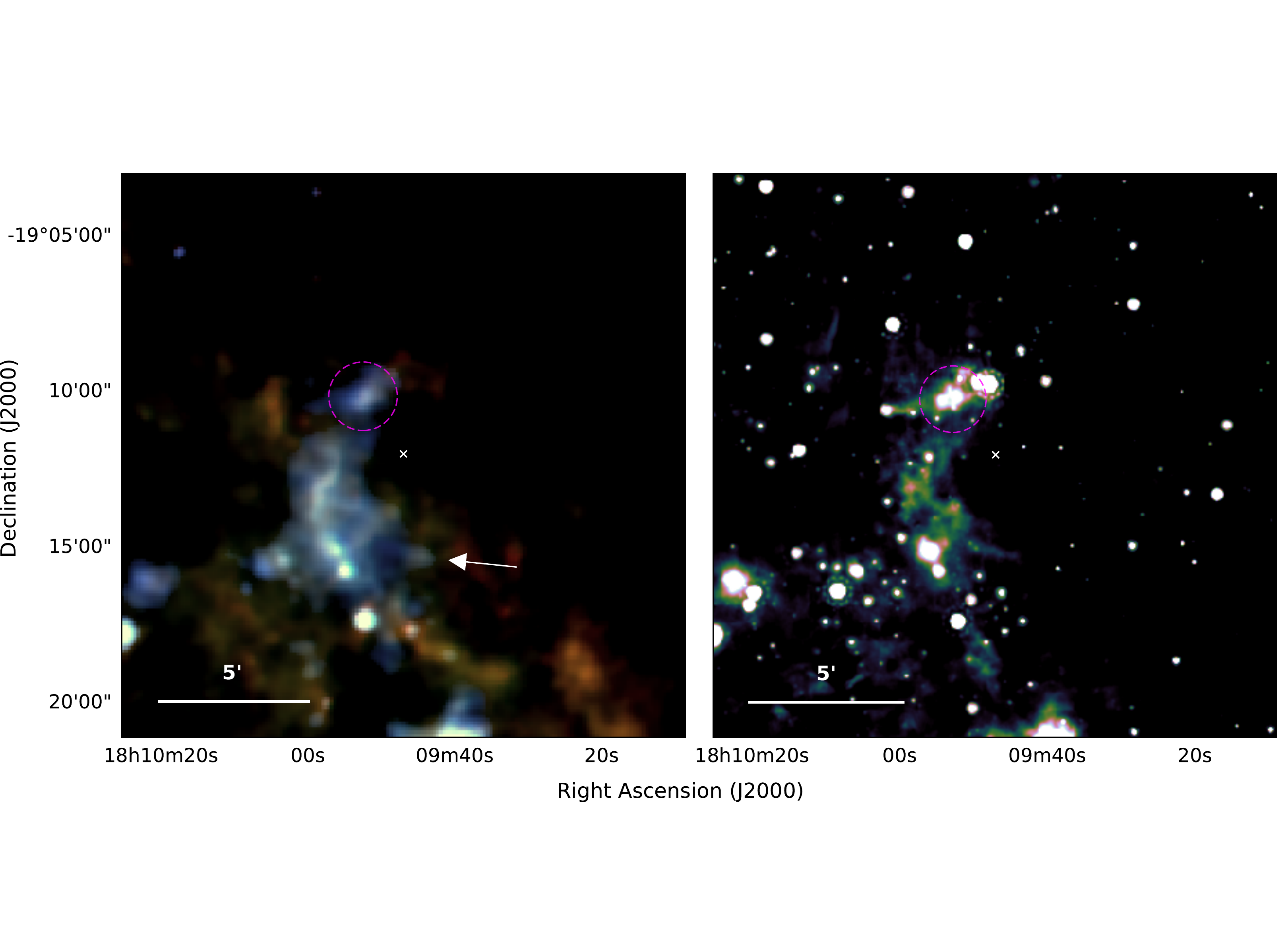}
	\caption{G11.1+0.1 - \textit{Left:} \textit{Herschel} three colour image.
	\textit{Right:}  \textit{Spitzer}  MIPS 24\,$\mu$m image. Dust is seen in an arc, with small clumps to the east of the remnant at the outer shock (blue in the \textit{Herschel} image).
	The white cross shows the radio coordinates of the SNR centre from \citet{Green2014}.}
	\label{fig:G11.1+0.1Image}
\end{figure*}

\begin{figure*}
	\centering
	\includegraphics[width=1.0\linewidth]{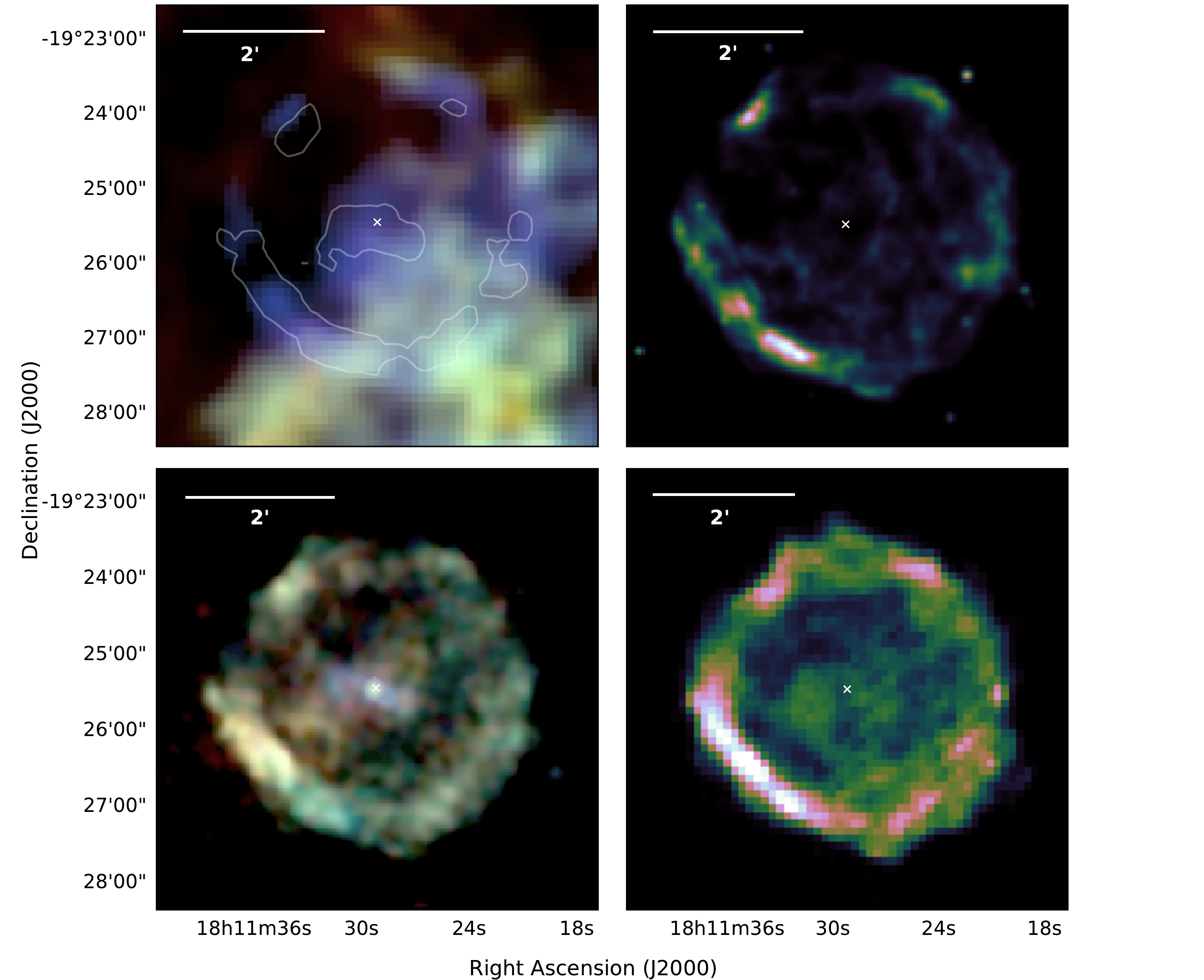}
	\caption{G11.2$-$0.3 - \textit{Top-left}: \textit{Herschel} three colour image with X-ray contours overlaid showing the location of the PWN and outer shocks, due to interaction with surrounding ISM / CSM. Dust is clearly seen in a bright ring and south of the pulsar and its nebula.
	\textit{Top-right}:  \textit{Spitzer}  MIPS 24\,$\mu$m image.
	\textit{Bottom-left}: \textit{Chandra} three colour image, colours are red = $0.8 - 1.2$ keV, green = $1.2 - 2.0$ keV, and blue = $2.0 - 10.0$ keV.
	\textit{Bottom-right}: VLA 20\,cm radio image.
	The white crosses show the X-ray coordinates of the SNR centre, which is at the location of the central pulsar.}
	\label{fig:G11.2-0.3Image}
\end{figure*}

\begin{figure}
	\centering
	\includegraphics[width=1.0\linewidth]{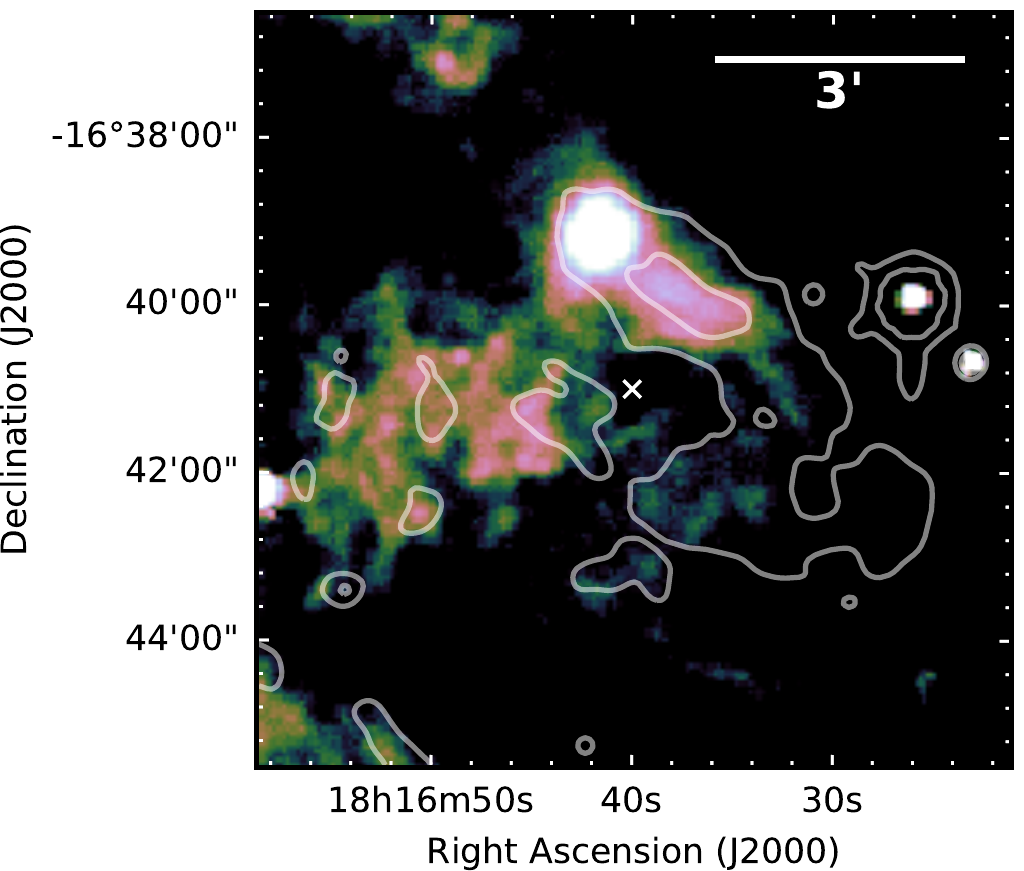}{}
	\caption{G14.1$-$0.1 - \textit{Herschel} 70\,$\mu$m image with 24\,$\mu$m contours overlaid.
	The magenta cross shows the radio coordinates of the SNR centre from \citet{Green2014}.}
	\label{fig:G14.1-0.1Image}
\end{figure}

\begin{figure}
	\centering
	\includegraphics[width=1.0\linewidth]{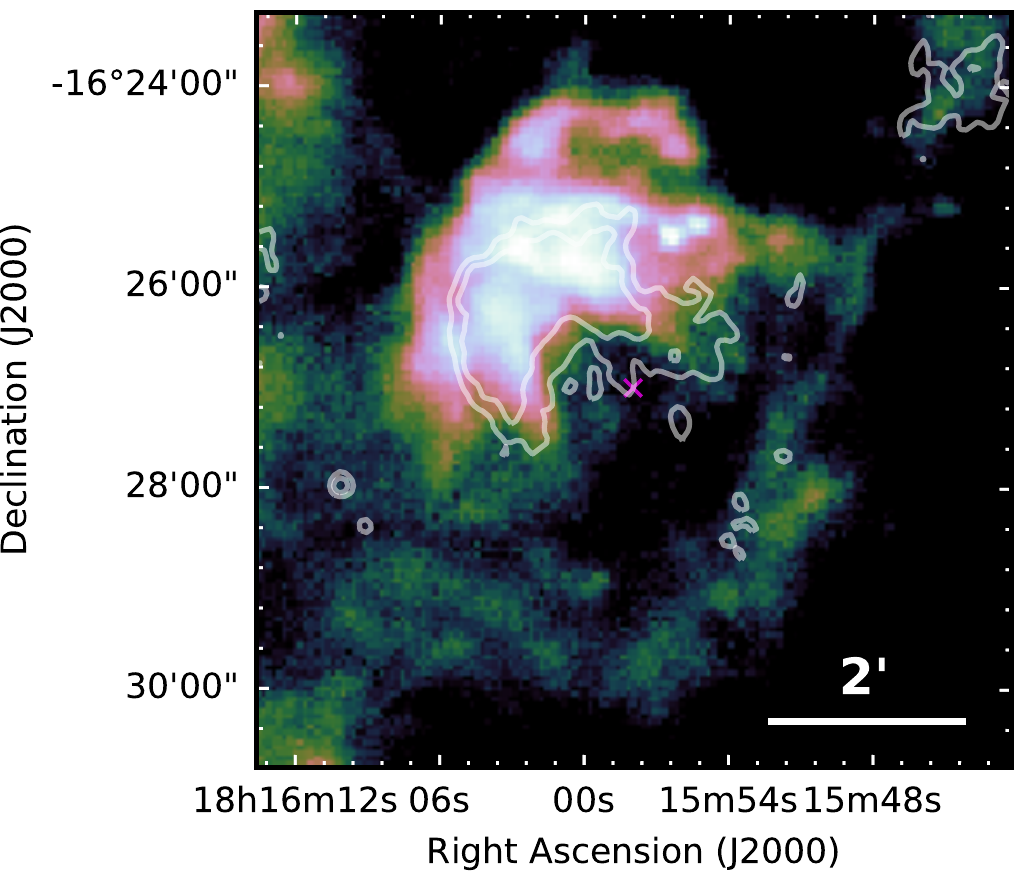}{}
	\caption{G14.3$+$0.1 - \textit{Herschel} 70\,$\mu$m image with 24\,$\mu$m contours overlaid.
	The magenta cross shows the radio coordinates of the SNR centre from \citet{Green2014}.}
	\label{fig:G14.3+0.1Image}
\end{figure}

\begin{figure}
	\centering
	\includegraphics[width=1.0\linewidth]{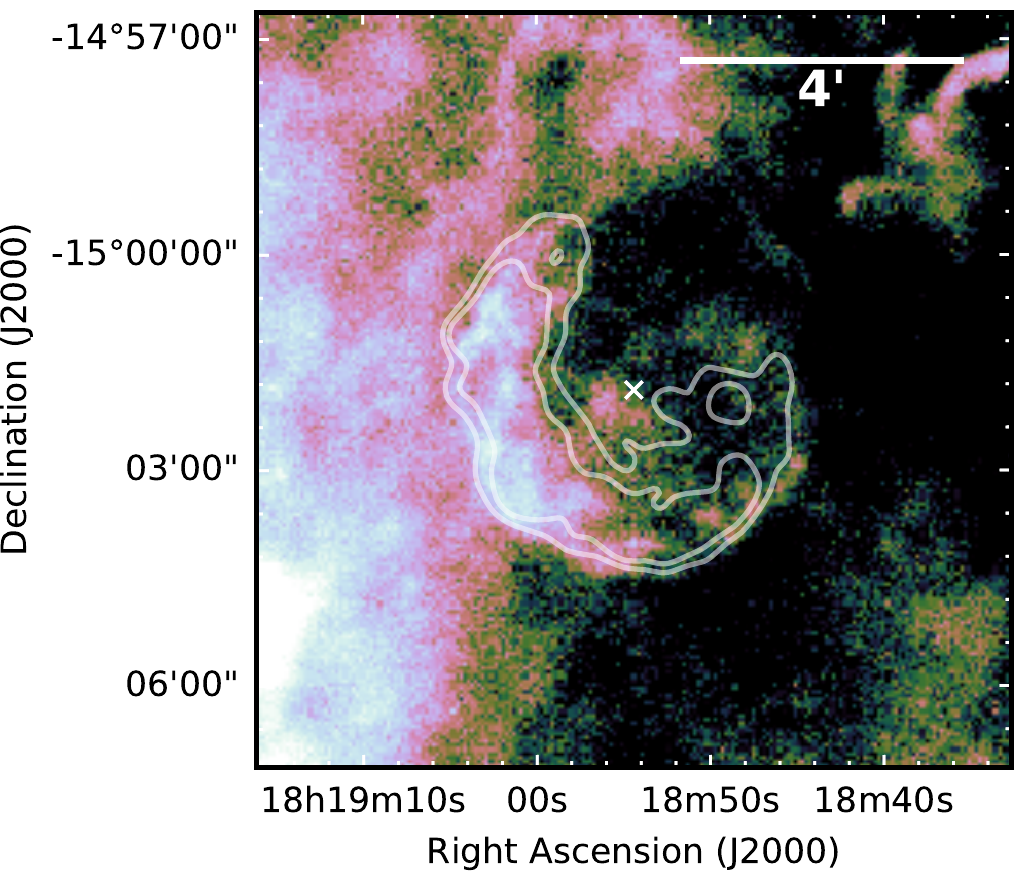}{}
	\caption{G15.9$+$0.2 - \textit{Herschel} 70\,$\mu$m image with X-ray contours overlaid. Dust is detected at 70\,$\mu$m in a partial shell which is brightest to the east.
	The white cross shows the radio coordinates of the SNR centre from \citet{Green2014}.}
	\label{fig:G15.9+0.2Image}
\end{figure}

\begin{figure}
	\centering
	\includegraphics[width=1.0\linewidth]{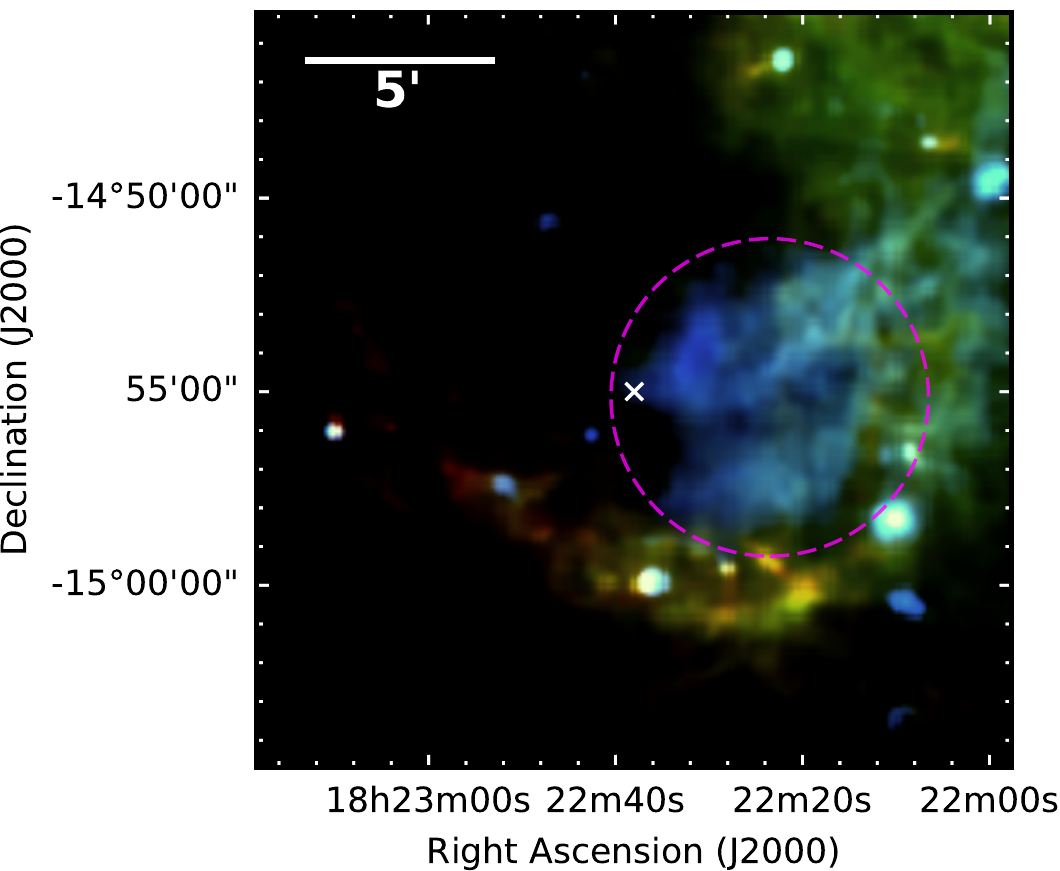}{}
	\caption{G16.4$-$0.5 - \textit{Herschel} three colour image. Diffuse dust emission is detected in the central region in blue as indicated by the magenta circle. A ridge of dust is detected at wavebands of 250\,$\mu$m and greater along the southern edge of the remnant which probably is not associated.
	The white cross shows the radio coordinates of the SNR centre from \citet{Green2014}.}
	\label{fig:G16.4-0.5Image}
\end{figure}

\begin{figure}
	\centering
	\includegraphics[width=1.0\linewidth]{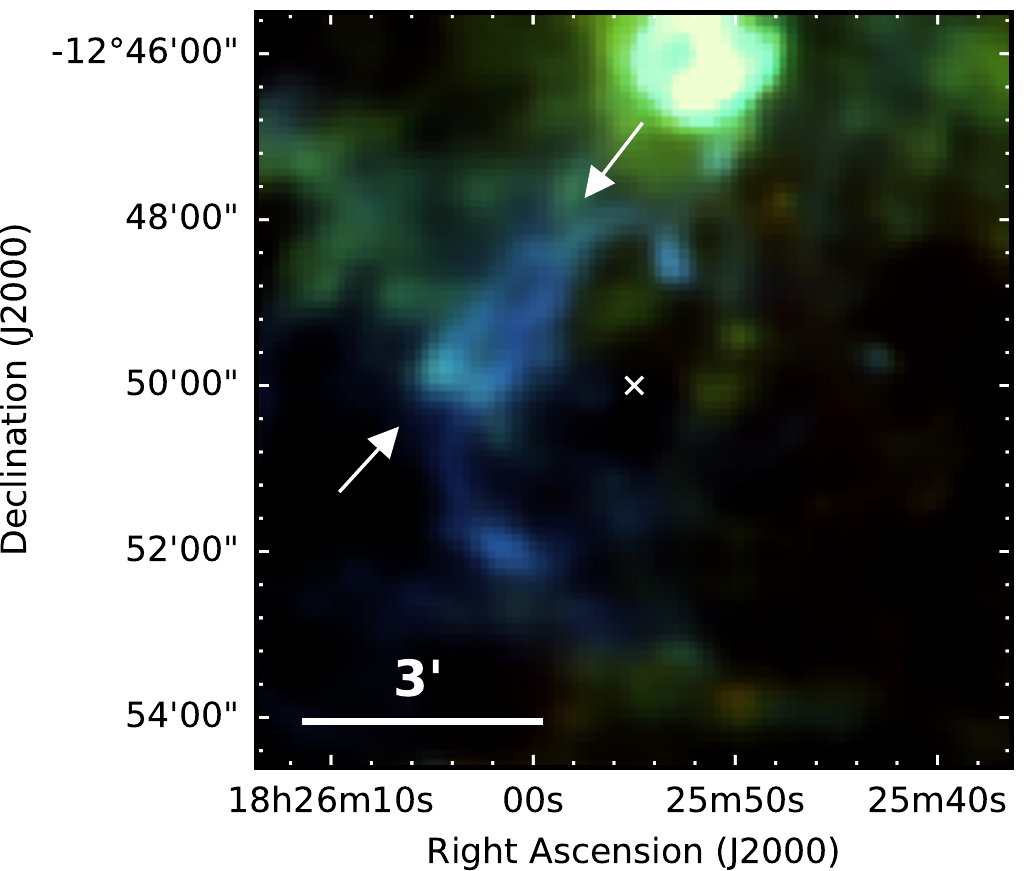}{}
	\caption{G18.6$-$0.2 - \textit{Herschel} three colour image. A partial shell of dust is detected at the eastern edge of this remnant, at the same location as radio structures. Two parallel filaments of dust are detected in the north-eastern shell, between the two arrows.
	The white cross shows the radio coordinates of the SNR centre from \citet{Green2014}.}
	\label{fig:G18.6-0.2Image}
\end{figure}

\begin{figure}
	\centering
	\includegraphics[width=1.0\linewidth]{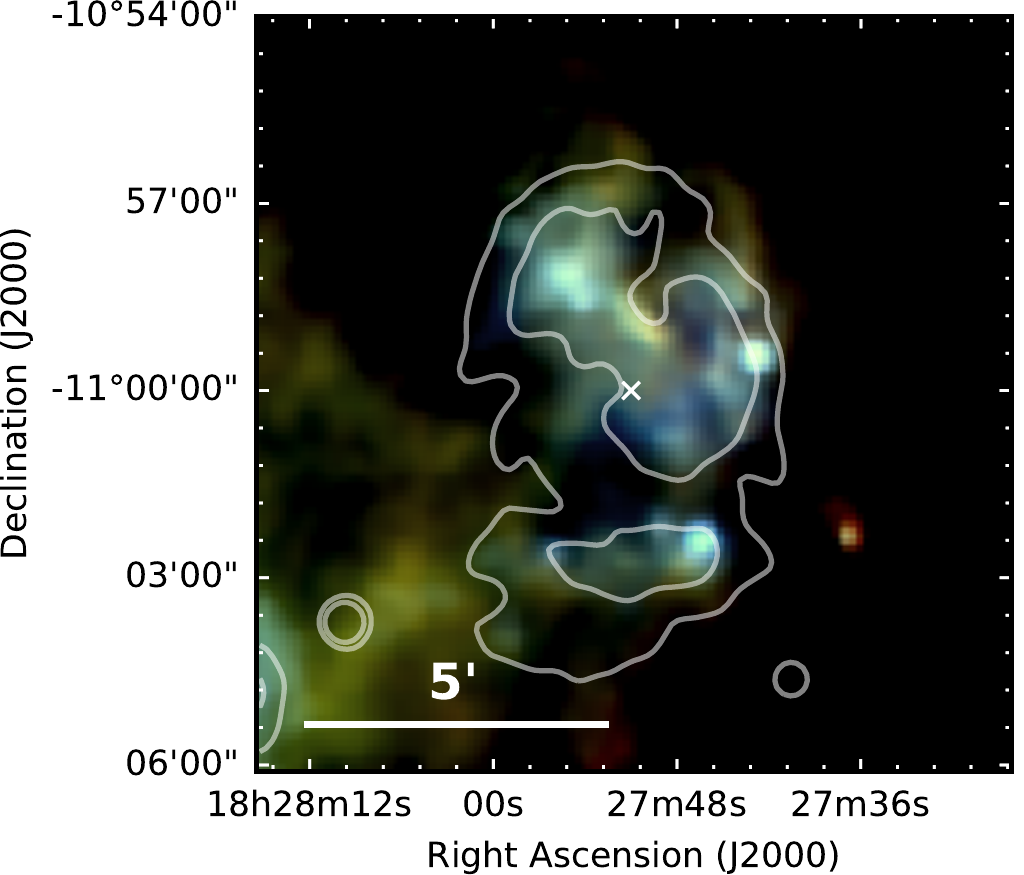}{}
	\caption{G20.4$+$0.1 - \textit{Herschel} three colour image with 20\,cm VLA contours overlaid. FIR emission is detected within the contours at all \textit{Herschel} wavelengths.
	The white cross shows the radio coordinates of the SNR centre from \citet{Green2014}.}
	\label{fig:G20.4+0.1Image}
\end{figure}

\begin{figure*}
	\centering
	\includegraphics[width=1.0\linewidth, trim = 0cm 5.7cm 0cm 6.8cm, clip]{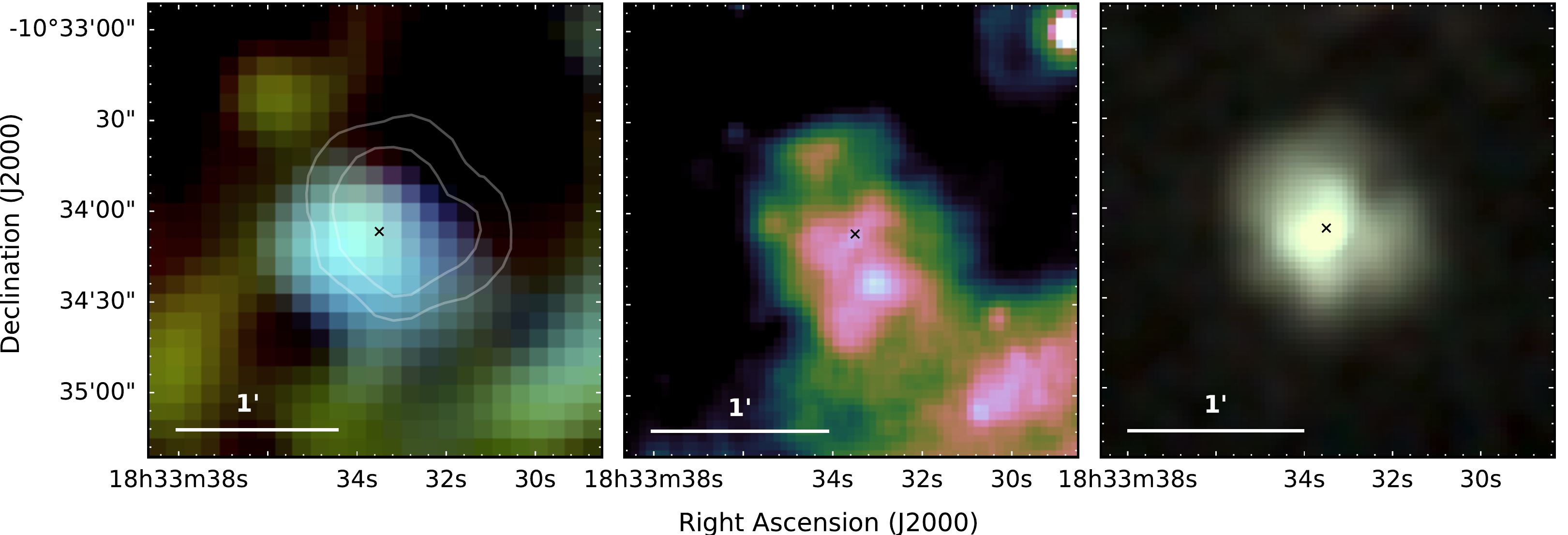}{}
	\caption{G21.5$-$0.9 -
	\textit{Left}: \textit{Herschel} three colour image with X-ray contours overlaid to show the location of the pulsar and PWN. Dust is observed in a clump in the central area, at the location of the PWN.
	\textit{Middle}:  \textit{Spitzer}  MIPS 24~$\mu$m image.
	\textit{Right}: \textit{Chandra} colour image (red = 1.0 - 2.1 keV, green = 2.1 - 4.0 keV, and blue = 4.0 - 10.0 keV).
	The black crosses shows the X-ray coordinates of the SNR centre.}
	\label{fig:G21.5-0.9Image}
\end{figure*}

\begin{figure*}
	\includegraphics[width=1.0\linewidth, trim = 0cm 3.2cm 0cm 3.5cm, clip]{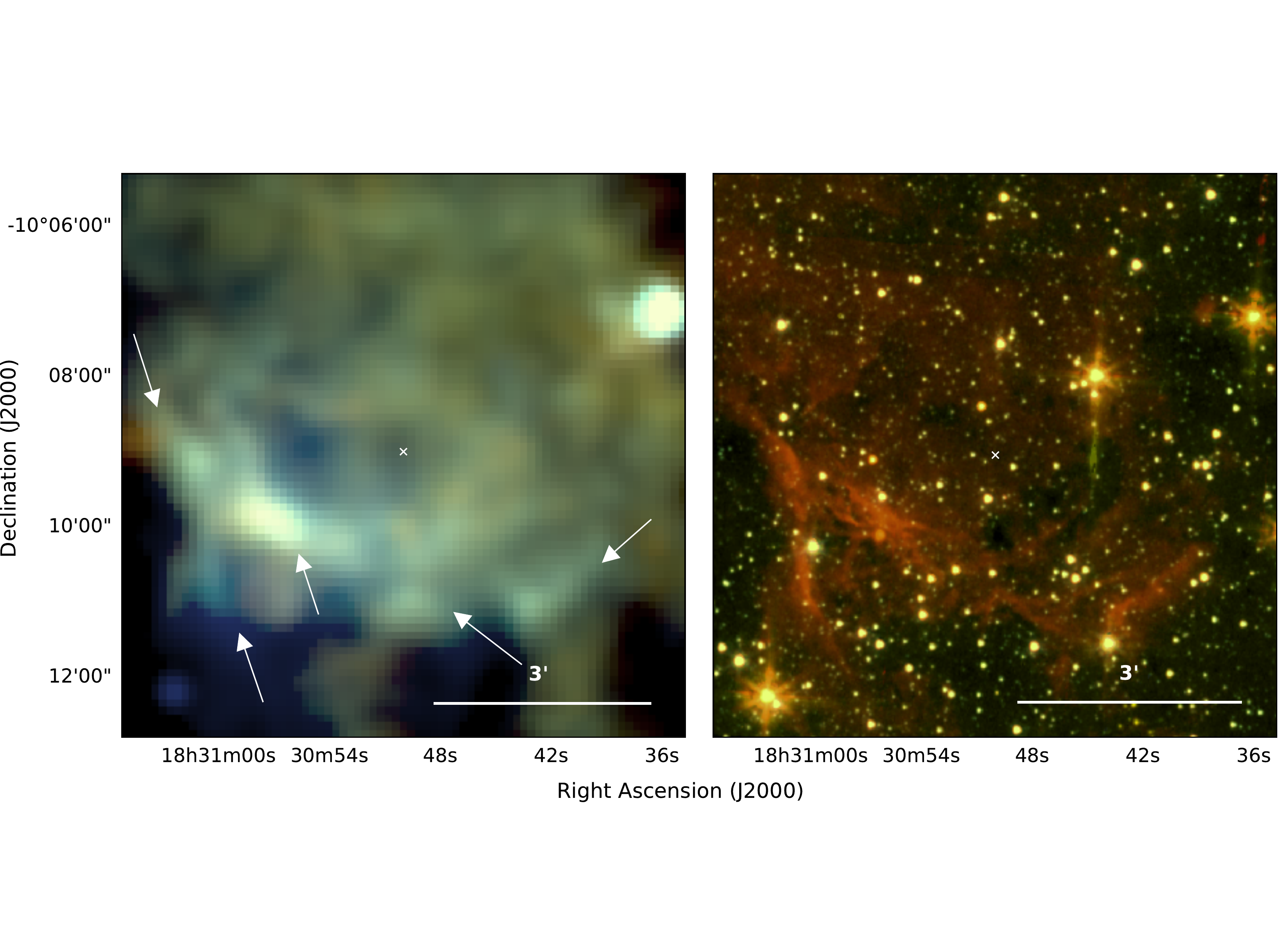}{}
	\caption{G21.5$-$0.1 - \textit{Left}: \textit{Herschel} three colour image. Dust is observed in the central region in a filled shell. The arrows indicate filaments of dust which correlate with the IRAC detection by \citet{Reach2006}.
	\textit{Right}:  \textit{Spitzer}  IRAC four colour image, colours are red = 8.0~$\mu$m, yellow = 5.8~$\mu$m, green = 4.5~$\mu$m, and blue = 3.6~$\mu$m.
	The white cross shows the radio coordinates of the SNR centre from \citet{Green2014}.}
	\label{fig:G21.5-0.1Image}
\end{figure*}

\begin{figure*}
	\centering
	\includegraphics[width=1.0\linewidth]{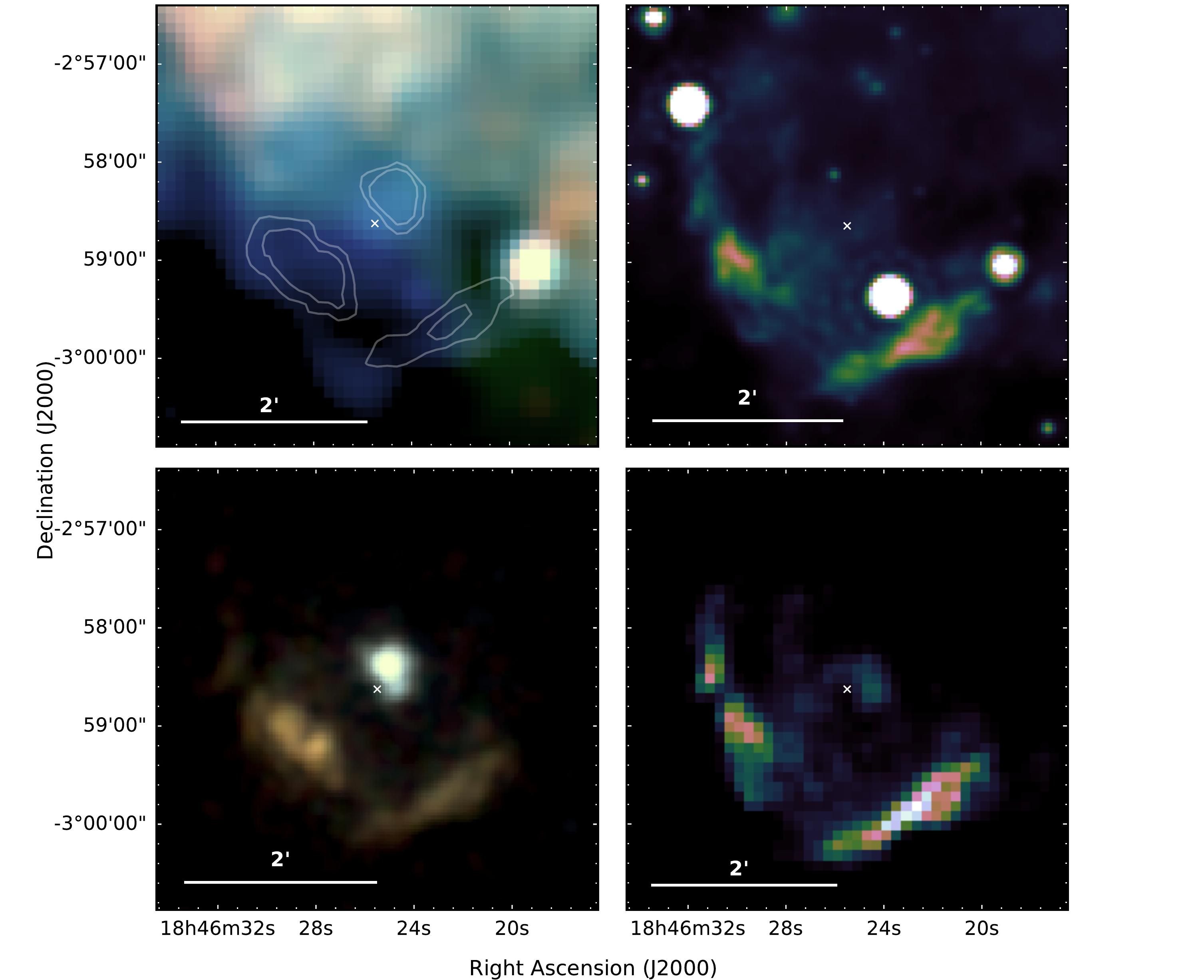}{}
	\caption{G29.7$-$0.3 (Kes 75) -\textit{Top-left}: \textit{Herschel} three colour image with X-ray contours from \textit{Chandra} overlaid. Dust is detected in a central clump at the same location as the pulsar and PWN.
	\textit{Top-right}:  \textit{Spitzer}  MIPS 24~$\mu$m image.
	\textit{Bottom-left}: \textit{Chandra} colour image (red = 1.0 - 1.7 keV, green = 1.7 - 2.6 keV, and blue = 2.6 - 8.0 keV).
	\textit{Bottom-right}: VLA 20\,cm radio image.
	The white crosses shows the X-ray coordinates of the SNR centre.}
	\label{fig:G29.7-0.3Image}
\end{figure*}

\begin{figure}
	\centering
	\includegraphics[width=1.0\linewidth]{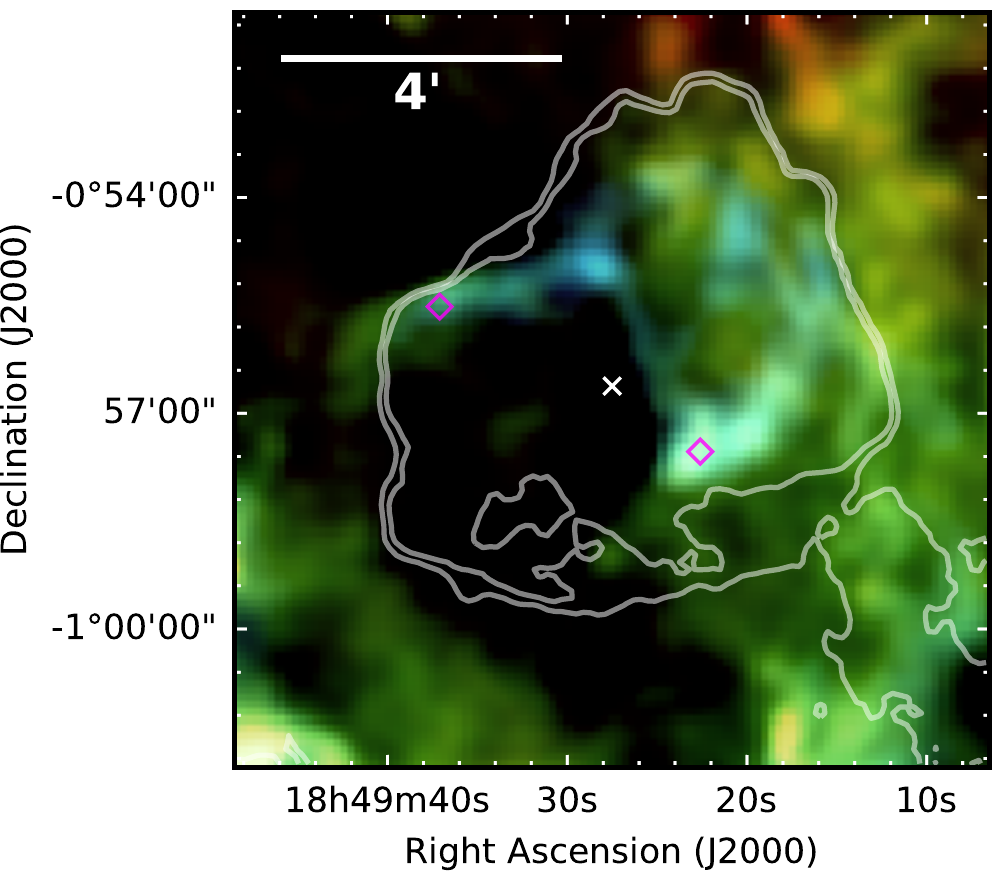}
	\caption{SNR 3C391 (G31.9$+$0.0) - \textit{Herschel} three colour image with X-ray contours overlaid. The two diamonds indicate the locations of the two OH masers \citep{Frail1996}. Dust is detected at the outer edges of the shock in a semi-circular shell.
	The white cross shows the X-ray coordinates of the SNR centre.}
	\label{fig:G31.9+0.0Image}
\end{figure}

\begin{figure}
	\includegraphics[width=1.0\linewidth]{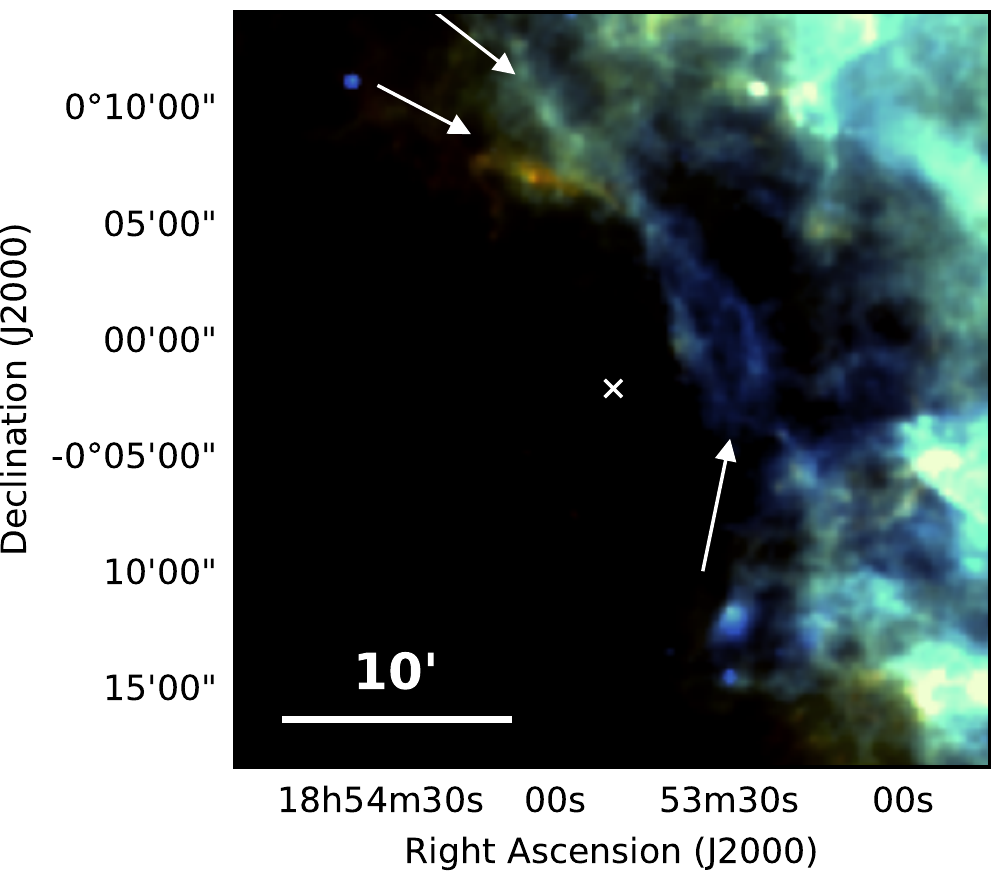}
	\caption{G33.2$-$0.6 - \textit{Herschel} three colour image. Dust is detected in filaments in an arc towards the western shell of the SNR, indicated by the lower arrow.
	Two filaments seem to extend to from this arc as indicated by the upper arrows; a filament extending to the north is detected in all \textit{Herschel} wavebands, and a cooler filament extending to the north-east is detected in all but 70\,$\mu$m.
	The white cross shows the radio coordinates of the SNR centre.}
	\label{fig:G33.2-0.2Image}
\end{figure}

\begin{figure}
	\centering
	\includegraphics[width=1.0\linewidth]{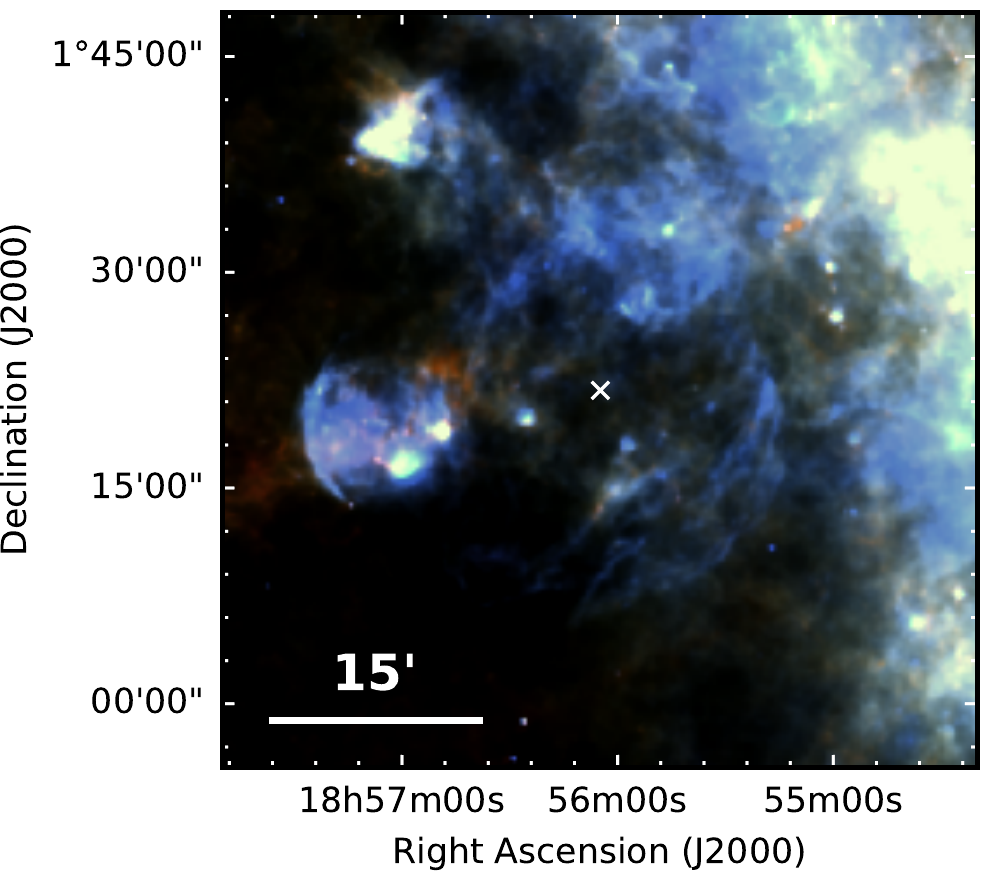}
	\caption{SNR W44 (G34.7$-$0.4) - \textit{Herschel} three colour image. Dust is seen in filaments at the outer edges of the shock and is brightest to the west.
	The bright patch of emission to the east is likely an H\,{\sc ii} region \citep{Rho1994}.
	The white cross shows the X-ray coordinates of the SNR centre. 
	}
	\label{fig:G34.7-0.4Image}
\end{figure}

\begin{figure}
	\centering
	\includegraphics[width=1.0\linewidth, trim = 0cm 0cm 2.5cm 0cm, clip]{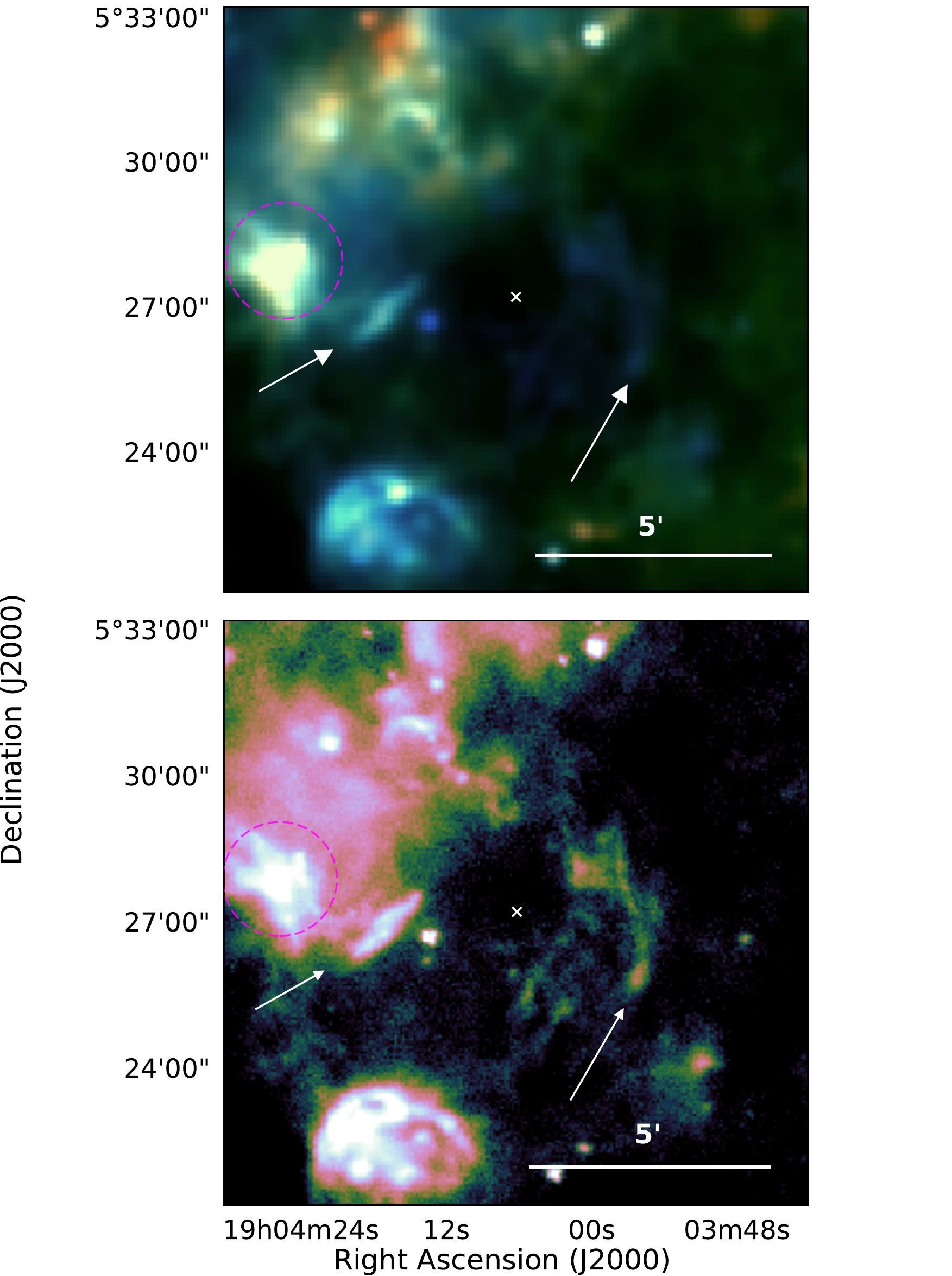}
	\caption{SNR 3C 396 (G39.2$-$0.3) -
	\textit{Top:} \textit{Herschel} three colour image with radio (20\,cm) contours overlaid.
	\textit{Bottom:} \textit{Herschel} 70\,$\mu$m image.
	Within the magenta circle, FIR emission is detected at the location of a blowout tail detected at 1465\,MHz (\citealp{Patnaik1990}).
	Filamentary dust is detected at the western outer edges of the shocks and in regions of very high radio polarisation, as indicated by the arrows.
	The white cross shows the X-ray coordinates of the SNR centre.}
	\label{fig:G39.2-0.3Image}
\end{figure}

\textit{G10.5-0.0:} \citet{Brogan2006} detected a partial radio shell at a wavelength of 90 cm from this remnant which has a potential X-ray counterpart \citep{Sugizaki2001}. PG11 classified this source as a level 1 MIR detection as there is 24~$\mu$m emission which roughly coincides with radio structures in places. There is FIR emission to the north of the 24~$\mu$m structure, close to a radio peak at $\alpha\,=\,18^\text{h}09^\text{m}02.4^\text{s}, \delta\,=\,-19^\circ48^\prime06.2''$ although we do not see a convincing likeness to the radio structure (Figure\,A2). There are also FIR peaks at $\alpha\,=\,18^\text{h}09^\text{m}07.4^\text{s}, \delta\,=\,-19^\circ46^\prime18''$ and $\alpha\,=\,18^\text{h}09^\text{m}06.8^\text{s}, \delta\,=\,-19^\circ47^\prime09.8''$ which do not correlate with any radio emission.
\bigskip

\textbf{G11.1$-$1.0} (Figure~\ref{fig:G11.1-1.0Image}): Unstudied by both R06 and PG11, we detect a shell of FIR dust emission in Figure~\ref{fig:G11.1-1.0Image}. This is brightest in the south-eastern region which is detected across all of the \textit{Herschel} bands. The structure is very similar to that in the 90 cm radio \citep{Brogan2006} and H$\alpha$ \citep{Stupar2011}, especially in the filaments near to $\alpha\,=\,18^\text{h}14^\text{m}29^\text{s}, \delta\,=\,-19^\circ43^\prime53''$, $\alpha\,=\,18^\text{h}14^\text{m}30^\text{s}, \delta\,=\,-19^\circ49^\prime00''$, and $\alpha\,=\,18^\text{h}13^\text{m}56^\text{s}, \delta\,=\,-19^\circ49^\prime38''$. There is also bright 24~$\mu$m and IRAC (5.8 and 8.0~$\mu$m) emission from the shell coinciding with the \textit{Herschel} emission, although the entire structure is not covered by the 4.5 and 8.0\,$\mu$m GLIMPSE bands or the 24\,$\mu$m band.
\bigskip

\textbf{G11.1$+$0.1} (Figure~\ref{fig:G11.1+0.1Image}): This region is very confused however, there is FIR dust emission in a partial shell structure detected to the east of this remnant which correlates with 90 cm radio emission detected by \citet{Brogan2004}. Emission is detected across all \textit{Herschel} bands from a clump centred at $\alpha\,=\,18^\text{h}09^\text{m}53^\text{s}, \delta\,=\,-19^\circ10^\prime21''$ and a much fainter extended filament near to $\alpha\,=\,18^\text{h}09^\text{m}46^\text{s}, \delta\,=\,-19^\circ15^\prime12''$. Unstudied by R06, this remnant was also a level 1 detection by PG11 who detected 24 $\mu$m emission at the locations of bright radio knots.
\bigskip

\textbf{G11.2$-$0.3} (Figure~\ref{fig:G11.2-0.3Image}): This core-collapse remnant has a composite radio morphology which is very similar to the X-ray shell \citep{Vasisht1996, Chevalier2005}. Also associated with the SNR is a central pulsar (AX J1811.5-1926) at $\alpha\,=\,18^\text{h}11^\text{m}29.22^\text{s}, \delta\,=\,-19^\circ25^\prime27.6''$ \citep{Kaspi2001} which has almost the same energy as expected at birth. X-ray morphology indicates that the surrounding PWN has been compressed and all ejecta reheated by the passage of the reverse shock \citep{Torii1997, Borkowski2016}.
Detection of 1.644\,$\mu$m [FeII] emission from the shell and knots surrounding the PWN indicates the presence of shocked CSM and ejecta material \citep{Koo2007, Moon2009}.

Expansion rates suggest that this is a young SNR with an estimated age of around 1400\,--\,2400 yrs \citep{Tam2003, Borkowski2016}. R06 suggested that, because of this young age, IRAC emission from shocked gas in filaments near $\alpha\,=\,18^\text{h}11^\text{m}35^\text{s}, \delta\,=\,-19^\circ26^\prime23''$ may originate from ejecta.
PG11 detected the same filaments as diffuse, unresolved MIPS emission, and a shell which correlates with X-ray structures. Almost identical to the X-ray and 24~$\mu$m structure, we detect a ring of dust emission which is especially bright at 70\,$\mu$m in Figure~\ref{fig:G11.2-0.3Image}. The southern rim of the SNR is the brightest region at 70\,$\mu$m, as well as in the X-ray and MIR (MIPS and IRAC). The emission in this region is much more confused at the longer \textit{Herschel} bands. There is also bright dust emission in the central region centred at $\alpha\,=\,18^\text{h}11^\text{m}29^\text{s}, \delta\,=\,-19^\circ25^\prime54''$ which coincides with X-ray emission from the central pulsar and its nebula. We suggest that this could be ejecta dust heated by the PWN. We checked \textsuperscript{12}CO (J = 3$\rightarrow$2) data from the CO High Resolution Survey (COHRS, \cite{Dempsey2013}) and found no detection from the SNR or surrounding ISM clouds.
\bigskip

\textbf{G14.1$-$0.1} (Figure~\ref{fig:G14.1-0.1Image}): PG11 detected a horse-shoe shape of 24\,$\mu$m emission roughly matching the radio shell structure \citep{Brogan2004}. There is diffuse 70\,$\mu$m emission to the north-east of the shell and a FIR peak coinciding with a radio peak at roughly $\alpha\,=\,18^\text{h}16^\text{m}41.6^\text{s}, \delta\,=\,-16^\circ39^\prime11''$.
\bigskip

\textbf{G14.3$+$0.1} (Figure~\ref{fig:G14.3+0.1Image}): We clearly observe 70\,$\mu$m emission coincident with the radio partial shell structure detected by \citet{Brogan2006}. There is emission in the longer \textit{Herschel} wavebands, however the structure is different and association with the SNR is unclear.
\bigskip

\textbf{G15.9$+$0.2} (Figure~\ref{fig:G15.9+0.2Image}): This relatively young source ($\leqslant$2400\,yr) is the remnant of a CCSN and contains the neutron star CXOU J1818 \citep{Reynolds2006, Klochkov2016}. Strong lines in the X-ray indicate the presence of ejecta. This source is undetected by IRAC; however, the 24\,$\mu$m structure closely correlates with the X-ray and radio. A partial shell of dust is detected around the eastern and south-eastern edge of this remnant at the location of the X-ray and radio emission.
\bigskip

\textbf{G16.4$-$0.5} (Figure~\ref{fig:G16.4-0.5Image}): Unstudied by R06, PG11 classified this SNR as a level 1 MIPS detection. This SNR has a partial radio shell morphology \citep{Brogan2006}. A region of diffuse dust emission at 70\,$\mu$m is detected towards the centre of the remnant, identified by a magenta circle in Figure\,A2, which corresponds to emission at 24~$\mu$m. A 4$^\prime$ long filament centred on $\alpha\,=\,18^\text{h}22^\text{m}17^\text{s}, \delta\,=\,-14^\circ52^\prime51''$ is detected across all five \textit{Herschel} wavebands and coincides with MIPS (PG11) and IRAC emission.

There is bright emission in all \textit{Herschel} bands along the southern ridge (red structure in Figure~\ref{fig:G16.4-0.5Image}). This emission seems to be of a similar temperature to, or cooler than, the surrounding ISM and association with the SNR is unlikely.
\bigskip

\textit{G17.4$-$0.1:} This SNR has a partial shell radio structure. The region is confused in the FIR and it is difficult to distinguish any SNR emission from that of the ISM. There is a bright FIR region to the west of the SNR which coincides with 24~$\mu$m emission and a radio peak detected by \citet{Brogan2006}, although the morphology is different and association is unclear.
\bigskip

\textbf{G18.6$-$0.2} (Figure~\ref{fig:G18.6-0.2Image}): \citet{Voisin2016} suggested that the pulsar PSR J1826$-$1256 may be associated with this remnant as their estimated distances are similar. Although there is dust emission in the region of the pulsar in Figure~\ref{fig:G18.6-0.2Image} at $\alpha\,=\,18^\text{h}26^\text{m}08.2^\text{s}, \delta\,=\,-12^\circ56^\prime46''$, it is indistinguishable from the local ISM and unclear as to whether any of this is associated to the PSR J1826$-$1256.
Dust emission is detected at 70\,$\mu$m from the eastern region of this shell-type SNR which is brightest from two parallel filaments detected near to $\alpha\,=\,18^\text{h}26^\text{m}00^\text{s}, \delta\,=\,-12^\circ49^\prime26''$ and $\alpha\,=\,18^\text{h}26^\text{m}02^\text{s}, \delta\,=\,-12^\circ49^\prime08''$. The morphology seems to correlate with 90 cm radio emission from \citet{Brogan2006} and the partial shell detected by PG11. At 160\,$\mu$m the region is too confused to determine if any emission is associated with the SNR, and in the longer \textit{Herschel} wavebands there is no evidence of SNR emission.
\bigskip

\textbf{G20.4$+$0.1} (Figure~\ref{fig:G20.4+0.1Image}): PG11 detected emission correlating with the radio shell of this SNR. We detect FIR emission at all \textit{Herschel} wavelengths which lies within the radio contours, as shown in Figure~\ref{fig:G20.4+0.1Image}.
\bigskip

\textbf{G21.5$-$0.9} (Figure~\ref{fig:G21.5-0.9Image}): This Crab-like remnant has a pulsar (PSR J1833-1034) at its centre with a non-thermal X-ray halo \citep{Camilo2005}. Properties of the PWN, the pulsar, and the shell suggest that the remnant is $\lesssim1000$ years old  \citep{Camilo2005} and H\,{\small I} and CO observations tell us that this SNR is at a distance of 4.8 kpc \citep{Tian2008}.
Emission at all \textit{Herschel} bands is detected at $\alpha\,=\,18^\text{h}33^\text{m}33.8^\text{s}, \delta\,=\,-10^\circ34^\prime14''$, slightly offset from the location of the central pulsar and its wind nebula as shown by X-ray contours in Figure~\ref{fig:G21.5-0.9Image}. This is more confused with the local environment at longer wavebands. PG11 also made a level 1 detection of the central region. We suggest that dust in this region is heated by the PWN. We checked \textsuperscript{13}CO (J = 1$\rightarrow$0) data from the Galactic Ring Survey (GRS, \cite{Jackson2006}) and found no CO detection towards the SNR.
\bigskip

\textbf{G21.5$-$0.1} (Figure~\ref{fig:G21.5-0.1Image}): Dust is observed in a filled shell from this remnant at all \textit{Herschel} wavelengths, which correlates with the 24~$\mu$m emission (PG11) and the 90 cm radio structure detected by \citet{Brogan2006}. Filaments of dust are indicated in Figure~\ref{fig:G21.5-0.1Image}. This includes a $\sim2.5^\prime$ bright filament detected to the east, centred near $\alpha\,=\,18^\text{h}31^\text{m}01^\text{s}, \delta\,=\,-10^\circ09^\prime54''$ which is also detected in the MIR. Another FIR detected filament is centred near $\alpha\,=\,18^\text{h}30^\text{m}41.5^\text{s}, \delta\,=\,-10^\circ10^\prime47''$ and is roughly 1.6$^\prime$ long. Although unstudied by R06, PG11 detected MIPS and IRAC emission along the southern ridge.

Nevertheless, the origin of this emission requires further study.
\citet{Anderson2017} suggests that this is a H\,{\sc ii} region which has been incorrectly classified as a SN structure.
This is because the entire structure coincides with the WISE H\,{\sc ii} region G21.560$-$0.108 and PG11 derived a high MIR to radio flux ratio indicative of H\,{\sc ii} regions
\bigskip

\textit{G23.6$+$0.3:} PG11 detected an elongated region of 24\,$\mu$m emission at the location of the SNR radio structure. However, they argue that the SNR morphology more closely resembles that of a H\,{\sc ii} region than a SNR, and the nature of this object should therefore be reconsidered.
There is FIR emission in the region of the radio structure, however this is offset (Figure\,A2).
\bigskip

\textit{G27.4$+$0.0:} This shell-type SNR is thought to have been produced by a very massive progenitor ($\gtrsim 20 M_\odot$) between 750 and 2100 years ago \citep{Kumar2014}. Although there is a very good detection at 24 $\mu$m, the region is confused at \textit{Herschel} wavelengths (Figure \ref{fig:G27.4+0.0Image}). We find some evidence of dust emission at 70$\mu$m in the region which may be associated with X-ray and 24$\mu$m SNR filaments. However, we cannot definitively conclude this as there is extensive interstellar dust emission to the west confusing the region at FIR wavelengths.
\bigskip

\textit{G29.6$+$0.1:} This is a young remnant \citep[< 8000 yrs,][]{Gaensler1999} with a non-thermal radio shell and an associated compact source (AX J1845$-$0258) which is likely a pulsar \citep{Gaensler1999, Vasisht2000}. Broad molecular lines have been detected towards the remnant, suggesting that it is interacting with a molecular cloud (\citealp{Kilpatrick2016}).
Numerous young stellar objects are detected in this field, including FIR bright sources at $\alpha\,=\,18^\text{h}44^\text{m}51^\text{s}, \delta\,=\,-02^\circ55^\prime18''$, $\alpha\,=\,18^\text{h}44^\text{m}53.3^\text{s}, \delta\,=\,-02^\circ56^\prime03''$, and $\alpha\,=\,18^\text{h}44^\text{m}49.2^\text{s}, \delta\,=\,-02^\circ58^\prime15''$ \citep{Veneziani2013}. There is also FIR emission across all \textit{Herschel} and  \textit{Spitzer}  bands coincident with the radio source at $\alpha\,=\,18^\text{h}44^\text{m}55.1^\text{s}, \delta\,=\,-02^\circ55^\prime36.9''$ \citep{Gaensler1999}. The radio shell is not detected by IRAC or MIPS, although at \textit{Herschel} wavelengths there is emission which may be associated (Figure\,A2). However, contamination from local ISM to the south-west makes it difficult to distinguish SNR emission.
\bigskip

\textbf{G29.7$-$0.3} (Figure~\ref{fig:G29.7-0.3Image}): It is possible that this SNR resulted from a Wolf-Rayet star which exploded as a type Ib/c SN  after clearing a $\sim\,10$ pc bubble \citep{Morton2007}. CO observations of an associated molecular cloud puts the remnant at a kinematic distance of $\sim\,10.6$ kpc, at the far side of the Sagittarius arm \citep{Su2009}. Although there is bright MIPS emission from this remnant in both a partial shell and central region (level 1 by PG11), there is no clear emission at IRAC wavelengths from the SNR (level 3 by R06). It was suggested that the lack of IRAC emission from the shell is due to shock destruction of small dust grains (\citealp{Morton2007}).
At 70\,$\mu$m in Figure~\ref{fig:G29.7-0.3Image} we detect a region of diffuse dust emission centred at $\alpha\,=\,18^\text{h}46^\text{m}24.9^\text{s}, \delta\,=\,-02^\circ58^\prime30''$, at the same location as X-ray emission from the PWN and the central pulsar \citep{Helfand2003}. At the longer \textit{Herschel} wavelengths there is too much confusion to distinguish if there is any cold dust in the SNR. There is FIR emission from the shell region coinciding with a bright 24~$\mu$m structure, however this is much less bright and difficult to detect in some regions.
The object detected at $\alpha\,=\,18^\text{h}46^\text{m}19^\text{s}, \delta\,=\,-02^\circ59^\prime03''$ is a young stellar object \citep{Veneziani2013}.
\bigskip

\textbf{3C 391, G31.9$+$0.0} (Figure~\ref{fig:G31.9+0.0Image}): This is a young mixed-morphology SNR with an incomplete radio shell structure \citep{Goss1979}.
Both R06 and PG11 classified this SNR as  a level 1 detection. R06 found MIR emission originating from shocked molecular gas at $\alpha\,=\,18^\text{h}49^\text{m}23^\text{s}, \delta\,=\,-00^\circ57^\prime38''$ and $\alpha\,=\,18^\text{h}49^\text{m}29^\text{s}, \delta\,=\,-00^\circ55^\prime00''$, at the ends of the SNR's bright semicircular radio shell. The southern patch is coincident with one of two 1720 MHz OH masers detected in the SNR where the remnant seems to be breaking into the edge of a molecular cloud (\citealp{Frail1996}).
The MIR patches are well detected by \textit{Herschel} in Figure~\ref{fig:G31.9+0.0Image}; the southern patch is detected at all \textit{Herschel} bands, however the northern patch is detected only at 70 and 160\,$\mu$m. Beyond these points, FIR emission extends into an arc around the northwestern shell in all \textit{Herschel} bands, although there is confusion at bands other than 70\,$\mu$m.

R06 discussed a bar of shocked, ionised gas from which both R06 and PG11 concluded there was a contribution of [Fe II] 5.34\,$\mu$m line emission. This bar coincides with the brightest part of the radio shell and is also detected in the FIR at 70 and 160\,$\mu$m near to $\alpha\,=\,18^\text{h}49^\text{m}16^\text{s}, \delta\,=\,-00^\circ55^\prime03''$. The FIR emission is likely associated with dust rather than line emission considering the high luminosity of dust at 70\,--\,160\,$\mu$m as in the cases of the Crab, Cas A \citep{Gomez2012b, DeLooze2017}, and three SNRs studied ourselves in Section\,\ref{DustMasses}.

\bigskip

\textbf{G33.2$-$0.6} (Figure~\ref{fig:G33.2-0.2Image}): A partial shell is detected to the western edge of this remnant which has a higher temperature than the surrounding medium as seen in Figure~\ref{fig:G33.2-0.2Image}. This arc corresponds to the 1465 MHz radio structure detected by \citet{Dubner1996}. Most noticeable are two filaments; the inner one is $\sim5^\prime$ with a midpoint near $\alpha\,=\,18^\text{h}53^\text{m}35^\text{s}, \delta\,=\,-00^\circ00^\prime36''$, and the outer filament is $\sim15.5^\prime$ near to $\alpha\,=\,18^\text{h}53^\text{m}43^\text{s}, \delta\,=\,+00^\circ04^\prime22''$. These filaments are detected at \textit{Spitzer} wavelengths, although neither R06 nor PG11 classified this source as a clear detection. FIR emission is also detected at the location of the compact IRAS source (IRAS 18509$-$0015) to the south-west of the SNR at $\alpha\,=\,18^\text{h}53^\text{m}29^\text{s}, \delta\,=\,-00^\circ12^\prime03''$. To the north of the remnant there are two filaments detected, one cooler which is visible in red in Figure~\ref{fig:G33.2-0.2Image}, and another warmer filament extending to the north which is visible in blue and is detected in all \textit{Herschel} wavebands. Both of these filaments are outside of the region of radio emission, we therefore cannot confidently determine whether these are associated with the remnant.
\bigskip

\textit{Kes 79, G33.6$+$0.1:} This mixed-morphology SNR is most likely in the Sedov-Taylor phase of evolution, with an age of 4.4\,--\,6.7 kyr \citep{Zhou2016}. CO observations suggest that the remnant is interacting with molecular clouds to the east \citep[e.g.][]{Green1992, Zhou2016}. PG11 detected 24~$\mu$m emission corresponding to the X-ray filaments to the east and the brightest X-ray contours.

Although there are FIR features which may correspond to the SNR (Figure\,A2) this isn't clear; the region is confused and there is extensive FIR emission from the local ISM.
A circle of FIR emission of radius $\sim$\,0.66$^\prime$ at $\alpha\,=\,18^\text{h}52^\text{m}39^\text{s}, \delta\,=\,+00^\circ41^\prime59''$ coincides with an IR bubble \citep{Simpson2012} and an X-ray dark region. There are also FIR point sources corresponding to MIR sources which may be protostars \citep{Reach2006} in an infrared dark cloud along the eastern edge. There is no dust emission in the region of the X-ray point source.
\bigskip

\textbf{W44, G34.7$-$0.4} (Figure~\ref{fig:G34.7-0.4Image}): This large core-collapse SNR has a well defined radio shell which is centrally filled by thermal X-ray emission \citep{Jones1993, Rho1994} and was a level 1 detection by both R06 and PG11. Similar to the MIR structure, the FIR dust emission traces the radio elliptical shell, although the FIR is fainter towards the south.
At MIR wavelengths shocked H$_2$ is detected along the eastern border \citep{Reach2005a} where the SNR is interacting with a molecular cloud \citep{Giacani1997, Yusef-Zadeh2003, Reach2006}. We detect 70 and 160\,$\mu$m emission at this edge which is possibly from shock-heated dust.
We also detect bright FIR filaments at the western edge, closely following the radio structure.
There is some potential detection of emission at the longer \textit{Herschel} wavebands, however the emission is much more confused. The patch of bright emission to the east, centred at $\alpha\,=\,18^\text{h}57^\text{m}05^\text{s}, \delta\,=\,+01^\circ18^\prime40''$ is probably an H\,{\sc ii} region \citep{Rho1994}. We do not detect emission in the region of the pulsar, PSR B1853$+$01, which is located at $\alpha\,=\,18^\text{h}56^\text{m}10.65^\text{s}, \delta\,=\,+01^\circ13^\prime21.3''$.
\bigskip

\textit{G35.6$-$0.4:} Recently re-classified as a SNR rather than a H\,{\sc ii} region, this shell-type remnant is likely around 30,000\,yrs old and at a distance of 3.6\,$\pm$\,0.4\,kpc \citep{Green2009, Zhu2013}. PG11 classed this as a level 1 detection although there is not a convincing likeness between IR and radio emission.
There is a FIR structure to the south-east of the SNR which does not have a similar morphology to radio emission \citep{Zhu2013}. A 3.5$^\prime$ rim of dust at $\alpha\,=\,18^\text{h}58^\text{m}12^\text{s}, \delta\,=\,+02^\circ08^\prime47''$ may be associated with the radio structure, although this is unclear.

FIR emission is detected from the planetary nebula PN G35.5$-$0.4 at $\alpha\,=\,18^\text{h}57^\text{m}59.5^\text{s}, \delta\,=\,+02^\circ07^\prime07''$ whose distance has been estimated as 3.8\,$\pm$\,0.4\,kpc \citep{Zhu2013}. There is also dust emission coinciding with a molecular clump towards the centre of the gamma ray source HESS J1858+020 at $\alpha\,=\,18^\text{h}58^\text{m}21^\text{s}, \delta\,=\,+02^\circ05^\prime12''$ \citep{HESS2008c}. A YSO embedded in this clump probably belongs to a larger molecular cloud which is interacting with G35.6$-$0.4 and a nearby H\,{\sc ii} region \citep{Paron2011}.
\bigskip

\textbf{3C 396, G39.2$-$0.3} (Figure~\ref{fig:G39.2-0.3Image}): There was a level 1 detection of this SNR by both R06 and PG11. Similar to the MIR detection described by R06, FIR emission from this SNR is mainly detected from three regions:

A bright region of emission in Figure~\ref{fig:G39.2-0.3Image} at $\alpha\,=\,19^\text{h}04^\text{m}26^\text{s}, \delta\,=\,+05^\circ27^\prime55''$ (within the magenta circle) coincides with the radio blowout tail (see \citealp{Patnaik1990} Figures 1 and 2 ) which extends out of the east shell and over the top of the SNR. FIR emission from the tail is detected in the five \textit{Herschel} bands. High radio polarisation and bright emission in the longer \textit{Herschel} channels is consistent with synchrotron dominated FIR emission in the region. \cite{Cruciani2016} found a significant correlation between the FIR and radio, 1.5\,GHz, emission in this region, although they could not rule out that this is due to diffuse interstellar emission.

Filamentary emission is detected on the western side of this remnant, near to $\alpha\,=\,19^\text{h}03^\text{m}56^\text{s}, \delta\,=\,+05^\circ25^\prime46$, across all of the \textit{Herschel} bands, although at bands longer than 70\,$\mu$m this is very confused with ISM emission. It is suggested that MIR emission at the same location originates from shocked, ionised gas \citep{Reach2006} suggesting that the FIR emission in this region could originate from warm, shocked dust.

FIR emission is detected at all \textit{Herschel} wavelengths to the east of the SNR in a region of very high radio polarisation. At 70 and 160\,$\mu$m this is resolved into two filaments at $\alpha\,=\,19^\text{h}04^\text{m}17^\text{s}, \delta\,=\,+05^\circ27^\prime07''$ and $\alpha\,=\,19^\text{h}04^\text{m}19^\text{s}, \delta\,=\,+05^\circ26^\prime33''$ (bottom panel of Figure\,\ref{fig:G39.2-0.3Image}).
\bigskip

\begin{figure}
	\centering
	\includegraphics[width=1.0\linewidth]{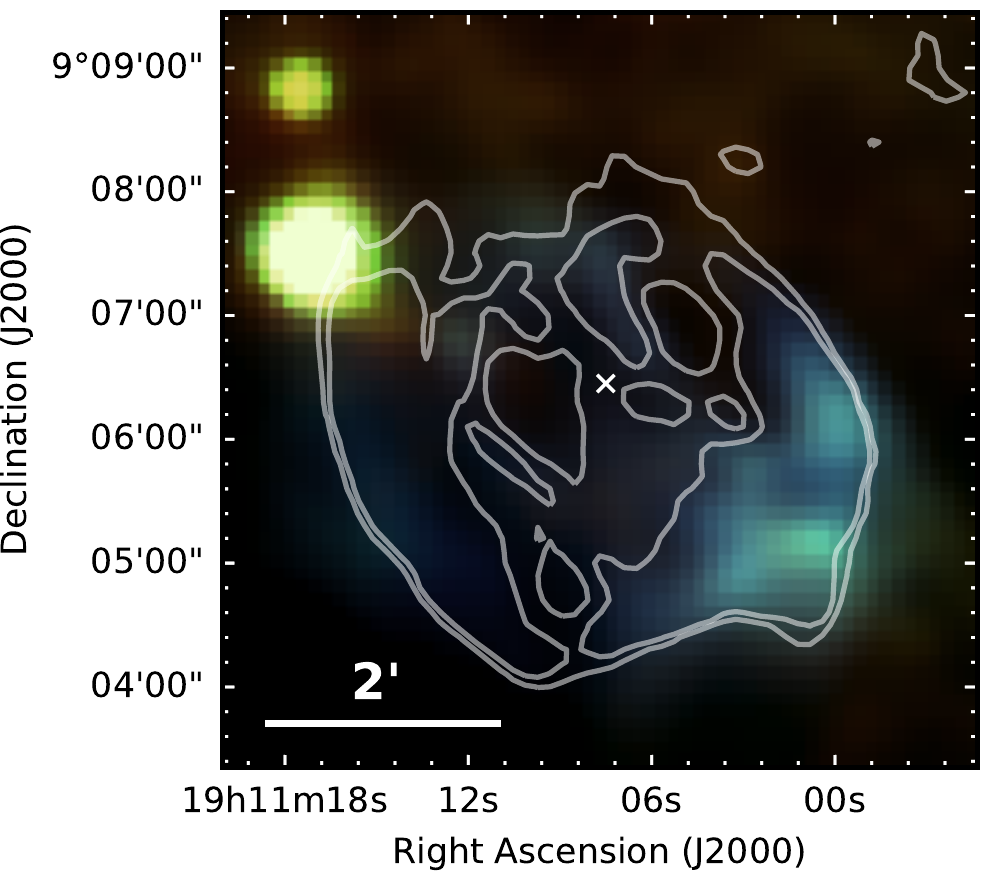}
	\caption{W49B (G43.3$-$0.2) - \textit{Herschel} three colour image overlaid with radio 20\,cm contours. Shock heated dust is seen in filaments in a barrel-hoop structure.
	The white cross shows the X-ray coordinates of the SNR centre.}
	\label{fig:G43.3-0.2Image}
\end{figure}

\begin{figure}
	\centering
	\includegraphics[width=1.0\linewidth]{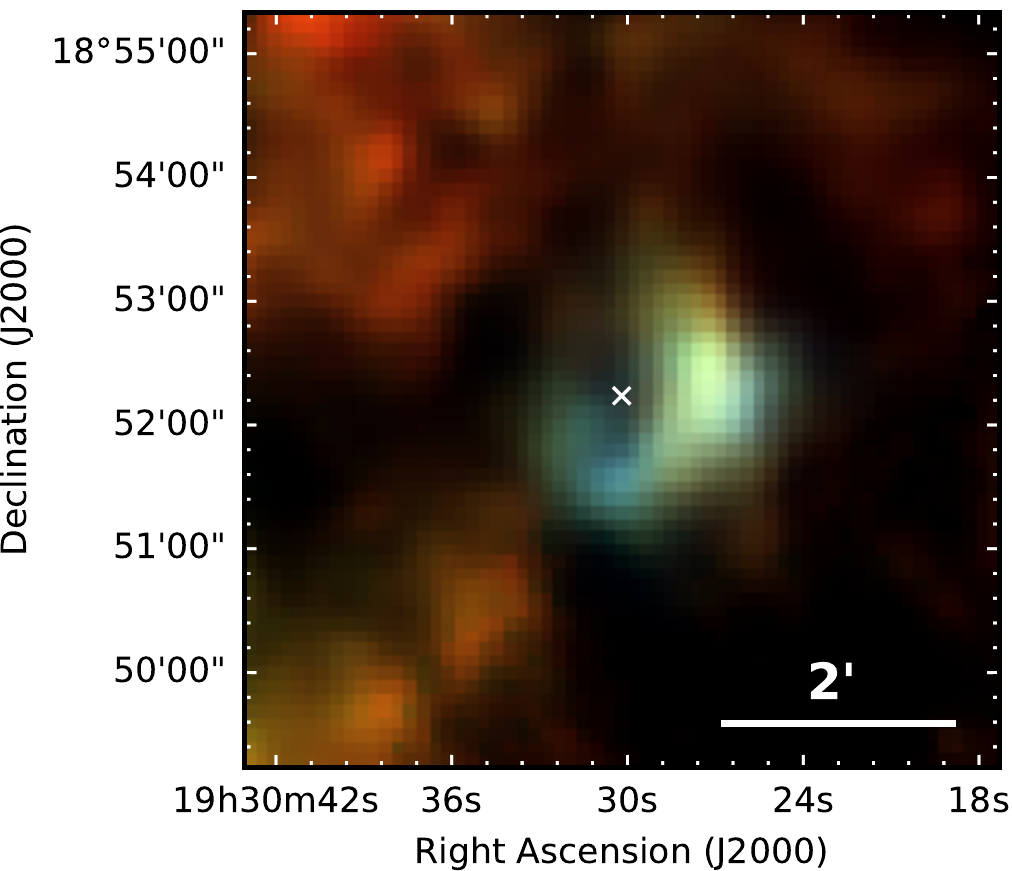}
	\caption{G54.1$+$0.3 - \textit{Herschel} three colour image with X-ray contours from \textit{Chandra} overlaid. Dust is observed in a central region at the location of the PWN.
	The white cross shows the X-ray coordinates of the SNR centre.
	For details on the discovery of dust in this source see \citet{Temim2017} and \citet{Rho2018}.
	}
	\label{fig:G54.1+0.3Image}
\end{figure}

\begin{figure}
	\centering
	\includegraphics[width=1.0\linewidth]{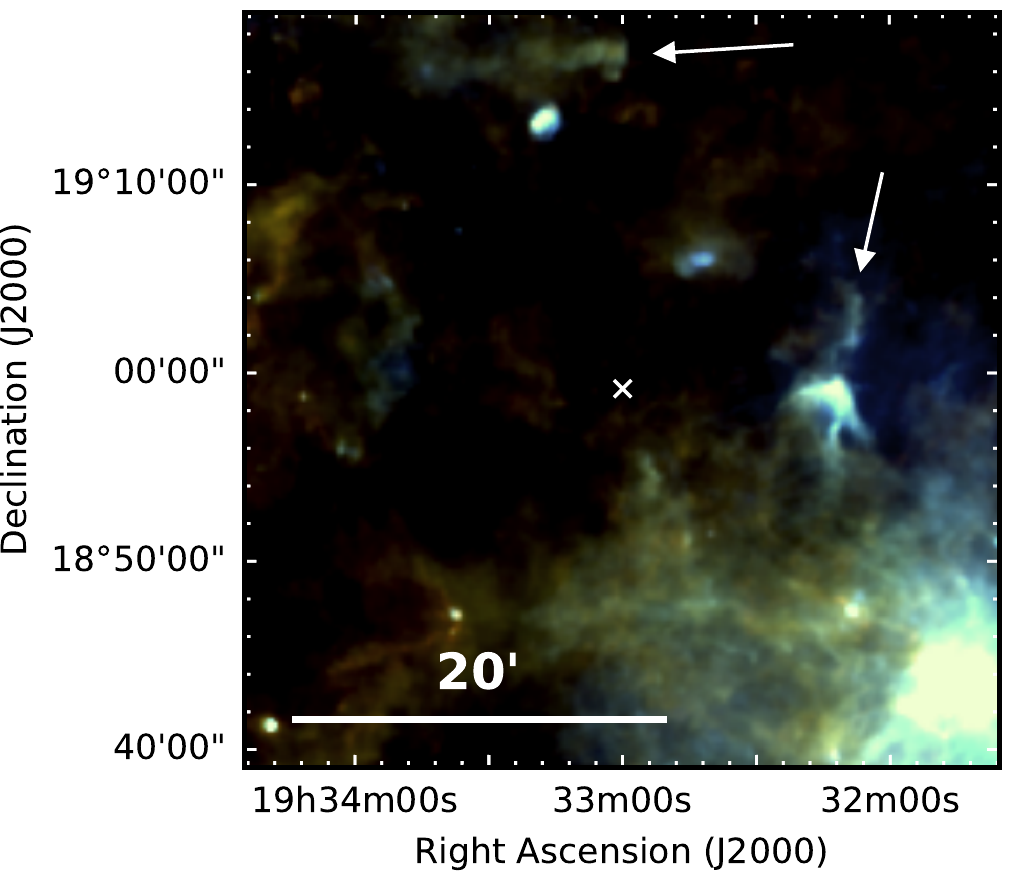}
	\caption{G54.4$-$0.3: \textit{Herschel} three colour image. Filaments of dust are observed at the outer edges of the shocks, as indicated by the white arrows.
	The white cross shows the X-ray coordinates of the SNR centre.}
	\label{fig:G54.4-0.3Image}
\end{figure}

\begin{figure}
	\centering
	\includegraphics[width=1.0\linewidth]{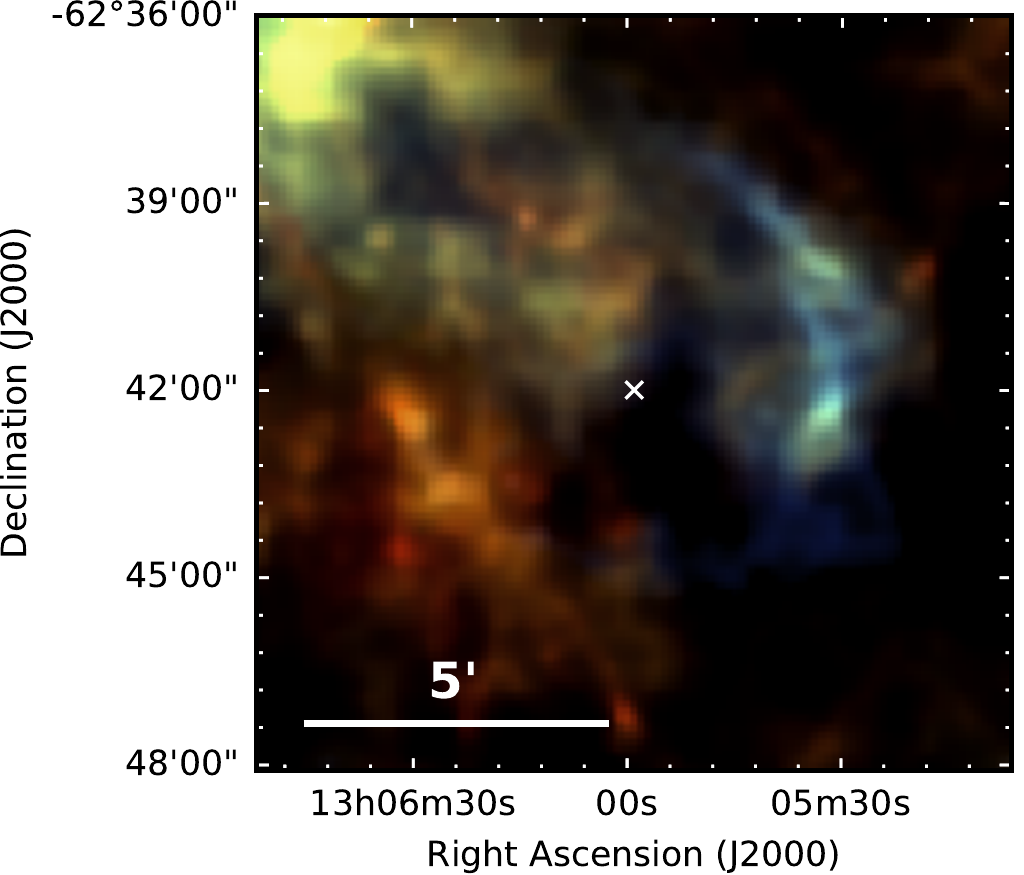}
	\caption{Kes 17 (G304.6$+$0.1) - \textit{Herschel} three colour image. An arc of dust is seen on filaments to the western edge of the outer shocks.
	The white cross shows the radio coordinates of the SNR centre from \citet{Green2014}.}
	\label{fig:G304.6+0.1Image}
\end{figure}

\textbf{W49B, G43.3$-$0.2} (Figure~\ref{fig:G43.3-0.2Image}): W49B is the first bipolar Type Ib/Ic SNR discovered in the Milky Way, which contributes to its rather unique radio barrel-hoop structure \citep{Moffett1994, Lopez2013}. X-ray emission is dominated by ejecta emission which suggests that the SNR is young \citep{Hwang2000}. The radio/IR morphologies seem anti-correlated with the X-ray emission.

Both R06 and PG11 detected this SNR to level 1. The FIR emission at 70, 160 and $250\mu$m in Figure~\ref{fig:G43.3-0.2Image} follows the MIR and radio morphology. The emission is especially bright to the southwest where there is a detection in all \textit{Herschel} wavebands although in the longer wavebands it is very confused. At 70\,--\,250\,$\mu$m we detect a $\sim1^\prime$ filament of emission centred at $\alpha\,=\,19^\text{h}11^\text{m}07^\text{s}, \delta\,=\,+09^\circ07^\prime01''$. R06 detected MIR emission due to ionic shocks from this filament, as we detect FIR continuum emission, it is likely that this is from shock-heated dust. R06 also established that MIR emission in the outer shell to the east and southwest is from shocked molecular gas where the SNR is interacting with a molecular cloud which encapsulates the wind-blown bubble surrounding the SNR \citep{Keohane2007}.
\bigskip

\textbf{G54.1$+$0.3} (Figure~\ref{fig:G54.1+0.3Image}): Described as a `close cousin' of the Crab Nebula, this remnant has a central PWN and an IR shell ~1.5$^\prime$ from the pulsar \citep{Koo2008}. Timing measurements of the central pulsar suggest a characteristic age of 2900 yr \citep{Camilo2002} and CO and HI observations suggest a distance of 6.2 kpc \citep{Leahy2008}.

Both R06 and PG11 classified this SNR as a level 3 detection, although there is bright 24 $\mu$m emission at the location of the PWN. \citet{Temim2017} detected an infrared shell to the south and west of the PWN. They completed a detailed analysis of MIR and FIR emission from the shell in the region of the PWN and found that their models require a minimum dust mass of $1.1\,\pm\,0.8\,M_\odot$.  \cite{Rho2018} fit the FIR-mm SED to derive a total dust mass of $0.08\,$-$\,0.9 M_{\odot}$ depending on the grain composition.  It is thought that the reverse shock has not yet reached this shell which might originate from dust in the SN ejecta potentially heated by early-type stars \citep{Temim2010}. However, the origin of the emission requires further investigation as \citet{Anderson2017} suggest that the structure is a mistakenly classified H\,{\sc ii} region, although faint, diffuse radio emission in the centre area may be associated with the SNR.
\bigskip

\textbf{G54.4$-$0.3} (Figure~\ref{fig:G54.4-0.3Image}): This SNR is located in an extended cluster of young population I objects from which molecular material was blown in to a shell by stellar winds before the supernova exploded \citep{Junkes1992}. R06 detected two MIR filaments likely corresponding to the location of shocks propagating into the bubble and/or molecular cloud, although their MIR colours are consistent with photo-dissociation regions (R06). The same filaments are detected by PG11 and across the five \textit{Herschel} wavebands as shown in Figure~\ref{fig:G54.4-0.3Image} at $\alpha\,=\,19^\text{h}32^\text{m}08^\text{s}, \delta\,=\,+19^\circ02^\prime56''$ and $\alpha\,=\,19^\text{h}33^\text{m}13^\text{s}, \delta\,=\,+19^\circ16^\prime20''$.
\bigskip

\textit{G55.0$+$0.3:} This shell-type remnant is at an estimated distance of 14 kpc and is highly evolved, aged around 1.9\,$\times\,10^6$ yrs (\citealp{Matthews1998}). Like R06 and PG11 at MIR wavelengths, we detect an arc of FIR emission near to $\alpha\,=\,19^\text{h}32^\text{m}05^\text{s}, \delta\,=\,+19^\circ46^\prime41''$ which could be associated with the eastern shell (Figure\,A2). However, it is difficult to disentangle SNR and ISM emission, making an association unclear.
\bigskip

\begin{figure}
	\centering
	\includegraphics[width=1.0\linewidth]{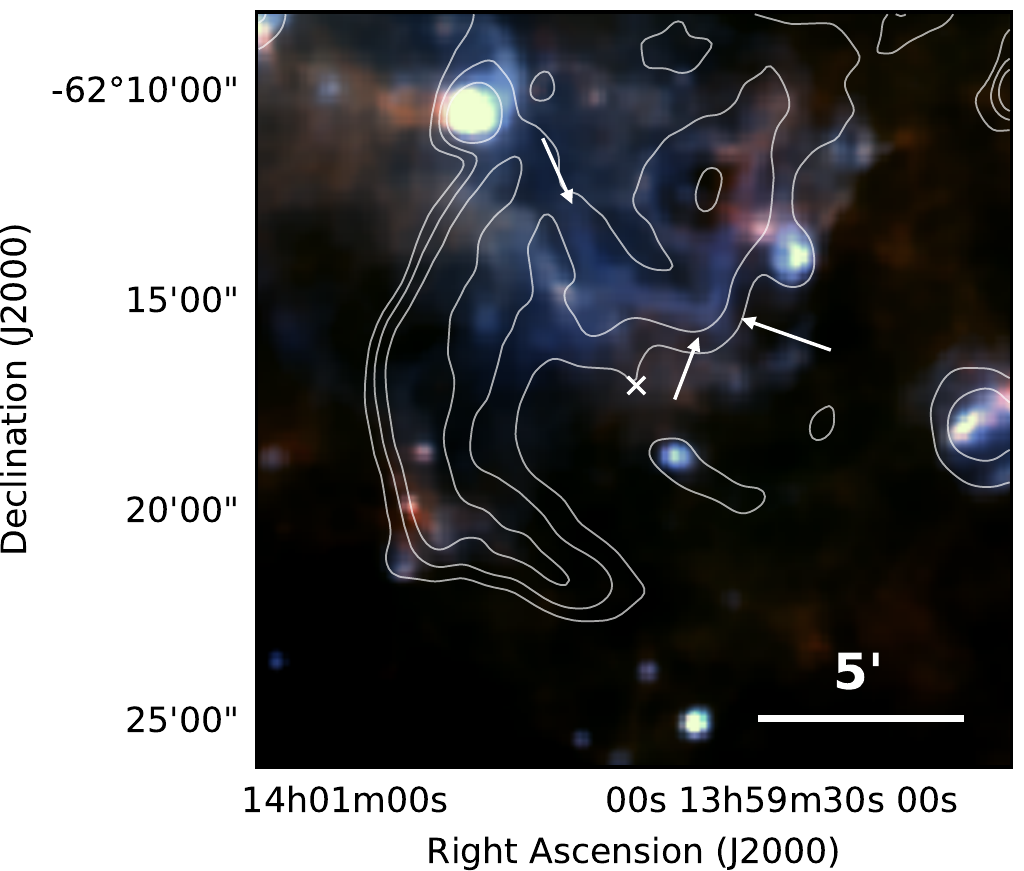}
	\caption{G310.8$-$0.4 - \textit{Herschel} three colour image with radio contours from MOST overlaid. The arrows indicate FIR emission which correlates with radio structure.}
	\label{fig:G310.8-0.4Image}
\end{figure}

\begin{figure}
	\centering
	\includegraphics[width=1.0\linewidth]{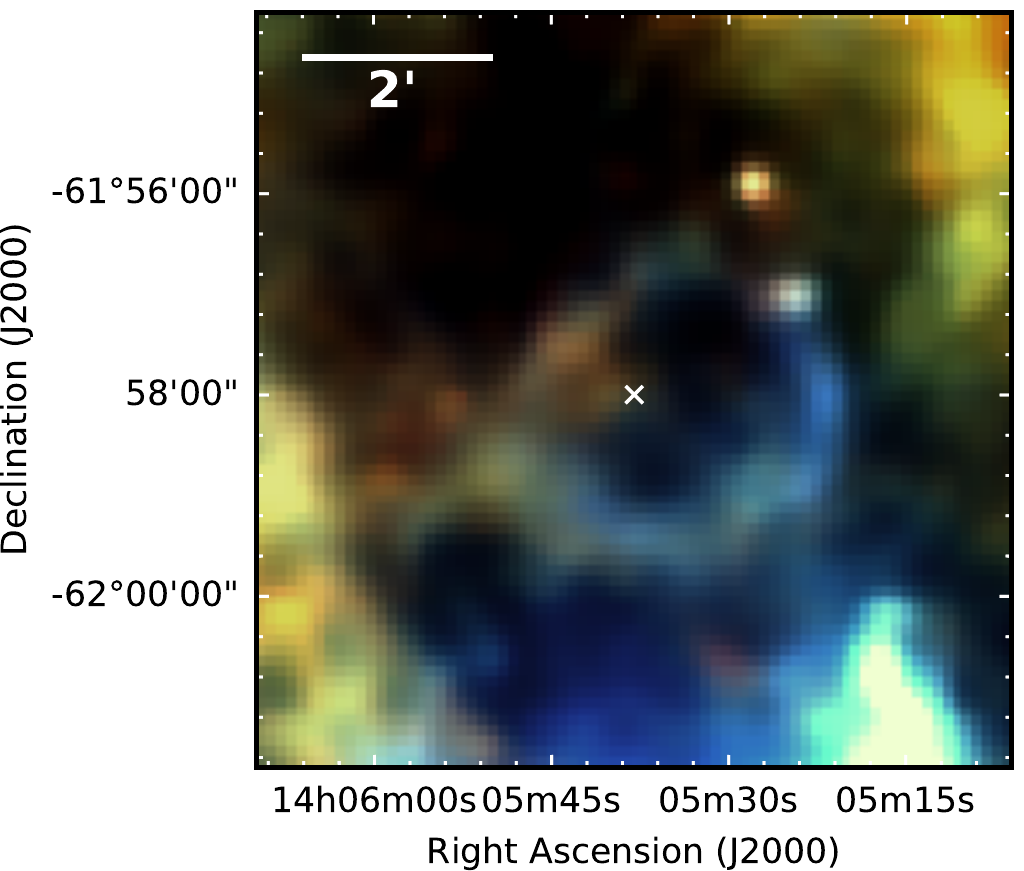}
	\caption{G311.5$-$0.3 - \textit{Herschel} three colour image. A ring of dust emission is detected at the outer edges of the shocks.
	The white cross shows the X-ray coordinates of the SNR centre.}
	\label{fig:G311.5-0.3Image}
\end{figure}

\begin{figure}
	\centering
	\includegraphics[width=1.0\linewidth, trim = 0cm 0cm 2.5cm 0cm, clip]{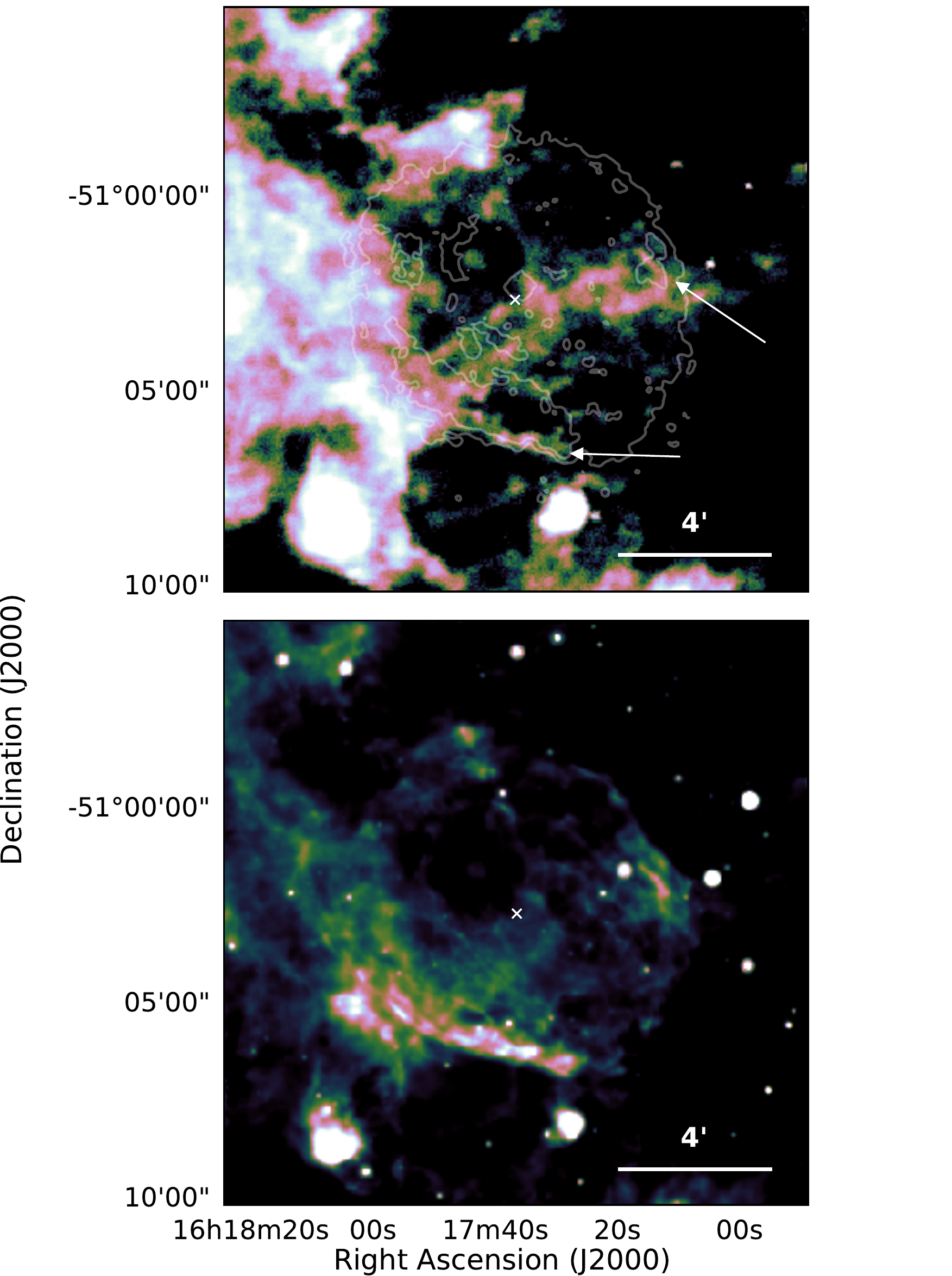}
	\caption{RCW 103, G332.4$-$0.4 - \textit{Top}: \textit{Herschel} 70\,$\mu$m image with X-ray contours overlaid.
	An arc of dust in the south coincides with X-ray structure and filaments to the north-west correspond with dust emission at 24\,$\mu$m, as indicated by the arrows.
	\textit{Bottom}:  \textit{Spitzer}  MIPS 24\,$\mu$m image.
	The white cross shows the X-ray coordinates of the SNR centre.}
	\label{fig:G332.4-0.4Image}
\end{figure}

\begin{figure}
	\centering
	\includegraphics[width=1.0\linewidth, trim = 0cm 0cm 2.5cm 0cm, clip]{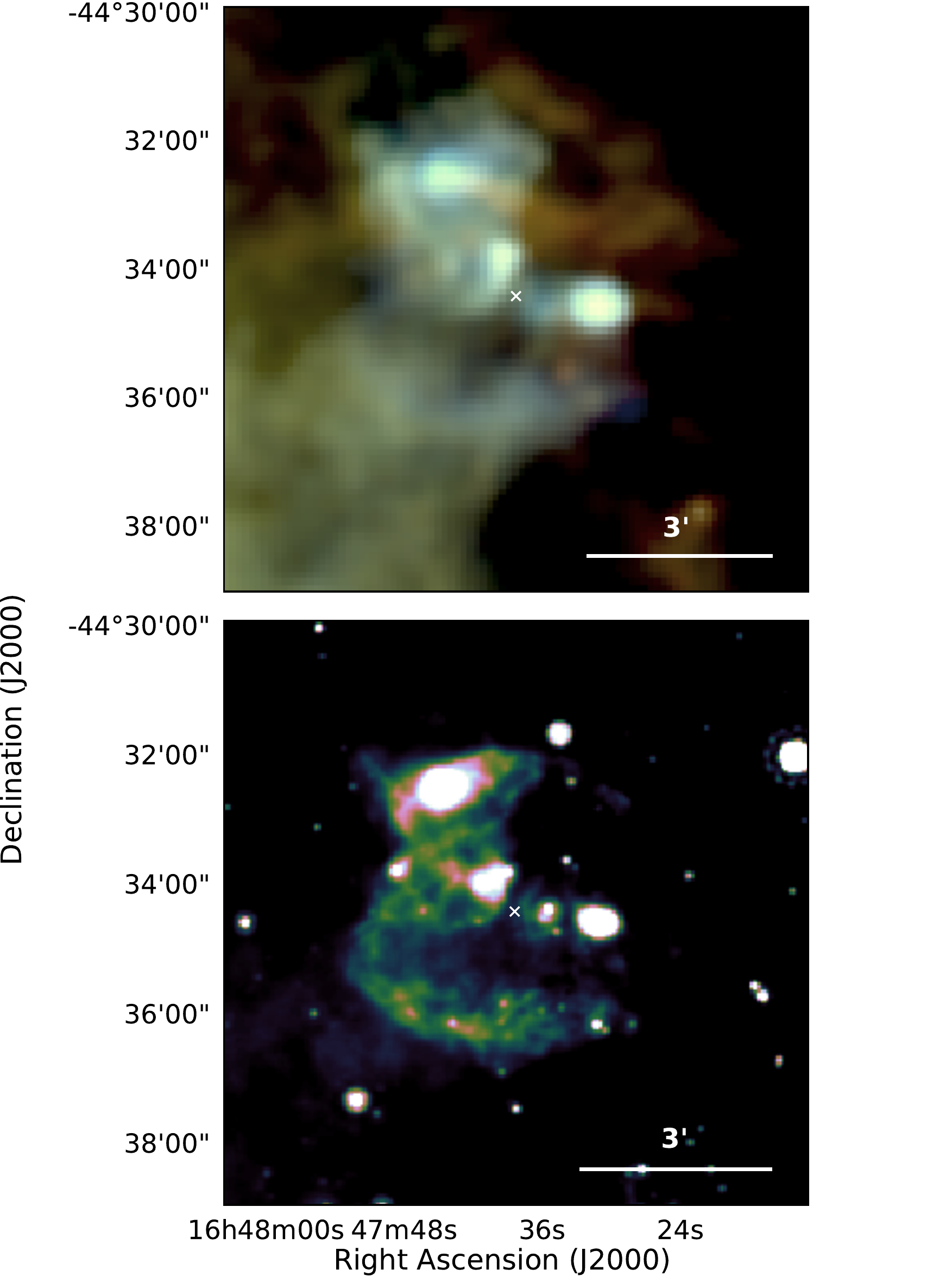}
	\caption{G340.6$+$0.3 - \textit{Top}: \textit{Herschel} three colour image. A shell of dust correlates is seen which is brightest along the southern edge.
	\textit{Bottom}:  \textit{Spitzer}  MIPS 24\,$\mu$m image.
	The white cross shows the X-ray coordinates of the SNR centre.}
	\label{fig:G340.6+0.3Image}
\end{figure}

\begin{figure}
	\centering
	\includegraphics[width=1.0\linewidth]{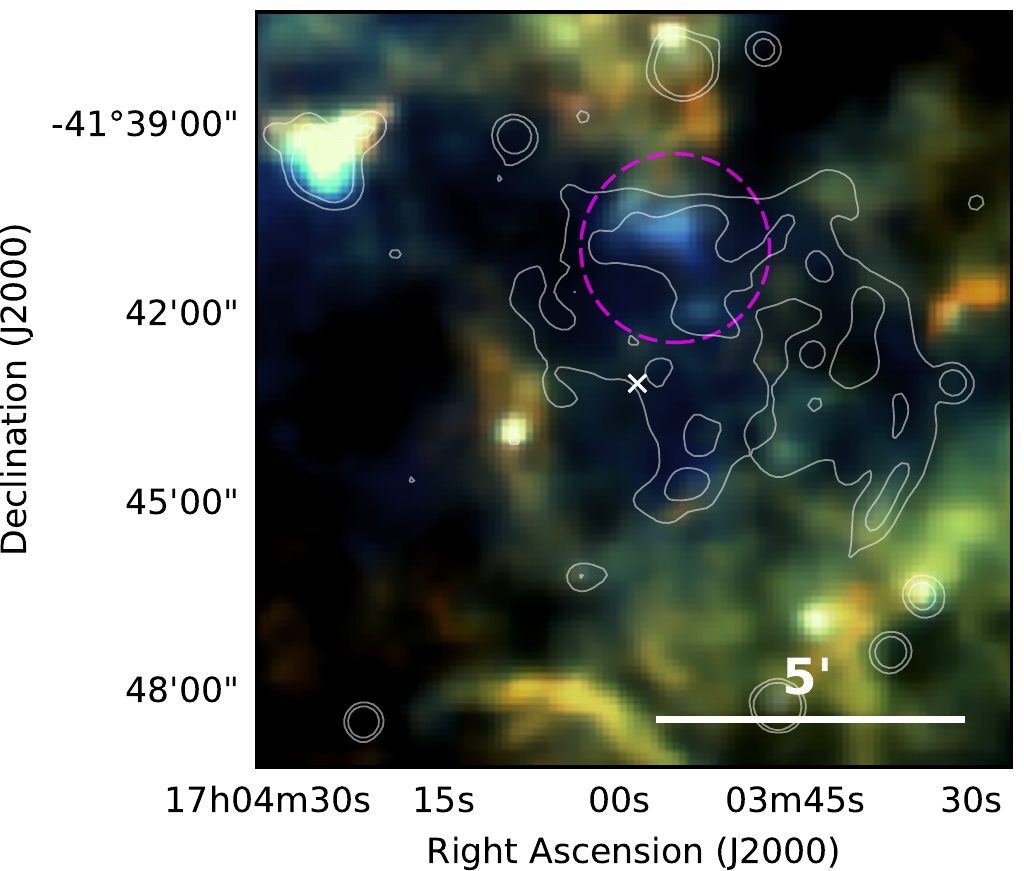}
	\caption{G344.7$-$0.1 - \textit{Herschel} three colour image. A region of dust emission is detected to the north of the remnant (within the magenta circle), centred at $\alpha\,=\,17^\text{h}03^\text{m}55^\text{s}, \delta\,=\,-41^\circ40^\prime43''$.
	MIPS 24\,$\mu$m contours are overlaid.
	The white cross shows the X-ray coordinates of the SNR centre.}
	\label{fig:G344.7-0.1Image}
\end{figure}

\textbf{Kes 17, G304.6$+$0.1} (Figure~\ref{fig:G304.6+0.1Image}): This middle-aged SNR \citep[28\,--\,64\,kyr,][]{Combi2010} is interacting with several massive molecular clouds causing bright filaments to the west by shock compression \citep{Combi2010}. R06 attributed MIR emission in the shell to molecular shocks. As shown in Figure~\ref{fig:G304.6+0.1Image}, there is bright FIR emission in this region across all of the \textit{Herschel} bands which is likely due to dust heated by shocked gas. There is also some diffuse 70\,$\mu$m emission along the southern ridge and longer wavelength emission towards the south east, although it is unclear whether this is associated with the SNR.
\bigskip

\textbf{G310.8-0.4} (Figure~\ref{fig:G310.8-0.4Image}) This SNR has a bright eastern radio shell at 0.843 GHz which is less defined to the west \citep{Whiteoak1996}. R06 and PG11 detected emission coinciding with radio contours to the south-east and part of the structure in the north-west. A 4.4$^\prime$ rim of diffuse FIR emission is detected near $\alpha\,=\,14^\text{h}00^\text{m}44^\text{s}, \delta\,=\,-62^\circ19^\prime03''$ which may be associated with the SNR (Figure\,A2). However this is unclear as the region is very confused in FIR and potentially related emission cannot be distinguished from the local ISM.
Filaments of FIR emission to the north of the SNR seems to correlate with radio emission, as indicated by the arrows.
\bigskip

\textbf{G311.5$-$0.3} (Figure~\ref{fig:G311.5-0.3Image}): This shell-type SNR is clearly detected by \textit{Herschel} in Figure~\ref{fig:G311.5-0.3Image} as a shell of dust emission similar to that detected by R06 and PG11. Detection of H$_2$ emission suggests that this SNR is interacting with molecular clouds \citep{Andersen2011} and R06 suggested that MIR emission is from shocked gas. At 70\,$\mu$m the FIR shell is brightest along the south west ridge, especially near $\alpha\,=\,14^\text{h}05^\text{m}22^\text{s}, \delta\,=\,-61^\circ58^\prime06''$.
However in the longer wavelength \textit{Herschel} images the brightest emission is towards the eastern and south-eastern edge of the shell. Like R06 we detect bright sources near $\alpha\,=\,14^\text{h}05^\text{m}24.3^\text{s}, \delta\,=\,-61^\circ57^\prime07''$ and $\alpha\,=\,14^\text{h}05^\text{m}23.5^\text{s}, \delta\,=\,-61^\circ56^\prime58''$, although at the \textit{Herschel} resolution these two sources are unresolved.
\bigskip

\textbf{RCW 103, G332.4$-$0.4} (Figure~\ref{fig:G332.4-0.4Image}): This shell-type remnant is at a distance of $\sim$3.1 kpc and the expansion velocity suggests an age of around 2000 yrs \citep{Carter1997}. The associated compact central object, 1E 161348-5055, which does not have a detected PWN \citep[e.g.][]{Tuohy1980, Reynoso2004} is most likely a magnetar with an extremely long period of 6.67\,hrs \citep{Rea2016}. X-ray emission across the SNR is dominated by shocked CSM with weaker emission from metal-rich ejecta and has been suggested to have had a $\sim$18\,--\,20$M_\odot$ progenitor with high mass-loss rate \citep{Frank2015}. The SNR is interacting with a molecular cloud on its southern side \citep{Oliva1990, Oliva1999}.

There is a $\sim$1.7$^\prime$ rim of 70\,$\mu$m emission along the southern edge of the SNR in Figure~\ref{fig:G332.4-0.4Image}, near $\alpha\,=\,16^\text{h}17^\text{m}36^\text{s}, \delta\,=\,-51^\circ06^\prime13''$, which is undetected in the other \textit{Herschel} bands. This coincides with X-ray and MIR structure detected by R06 and PG11 and is at a higher temperature than FIR structures to the east. The more resolved IRAC image suggests that this region consists of two emitting areas dominated by ionic and molecular shocks.

\bigskip

\textit{G337.2$-$0.7}: PG11 detected 24\,$\mu$m emission from filaments in this SNR which form two shell type features, corresponding to X-ray structure. Although there is some emission at 70\,$\mu$m (Figure\,A2) it is unclear whether this is associated with the SNR. There is a FIR region to the north-east coinciding with 24\,$\mu$m emission, however this does not correspond to the X-ray structure and it is unlikely to be associated with the SNR.

\bigskip

\textbf{G340.6$+$0.3} (Figure~\ref{fig:G340.6+0.3Image}): This shell-type SNR is at a distance of 15 kpc, on the other side of the Galaxy \citep{Kothes2007}. Like PG11, we detect a shell of dust, as seen in Figure~\ref{fig:G340.6+0.3Image}, which correlates with the 1.4 GHz radio shell \citep{Caswell1983}. This emission is not seen in the IRAC wavebands (Figure\,A2); some MIR emission is detected in the region but is not clearly associated with the SNR.
\bigskip

\begin{figure}
	\centering
	\includegraphics[width=1.0\linewidth]{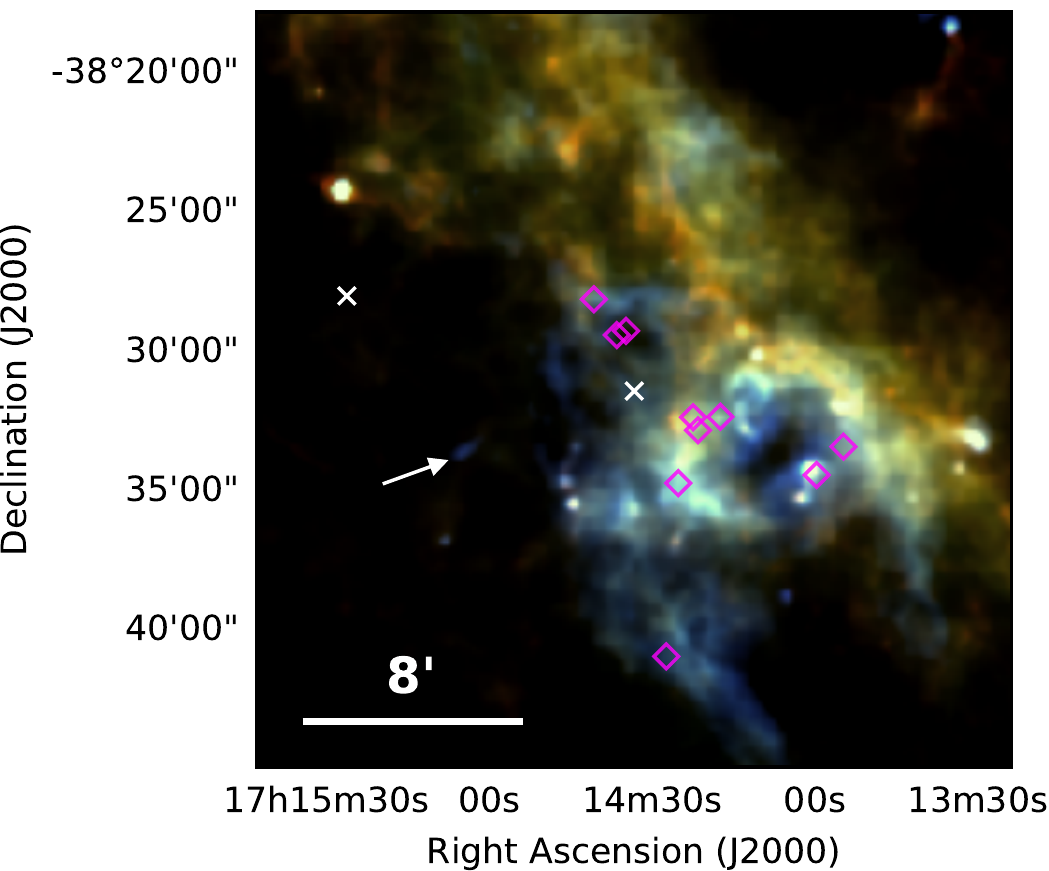}
	\caption{G348.5$-$0.0 and CTB 37A, G348.5$+$0.1 - \textit{Herschel} three colour image. The diamonds indicate the locations of OH 1720 MHz masers \citep{Frail1996}. Dust emission from G348.5$-$0.0 is detected as indicated by the arrow, and emission from CTB 37A is detected along the north-western edges of the remnant.
	The white crosses show the radio and X-ray coordinates of the centres of G348.5$-$0.0 (north-east) and G348.5$+$0.1 (south-west) respectively.}
	\label{fig:G348.5+0.1Image}
\end{figure}

\begin{figure}
	\centering
	\includegraphics[width=1.0\linewidth]{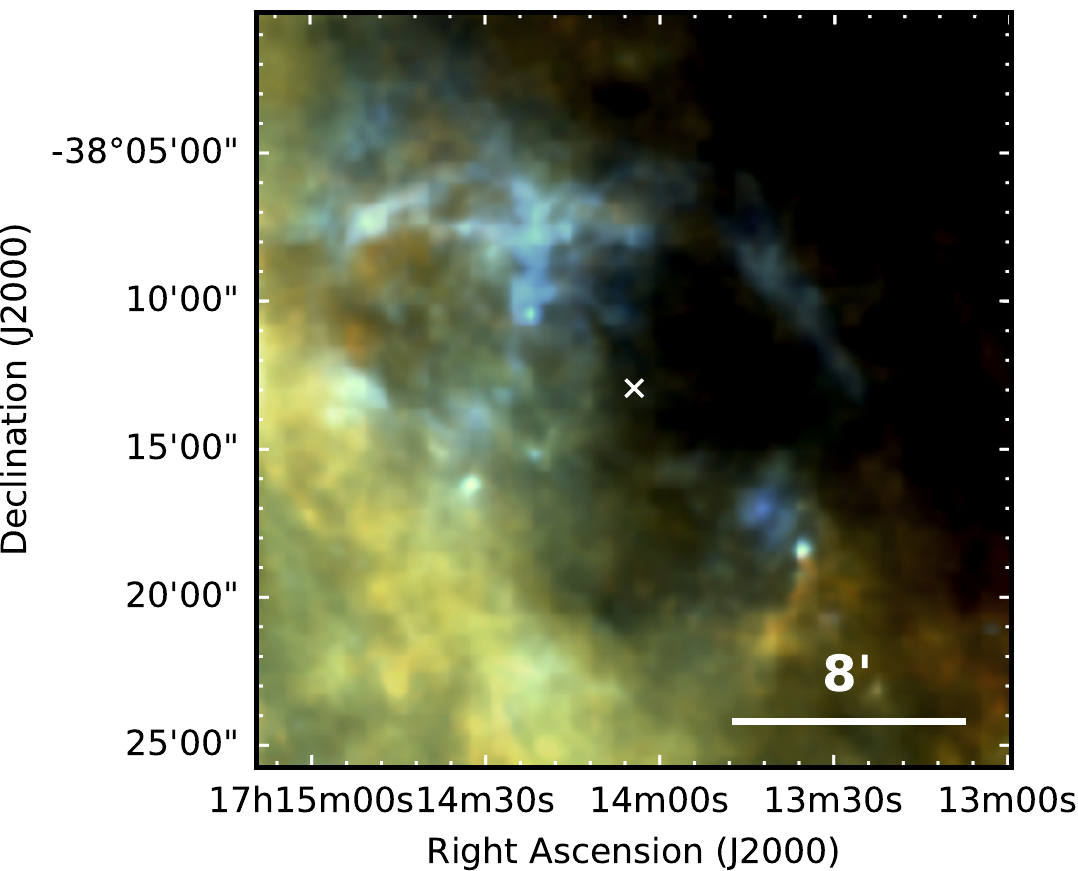}
	\caption{CTB 37B, G348.7$+$0.3 - \textit{Herschel} three colour image. Dust emission is detected in a partial shell structure around the northern edge of the remnant.
	The white cross indicates the X-ray coordinates of the SNR centre.}
	\label{fig:G348.7+0.3Image}
\end{figure}

\begin{figure}
	\centering
	\includegraphics[width=1.0\linewidth]{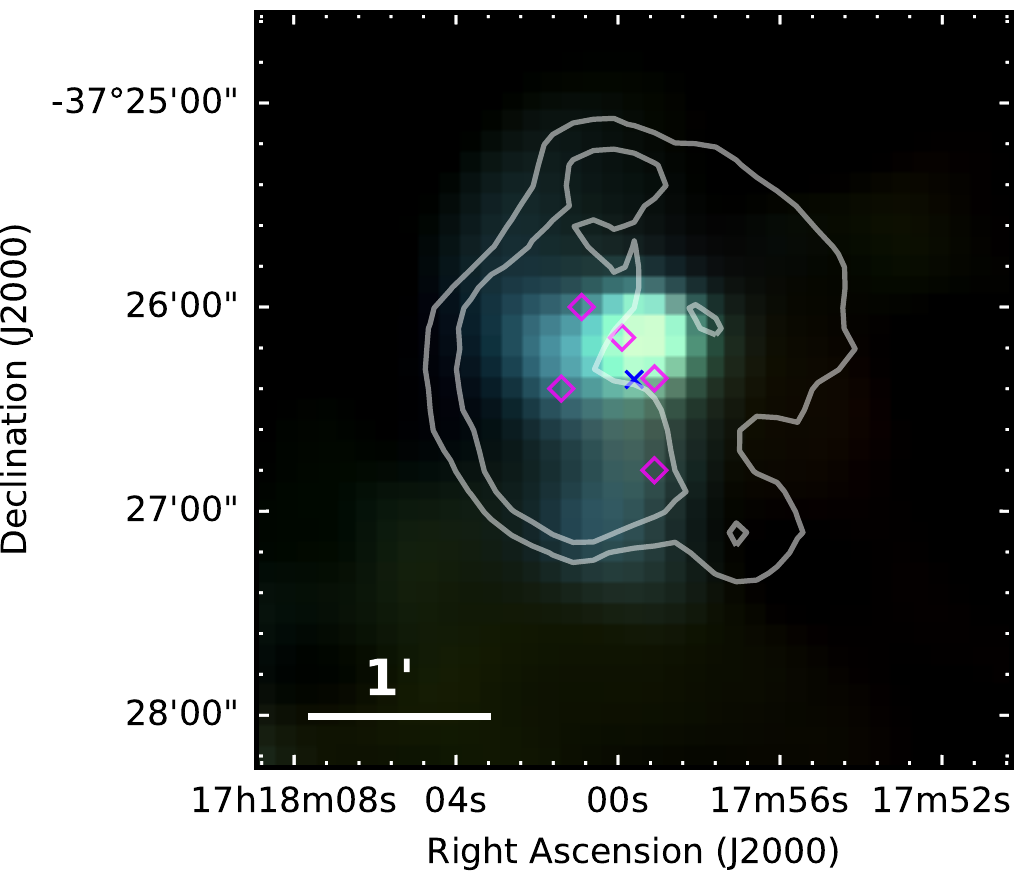}
	\caption{G349.7$+$0.2 - \textit{Herschel} three colour image with X-ray contours overlaid. The five OH masers are indicated by diamonds (\citealp{Frail1996}). Dust emission is observed at the western edge, where the remnant is interacting with a dense cloud.
	The blue cross shows the X-ray coordinates of the SNR centre.}
	\label{fig:G349.7+0.2Image}
\end{figure}

\textbf{G344.7$-$0.1} (Figure\,\ref{fig:G344.7-0.1Image}): This relatively young SNR ($\sim$3000 yrs) has an estimated distance of 14\,kpc \citep{Yamaguchi2012} and is classified as a mixed-morphology remnant, having thermal X-ray emission which completely fills the radio shell \citep{Yamauchi2005, Combi2010a, Giacani2011}. The radio emission is brightest towards the northwestern side where the remnant is expected to be interacting with a molecular cloud \citep{Combi2010}. Strong Fe K-shell ejecta emission and the abundance pattern of $\alpha$-elements suggests that this is a Type Ia SNR \citep{Yamaguchi2012}. The SNR contains a large mass of hydrogen ($\gtrsim$150\,M$_\odot$) which implies that the SNR is dominated by swept up ISM.

R06 detected an irregular MIR structure to the north of the SNR originating from shocked ionised gas, which coincides with a central radio peak \citep{Giacani2011}. We also detect FIR emission from this structure in Figure\,\ref{fig:G344.7-0.1Image}, centred at $\alpha\,=\,17^\text{h}03^\text{m}55^\text{s}, \delta\,=\,-41^\circ40^\prime43''$, although the \textit{Herschel} detection is less resolved.
It is likely that this emission arises from the interaction between the SN shock and a molecular cloud in front of the SNR \citep{Giacani2011}.
\bigskip

\textit{G345.7$-$0.2:} This SNR has a faint disk radio morphology and a peak close to the pulsar PSR J1707-4053, which is probably unrelated \citep{Taylor1993, Whiteoak1996}. We detect a diffuse region of 70\,$\mu$m emission (Figure\,A2) centred at $\alpha\,=\,17^\text{h}07^\text{m}39^\text{s}, \delta\,=\,-40^\circ54^\prime31''$, extending roughly 2$^\prime$ which correlates with MIPS 24~$\mu$m and 0.843 GHz radio emission \citep{Whiteoak1996}. We find no other evidence of SNR-related emission in the region. We suggest that the SNR centre is offset from that of \citet{Green2014} and the bar detected could be part of a structure extending to the east.
\bigskip

\textit{G346.6$-$0.2:} R06 detected a narrow rim of emission connecting three OH 1720 MHz masers along the southern shell. There is FIR emission which potentially corresponds to this structure (Figure\,A2), however the region is very confused as there is extensive dust emission to the north and west of the SNR.
\bigskip

\textbf{G348.5$-$0.0} (Figure~\ref{fig:G348.5+0.1Image}): Originally thought to be a jet associated with CTB 37A, this was classified by \citet{Kassim1991} as a separate partial shell remnant.
H\,{\small I} 21-cm absorption measurements suggest that this remnant is at a distance of $\leq$6.3 kpc (\citealp{Tian2012}). In Figure~\ref{fig:G348.5+0.1Image} we detect a $\sim$1.7$^\prime$ arc of dust emission centred at $\alpha\,=\,17^\text{h}15^\text{m}03.5^\text{s}, \delta\,=\,-38^\circ33^\prime30''$. This is detected across all \textit{Herschel} wavebands, although is very confused for $\lambda\,\textgreater\,160\,\mu$m. R06 suggested that this MIR emission is dominated by emission lines from shocked gas.

The OH 1720 MHz masers at $\alpha\,=\,17^\text{h}14^\text{m}34.8^\text{s}, \delta\,=\,-38^\circ29^\prime16''$ and $\alpha\,=\,17^\text{h}14^\text{m}36.5^\text{s}, \delta\,=\,-38^\circ29^\prime25''$ have a significantly different velocity to the other eight in the region which are associated with CTB 37A \citep{Frail1996}. It is suggested that these two are related to a nearby molecular cloud \citep{Reynoso2000} lying to the west of G348.5$-$0.0 with which the remnant is likely to be interacting. We do not detect FIR emission at the location of these masers.
\bigskip

\textbf{CTB 37A, G348.5$+$0.1} (Figure \ref{fig:G348.5+0.1Image}): Recently classified as a mixed morphology remnant \citep{Sezer2011, Yamauchi2014}, this SNR has a radio shell-like structure in the north and a ``breakout" to the south-west, and has thermally dominated central X-ray emission. At a distance in the range of 6.3\,--\,9.5 kpc, its proximity to both G348.5$-$0.0 and CTB 37B is coincidental \citep{Tian2012}. There are eight OH 1720 MHz masers associated with the SNR which have a similar velocity to incident CO\,(1--0) clouds detected by \citet{Reynoso2000} which are to the north-west and centre of the SNR.

The \textit{Herschel} emission in Figure \ref{fig:G348.5+0.1Image} is very confused. Similar to R06, we detect dust emission along the northern edge in a $\sim4.3^\prime$ arc centred at $\alpha\,=\,17^\text{h}14^\text{m}35^\text{s}, \delta\,=\,-38^\circ28^\prime22''$ and filaments along the western edge near to $\alpha\,=\,17^\text{h}14^\text{m}46^\text{s}, \delta\,=\,-38^\circ32^\prime33''$. Emission in the northern region coincides with an OH 1720 MHz maser and is suggested by R06 to originate from shocked molecular gas where the SNR has encountered very dense gas. We also detect dust emission in the region of the ``breakout" structure to the south of the SNR.
\bigskip

\textbf{CTB 37B, G348.7$+$0.3} (Figure~\ref{fig:G348.7+0.3Image}): This relatively young ($\sim$4900 yrs, \citealp{HESSCollaboration2008b}) SNR has a non-thermal radio shell (\citealp{Whiteoak1996}) and a magnetar which is slightly off-centre (\citealp{Halpern2010}). 21\,cm H\,{\small I} absorption indicates a distance of $\sim$13.2 kpc (\citealp{Tian2012}).

The region around the SNR is confused due to bright emission from IR bubbles at $\alpha\,=\,17^\text{h}14^\text{m}38.2^\text{s}, \delta\,=\,-38^\circ10^\prime18''$ and $\alpha\,=\,17^\text{h}13^\text{m}40.3^\text{s}, \delta\,=\,-38^\circ17^\prime33''$. We detect dust emission in Figure~\ref{fig:G348.7+0.3Image} around the northern edge of the SNR which coincides with the radio structure and is at a higher temperature than the surrounding medium.
\bigskip

\textbf{G349.7$+$0.2} (Figure~\ref{fig:G349.7+0.2Image}): The radial velocities of OH 1720 MHz masers suggest that this shell-type SNR is at a large distance of ~22 kpc (\citealp{Frail1996}), making it one the most X-ray luminous Galactic remnants (\citealp{Slane2002}). X-ray temperature fits give an age of ~2800 yrs. The presence of OH 1720 MHz masers and shocked molecular gas imply that the SNR is interacting with a molecular cloud. The MIR SNR shell is much brighter to the eastern side where a shock is propagating into the edge of a roughly cylindrical cloud (\citealp{Reach2006}). Like the MIR, the FIR emission in Figure~\ref{fig:G349.7+0.2Image} peaks in the region of the OH 1720 MHz masers, near $\alpha\,=\,17^\text{h}18^\text{m}00^\text{s}, \delta\,=\,-37^\circ26^\prime09''$.

\section{The mass of dust in G11.2$-$0.3, G21.5$-$0.9, and G29.7$-$0.3} \label{DustMasses}
\begin{table}
	\begin{tabular}{c c c c}
	\hline
	\multirow{2}{*}{Wavelength, $\mu$m} & \multicolumn{3}{c}{Flux, Jy} \\
	    & G11.2$-$0.3     & G21.5$-$0.9     & G29.7$-$0.3 \\ \hline\hline
	24  & $5.6 \pm 0.3$   & $0.19 \pm 0.02$ & $0.20 \pm 0.03$ \\
	70  & $47.7 \pm 6.7$  & $4.1 \pm 0.3$   & $5.6 \pm 1.0$ \\
	160 & $71.9 \pm 15.7$ & $6.6 \pm 0.7$   & $3.2 \pm 4.0$ \\
	250 & $26.6 \pm 5.5$  & $3.0 \pm 0.9$   & $0.55 \pm 2.10$ \\
	350 & $10.1 \pm 3.0$  & $1.5 \pm 0.8$   & $0.19 \pm 1.15$ \\
	500 & $2.3 \pm 0.9$   & $1.2 \pm 0.4$   & $0.02 \pm 0.20$ \\ \hline

	\end{tabular}
	\caption{Background subtracted flux measured for each SNR at FIR wavelengths.}
	\label{tab:FIRFlux}
\end{table}

\begin{table*}
	\begin{tabular}{c c c c c c c}\\
	\hline
	SNR & Frequency (GHz) & Flux (Jy) & Ref & Spectral Index \textsuperscript{a} & Wavelength ($\mu$m) & Estimated Synchrotron Flux (Jy)\textsuperscript{b} \\\hline\hline
	G11.2$-$0.3 & 1.4 & $0.11 \pm 0.02$ & 1 & $-0.10 \pm 0.08$ & 24 & $(43 \pm 40) \times 10^{-3}$ \\
	& 32 & $0.08 \pm 0.02$ & 1 & & 70 & $(46 \pm 44) \times 10^{-3}$ \\
	& & & & & 160 & $(48 \pm 44) \times 10^{-3}$ \\
	& & & & & 250 & $(50 \pm 44) \times 10^{-3}$ \\
	& & & & & 350 & $(51 \pm 44) \times 10^{-3}$ \\
	& & & & & 500 & $(52 \pm 43) \times 10^{-3}$ \\ \hline

	G21.5$-$0.9 & $327 \times 10^{-3}$ & $7.3 \pm 0.7$ & 2 & $-0.032 \pm 0.038$ & 24 & $0.23 \pm 0.05$\\
	& 1.43 & $7.0 \pm 0.4$ & 2 & & 70 & $0.40 \pm 0.08$ \\
	& 5   & $6.7 \pm 0.3$ & 3 & & 160 & $0.63 \pm 0.13$ \\
	& 32  & $5.6 \pm 0.3$ & 4 & $-0.56 \pm 0.02$ & 250 & $0.80 \pm 0.16$ \\
	& 70  & $4.3 \pm 0.6$ & 5 & & 350 & $0.95 \pm 0.18$ \\
	& 84.2 & $3.9 \pm 0.7$ & 6 & & 500 & $1.16 \pm 0.22$ \\
	& 90.7 & $3.8 \pm 0.4$ & 7 & & & \\
	& 100 & $2.7 \pm 0.5$ & 5 & & & \\
	& 141.9 & $2.5 \pm 1.2$ & 7 & & & \\
	& 143 & $3.0 \pm 0.4$ & 5 & \\ \hline

	G29.7$-$0.3 & 1.4 & 0.35 & 6 & -0.43 & 24 & $9.64 \times 10^{-3}$ \\
	& 4.9 & 0.25 & 6 & & 70 & $15.2 \times 10^{-3}$ \\
	& 5.0 & 0.28 & 8 & & 160 & $21.7 \times 10^{-3}$ \\
	& 15 & 0.17 & 6 & & 250 & $26.3 \times 10^{-3}$ \\
	& 89 & 0.08 & 9 & & 350 & $30.4 \times 10^{-3}$ \\
	& & & & & 500 & $35.4 \times 10^{-3}$ \\ \hline

	\end{tabular}
	\caption{
	The synchrotron flux measured for each SNR at radio wavelengths, and estimated synchrotron flux at Spitzer and Herschel bands.
	\textsuperscript{a}The spectral index is estimated by fitting a power law to the radio fluxes on the same line and below.
	\textsuperscript{b}The FIR synchrotron flux is estimated by extrapolating the fitted power law to FIR wavelengths.
	Radio fluxes for the core are taken from:
	\textsuperscript{1} \citet{Kothes2001};
	\textsuperscript{2} \citet{Bietenholz2011};
	\textsuperscript{3} \citet{Bietenholz2008};
	\textsuperscript{4} \citet{Morsi1987};
	\textsuperscript{5} \citet{Planck2016beta};
	\textsuperscript{6} \citet{Salter1989ApJ};
	\textsuperscript{7} \citet{Salter1989AA};
	\textsuperscript{8} \citet{Becker1984};
	\textsuperscript{9} \citet{Bock2005}.}
	\label{tab:Flux}
\end{table*}

As discussed in Section \ref{CatalogueResults}, we detected 4 SNRs with dust in the central region associated with PWN-heated ejecta. The dust mass in G54.1$+$0.3 has been studied in detail, and so in this section we determine the dust masses for the three new PWNe detected in this paper (G11.2$-$0.3, G21.5$-$0.9, and G29.7$-$0.3).

The flux density for the central region (PWN) of each SNR was estimated using aperture photometry on MIR-FIR images from  \textit{Spitzer} MIPS (24\,$\mu$m) and \textit{Herschel}. The IRAC images are not used as we do not see emission from the SNRs in the region of the PWN in these wavebands and the measured emission is dominated by unrelated point sources. Apertures of 1.9$^\prime$, 1.7$^\prime$, and 1.2$^\prime$ in diameter were centred at $\alpha\,=\,18^\text{h}11^\text{m}29^\text{s}, \delta\,=\,-19^\circ25^\prime54''$, $\alpha\,=\,18^\text{h}33^\text{m}34.2^\text{s}, \delta\,=\,-10^\circ34^\prime18.5''$, and $\alpha\,=\,18^\text{h}46^\text{m}25^\text{s}, \delta\,=\,-02^\circ58^\prime30''$ for G11.2$-$0.3, G21.5$-$0.9, and G29.7$-$0.3 respectively.

The background ISM level for G21.5$-$0.9 was estimated using an annulus around the source as the local environment seems reasonably uncluttered. For G11.2$-$0.3 and G29.7$-$0.3, 8 apertures were placed around the SNRs so as to avoid the shell and the bright cloud to the south of G11.2$-$0.3, and to avoid the young stellar object to the west of G29.7$-$0.3. Areas were selected to cover a range of flux levels to account for fluctuations in the ISM within the image cut-out. The background-subtracted fluxes are given in Table~\ref{tab:FIRFlux}. Due to large variations in the ISM level, the background subtraction is the largest source of uncertainty in our estimation of dust mass.

Flux calibration uncertainties are assumed as a percentage of the measured flux in each band; that is 4\% for MIPS \citep{Engelbracht2007}, 7\% for PACS \citep{Balog2014}, and 5.5\% for SPIRE data \citep{Bendo2013}. Uncertainty in the ISM level is estimated as the standard deviation of the background ISM values.

\subsection{Synchrotron Emission} \label{SynchrotronEmission}
A power-law synchrotron radiation spectrum, from charged particles accelerated in a magnetic field, can be detected across the electromagnetic spectrum. The contribution varies across SNR type. PWNe and Crab-like remnants tend to have a very flat radio spectrum and a steeper X-ray spectrum due to synchrotron losses, whereas shell-type SNRs have a much steeper radio spectrum.

In order to estimate the dust mass, first non-thermal synchrotron emission must be removed from the IR fluxes. The synchrotron flux density at a frequency, $\nu$, can be fitted by equation \ref{eqn:synch},

\begin{equation} \label{eqn:synch}
	S_\nu = S_{\nu_0}\bigg(\frac{\nu}{\nu_0}\bigg)^\alpha
\end{equation}

where $S_{\nu_0}$ is the synchrotron flux density at frequency $\nu_0$, and $\alpha$ is the spectral index which describes the flux density dependence on frequency. We use a least squares fitting routine to estimate $\alpha$. Like the Crab Nebula, the power law slope may break in the FIR region \citep[e.g.][]{Arendt2011, Gomez2012b}, in which case we would be overestimating the synchrotron contribution to the MIR and FIR fluxes.

The uncertainty in $\alpha$ is estimated using a Monte Carlo technique by producing 1000 sets of normally distributed radio flux values, using the measured flux at each frequency as mean and flux uncertainty as standard distribution. Fitting a power law to each set gives 1000 values of $\alpha$ and the standard deviation is used as the uncertainty in the estimated value.

The integrated synchrotron flux values for the SNR PWN regions and spectral index from the fit are given in Table~\ref{tab:Flux}. A least squares fit to G11.2$-$0.3 radio data \citep{Kothes2001} of the central compact object gives a spectral index of $\alpha\,=\,-0.10 \pm 0.08 $.
A single power law cannot fit the spectra for G21.5$-$0.9 (\citealp{Salter1989ApJ}) as the slope breaks at 40 GHz due to synchrotron losses. Fitting to radio data between 70 and 143 GHz (\citealp{Salter1989ApJ}; \citealp{Salter1989AA}; \citealp{Planck2016mic}),  \textit{Spitzer}  MIPS (24 $\mu$m) and \textit{Herschel} (500 $\mu$m) data, gives a spectral index of $\alpha\,=\,-0.53 \pm 0.01$. Whereas at lower frequencies \citep{Morsi1987, Bietenholz2008}, the fitting routine gives a flat spectral index of $\alpha\,=\,-0.032 \pm 0.034$.
A least squares fit to radio data of the Crab-like component of G29.7$-$0.3 \citep{Salter1989ApJ} gives a spectral index of $\alpha\,=\,-0.43$. It has been suggested that there may be a spectral break at around 55\,GHz, in which case our synchrotron flux in FIR wavebands will be overestimated \citep{Bock2005}.

The synchrotron contribution to the SNR SEDs is only significant for G21.5$-$0.9, for which it contributes between 9.5 and 96.7 per cent (160 and 500\,$\mu$m respectively) of the background subtracted flux at each wavelength. For G29.7$-$0.3, the contribution at 500\,$\mu$m is $\sim$ 59 per cent, however for the other wavebands it is less than 9 per cent, and for G11.2$-$0.3 the contribution in all wavebands is less than 3.5 per cent.

\subsection{Dust Emission} \label{DustEmission}
After removing the synchrotron contribution from the SNR SEDs, we assume that the remaining thermal FIR flux takes the form of a modified blackbody:

\begin{equation} \label{eqn:Greybody}
	\centering
	F_\nu = \frac{M_{\rm dust} B_\nu (T) \kappa_\nu}{D^2},
\end{equation}

where $F_\nu $ is the flux at a given wavelength, $M_{\rm dust}$ is the mass of dust, $B_\nu (T)$ is the Planck function at temperature $T$, $\kappa_\nu$ is the dust mass absorption coefficient, and $D$ is the distance to the source. The dust mass absorption coefficient, $\kappa_\nu$, describes the effective surface area for extinction per unit mass:

\begin{equation} \label{eqn:MAC1}
	\kappa_\nu = \frac{3 Q_\nu}{4 a\rho},
\end{equation}

where $a$ is the dust grain radius, $\rho$ is the grain material density, and $Q_\nu$ describes the emission efficiency and is dependent on both the frequency and the material type. The frequency dependence of $\kappa_\nu$ is given by

\begin{equation} \label{eqn:MAC2}
	\centering
	\kappa_\nu = \kappa_{\nu_0}\bigg(\frac{\nu}{\nu_0}\bigg)^{\beta},
\end{equation}

where $\beta$ is the dust emissivity index, which describes how the emissivity varies with frequency. In this study we assume $\kappa_{\lambda_0}\,=\,0.07 \pm 0.02$~m$^2 kg^{-1}$ for $\lambda_0\,=\,850$~$\mu$m \citep{Dunne2000, James2002, Clark2016}. In this function we assume that $\kappa_\nu$ can be estimated using a constant emissivity index of $\beta\,=\,1.9$ (indicative of normal interstellar dust grains; \citealp{Planck2014}).

The mass and temperature of the dust in the remnants is evaluated by a least squares fit to the flux values at wavelengths between 24 and 500 $\mu$m; we attempt both a single and double temperature modified blackbody fit. In order to derive physical estimates, the temperature and mass are constrained. A cold dust component must have a temperature of $15 < T_c < 50$K, a warm component must have a temperature of $50 < T_w < 150$K, and there must be a positive dust mass. We assume distances of 4.4 \citep{Green2004}, 4.7 \citep{Camilo2005}, and 10.6\,kpc \citep{Su2009} for G11.2$-$0.3, G21.5$-$0.9, and G29.7$-$0.3 respectively.
The best fit SEDs for all three PWNe are shown in Figures \ref{fig:G11.2-0.3SED} to \ref{fig:G29.7-0.3SED}.

The 1-$\sigma$ uncertainties in the derived mass and temperature are estimated in a similar way to the uncertainty in $\alpha$ in section \ref{SynchrotronEmission}. A modified blackbody is fit to 1000 sets of flux values to give distributions dust mass and temperature estimates. The uncertainties in the fit parameters are taken as the 16th and 84th percentiles. The largest source of uncertainty is the background subtraction; there is a large variation in the ISM level across the regions surrounding the SNRs making an estimate of the ISM contribution very uncertain.
The best fit and median values derived for dust temperature and mass are shown in Table~\ref{tab:TempMass}.
\bigskip

\begin{figure}
	\centering
	\includegraphics[width=1.0\linewidth]{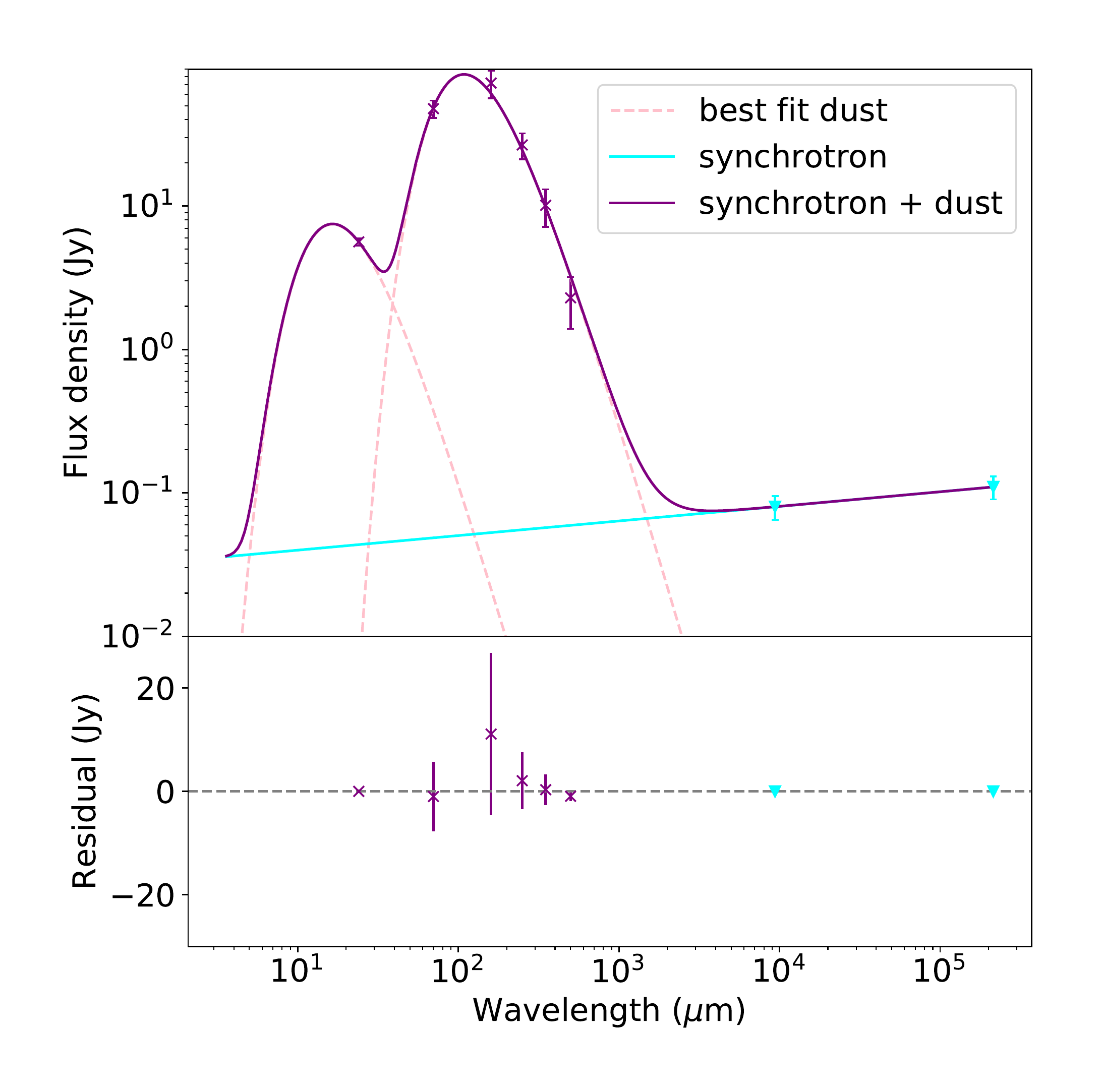}
	\includegraphics[width=1.0\linewidth]{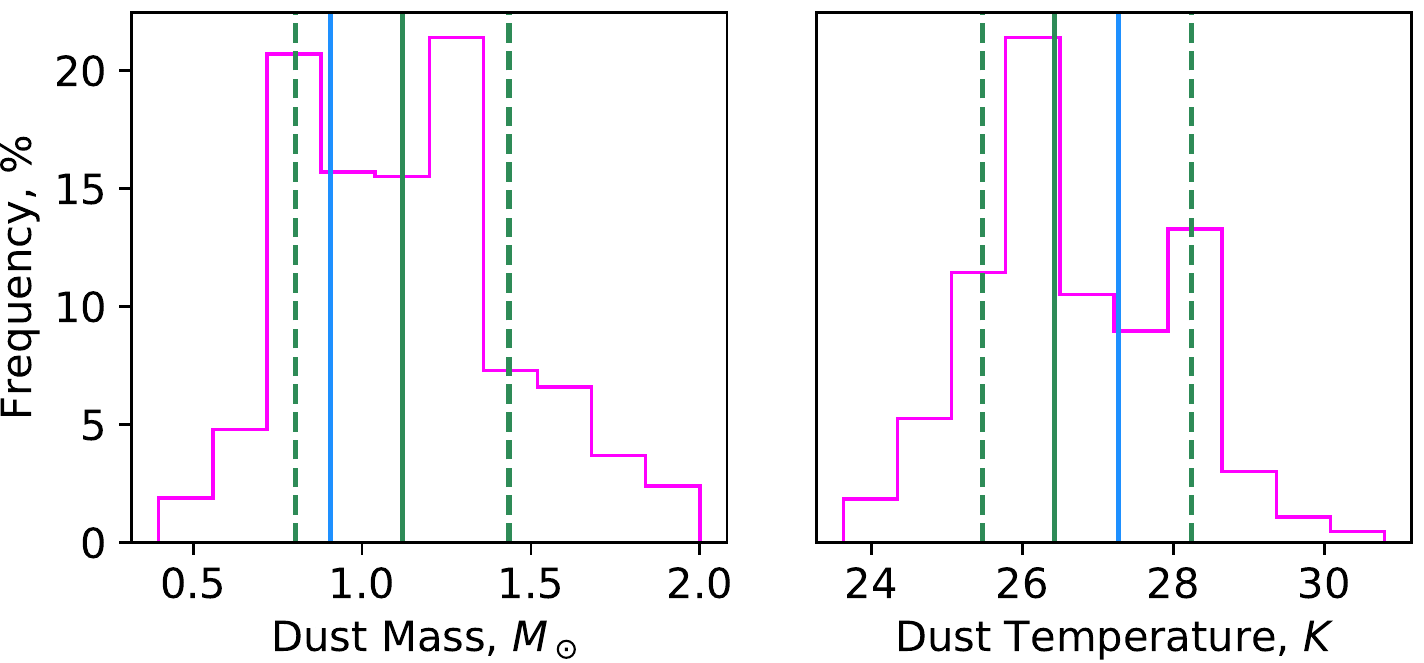}
	\caption{\textit{Top}: Far infrared SED of G11.2$-$0.3. The dashed pink lines are the two temperature components fitted to the FIR fluxes. The solid cyan line is the synchrotron component fit to  the radio flux of the compact object (cyan points). The solid purple line is the total SED of all components.
	\textit{Bottom}: The distribution of source masses obtained by the Monte Carlo routine where SEDs were fit to 1000 normally distributed samples of flux values. The dashed green lines indicate the 1-sigma uncertainties and the solid green line is the median value. The solid blue line indicates the best fit value.
	The best fit and median values derived for dust temperature and mass are shown in Table~\ref{tab:TempMass}.}
	\label{fig:G11.2-0.3SED}
\end{figure}

\begin{figure}
	\centering
	\includegraphics[width=1.0\linewidth]{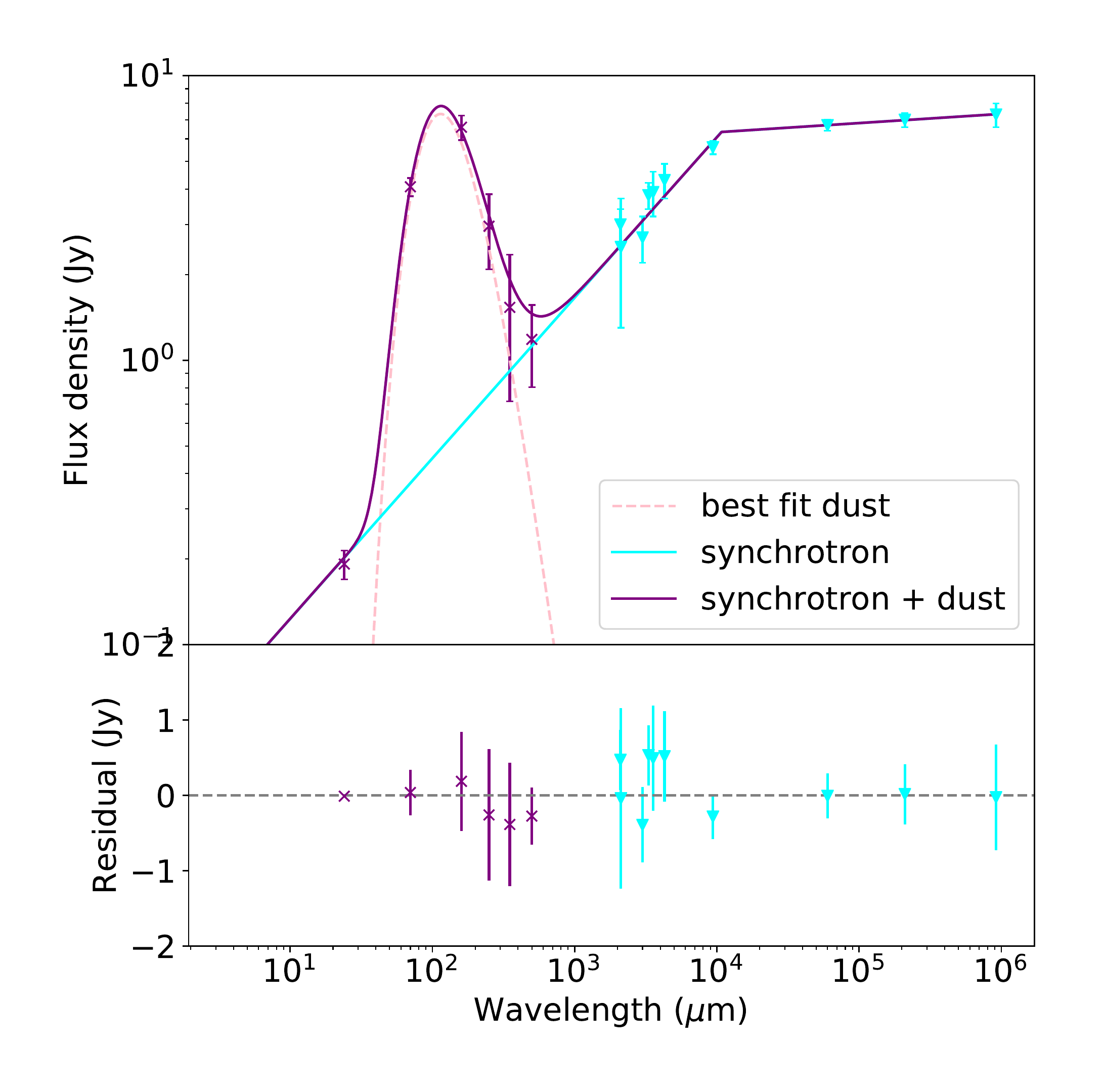}
	\includegraphics[width=1.0\linewidth]{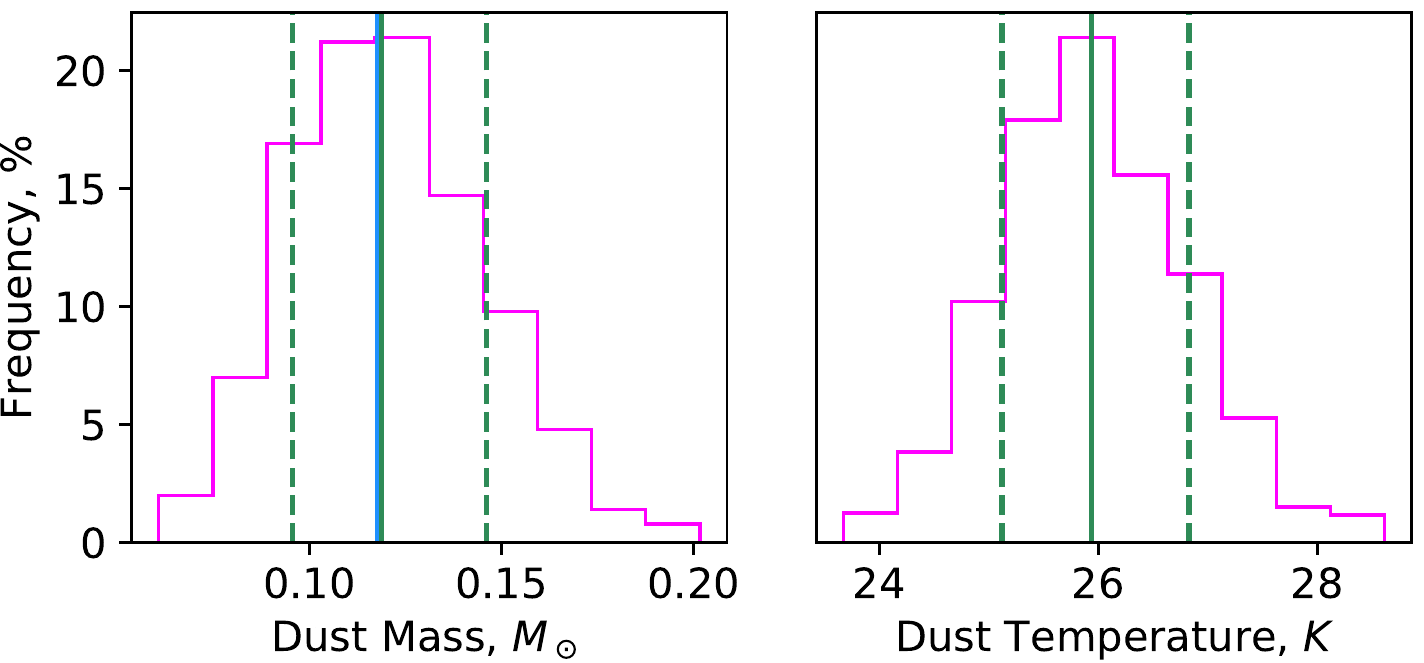}
	\caption{\textit{Top}: Far infrared SED of G21.5$-$0.9.
	\textit{Bottom}: The distribution of source masses obtained by the Monte Carlo routine where SEDs were fit to 1000 normally distributed samples of flux values.
	The colour schemes are the same as for Figure~\ref{fig:G11.2-0.3SED}.
	The best fit and median values derived for dust temperature and mass are shown in Table~\ref{tab:TempMass}.}
	\label{fig:G21.5-0.9SED}
\end{figure}

\begin{figure}
	\centering
	\includegraphics[width=1.0\linewidth]{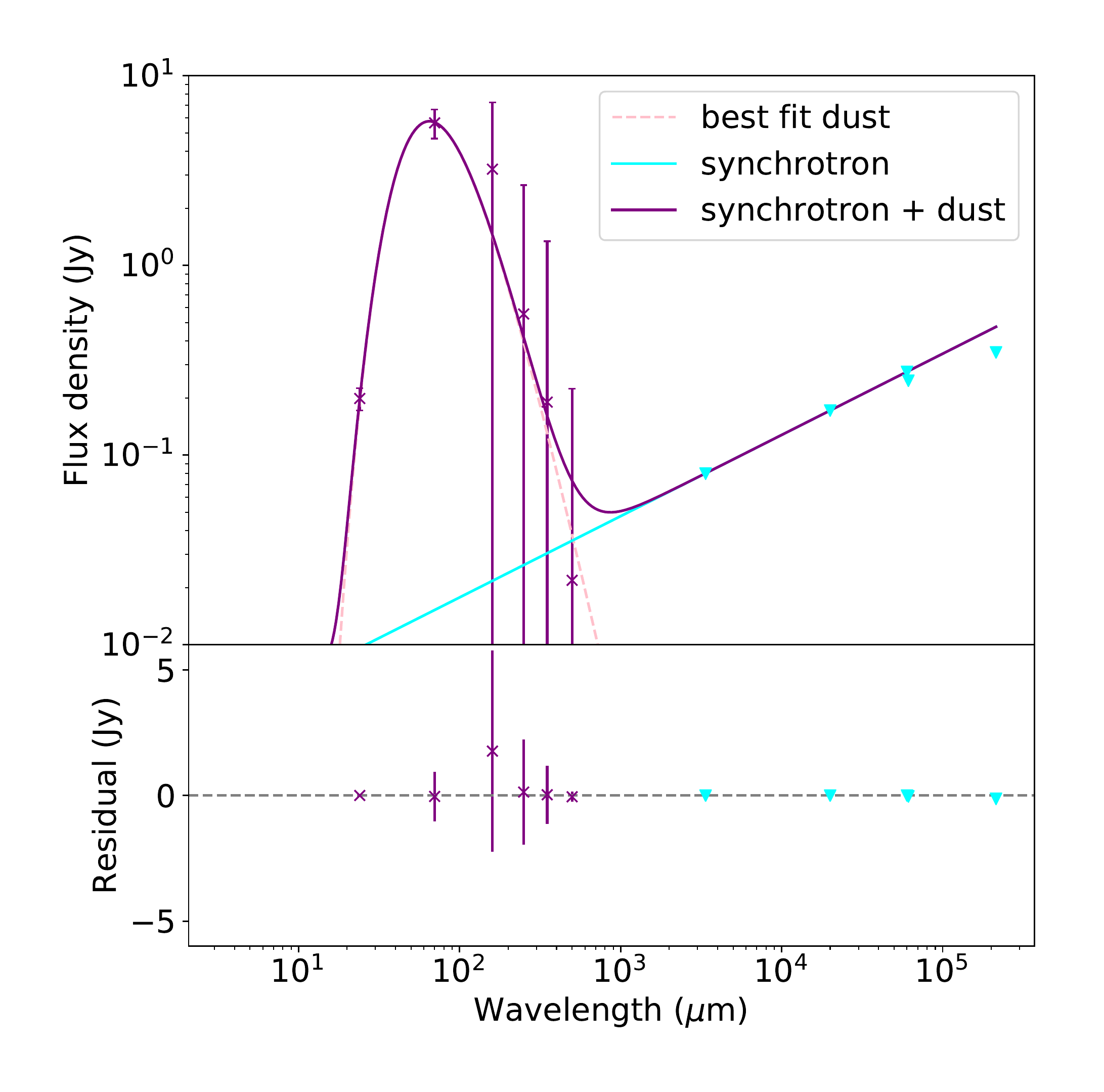}
	\includegraphics[width=1.0\linewidth]{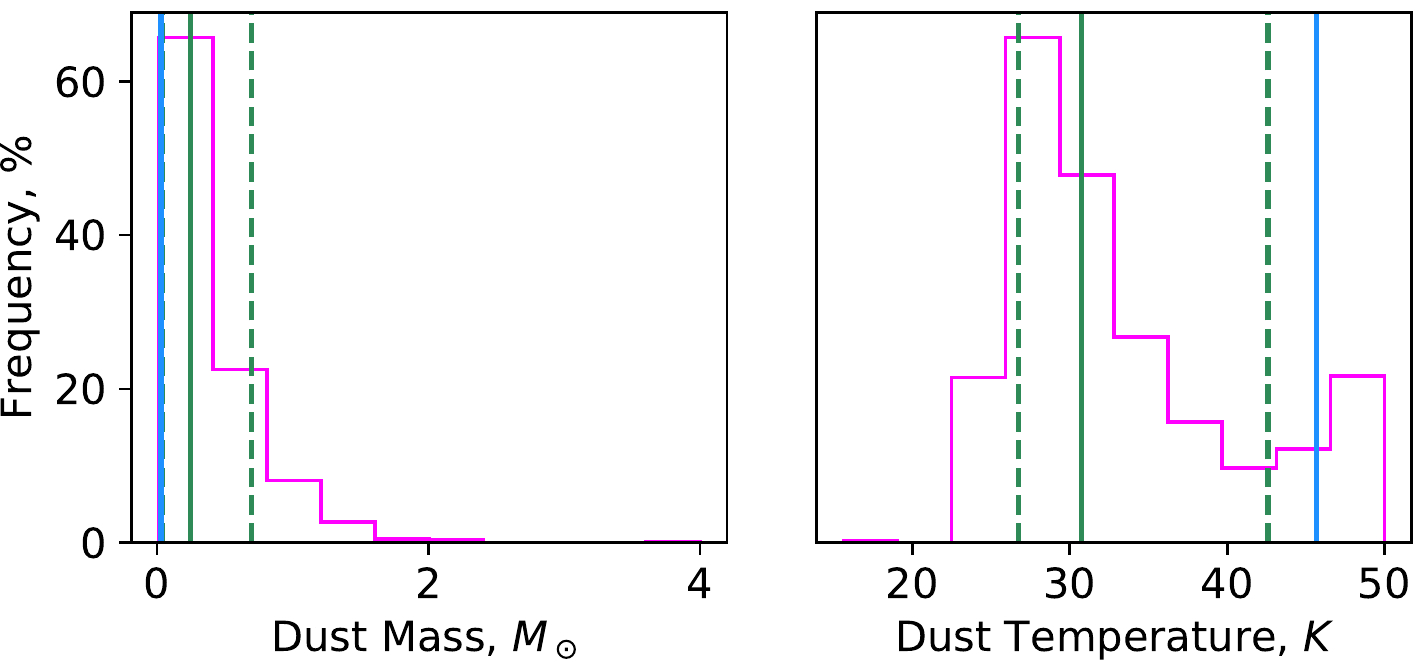}
	\caption{\textit{Top}: Far infrared SED of G29.7$-$0.3.
	\textit{Bottom}: The distribution of source masses obtained by the Monte Carlo routine where SEDs were fit to 1000 normally distributed samples of flux values.
	The colour schemes are the same as for Figure~\ref{fig:G11.2-0.3SED}.
	The best fit and median values derived for dust temperature and mass are shown in Table~\ref{tab:TempMass}.}
	\label{fig:G29.7-0.3SED}
\end{figure}

\textbf{G11.2$-$0.3:} We find that the thermal FIR emission between 24 and 350\,$\mu$m for this SNR is best described by the sum of two modified blackbodies with best-fit parameters for the cold temperature T$_c$\,=\,\ColdTempaBest\,K, and warm temperature T$_w$\,=\,\WarmTempaBest\,K. The cold component requires a best-fit mass of M$_d$\,=\,\ColdMassaBest\,M$_\odot$ and the warm component requires a mass of M$_d$\,=\,\WarmMassaBest\,M$_\odot$, as seen in Figure~\ref{fig:G11.2-0.3SED}.

There is large uncertainty in the mass and temperature of the warmer dust component as this peak mostly depends only on the 24\,$\mu$m flux. This makes the position of the SED peak extremely uncertain, giving a wide range of possible temperatures and masses. The Monte Carlo method gives a skewed distribution of possible values for the cold dust mass and temperature. There is a tail of high dust masses, skewing the median to values $>1\,\rm\,M_{\odot}$.
\bigskip

\textbf{G21.5$-$0.9:} We find that the global thermal emission in this SNR is well described by a single temperature dust component with a best-fit temperature of T\,=\,\ColdTempbBest\,K and best-fit mass of M$_d$\,=\,\ColdMassbBest\,M$_\odot$ as seen in Figure~\ref{fig:G21.5-0.9SED}.
The Monte Carlo method gives roughly Gaussian-shaped distributions of dust masses and temperatures and the result from this analysis is in agreement with the best fit result.
\bigskip

\textbf{G29.7$-$0.3:} The thermal emission in this SNR can be fitted by a single temperature dust component with a best-fit temperature of T = \ColdTempcBest\,K and dust mass of M$_d$\,=\,\ColdMasscBest\,M$_\odot$ as seen in the top panel of Figure~\ref{fig:G29.7-0.3SED}. There is a large difference in the ISM flux level between the north and south of this remnant making it difficult to obtain an accurate value for the ISM level, as indicated by the large error bars in the SNR fluxes. At wavelengths longer than 70\,$\mu$m the error bars indicate that a dust mass cannot be determined as the source flux is at a similar level to that of the ISM.
The median dust temperature from the Monte Carlo analysis is low compared to that of the best fit, resulting in a larger estimate for the dust mass. This discrepancy is caused by large uncertainty in the flux at long wavelengths, as indicated by the error bars in Figure~\ref{fig:G29.7-0.3SED}. A large fraction of the simulated fluxes at long wavelengths are much greater than the measured value, forcing the SED to peak at a longer wavelength. Therefore the Monte Carlo analysis is unable to give a constrained dust mass in these cases, although we clearly detect SN dust at a warmer temperature than the local ISM. This will be revisited in Section~\ref{ppmap} where the dust temperature and mass will be analysed with a more rigorous routine.

\bigskip

\begin{table*}
	\centering
	\begin{tabular}{c c c c c c c c c c c}
	\hline
	& & \multicolumn{4}{|c}{Best Fit\textsuperscript{a}}
	& \multicolumn{4}{|c}{Monte Carlo Median\textsuperscript{b}}
	& PPMAP\textsuperscript{c} \\

	& Distance & \multicolumn{2}{|c}{Cold dust} & \multicolumn{2}{|c}{Warm dust} & \multicolumn{2}{|c}{Cold dust} & \multicolumn{2}{|c}{Warm dust} & Cold Dust\\

	SNR & (kpc) & T\textsubscript{d} & M\textsubscript{d} & T\textsubscript{d} & M\textsubscript{d} & T\textsubscript{d} & M\textsubscript{d} & T\textsubscript{d} & M\textsubscript{d} & M\textsubscript{d} \\ \hline\hline
	G11.2$-$0.3 & 4.4 \textsuperscript{1}
	& \ColdTempaBest & \ColdMassaBest & \WarmTempaBest & \WarmMassaBest
	& \ColdTempa\,$^{+\ColdTempaP} _{-\ColdTempaN}$
	& \ColdMassa\,$^{+\ColdMassaP} _{-\ColdMassaN}$
	& \WarmTempa\,$^{+\WarmTempaP} _{-\WarmTempaN}$
	& (\WarmMassa\,$^{+\WarmMassaP} _{-\WarmMassaN}) \times 10^{-4}$
	& \PPMAPMassa\,$\pm$\,\PPMAPMassErra \\

	G21.5$-$0.9 & 4.7 \textsuperscript{2}
	& \ColdTempbBest & \ColdMassbBest & $-$ & $-$
	& \ColdTempb\,$\pm$\,\ColdTempbP
	& \ColdMassb\,$\pm$\,\ColdMassbP
	& $-$ & $-$
	& \PPMAPMassb\,$\pm$\,\PPMAPMassErrb \\

	G29.7$-$0.3 & 10.6 \textsuperscript{3}
	& \ColdTempcBest & \ColdMasscBest & $-$ & $-$
	& \ColdTempc\,$^{+\ColdTempcP} _{-\ColdTempcN}$
	& \ColdMassc\,$^{+\ColdMasscP} _{-\ColdMasscN}$
	& $-$ & $-$
	& \PPMAPMassc\,$\pm$\,\PPMAPMassErrc \\ \hline

	\end{tabular}
	\caption{
	The estimated mass (M$_\odot$) and temperature (K) of dust within each SNR derived from:
	\textsuperscript{a} a least squares SED fit to the data;
	\textsuperscript{b} median results from a Monte Carlo routine as described in Section~\ref{DustMasses};
	\textsuperscript{c} the PPMAP analysis as described in Section~\ref{ppmap}, where we do not assume a single dust temperature.
	Source distances are from: \textsuperscript{1}the near distance from HI absorption\citep{Green2004}; \textsuperscript{2}HI and CO observations \citep{Camilo2005}; \textsuperscript{3}the kinematic distance of an associated molecular cloud \citep{Su2009}.
	}
	\label{tab:TempMass}.
\end{table*}

\section{Analysing the dust properties with point process mapping (PPMAP)} \label{ppmap}
Although SED fitting, as described in Section \ref{DustMasses}, can provide approximate dust mass and temperature estimates, as well as the synchrotron contribution, it is limited. We assume that the dust has a uniform temperature and a fixed emissivity index of 1.9. 
Next, we employ point process mapping (PPMAP) \citep{Marsh2015, Marsh2017} in order to overcome these shortfalls and produce maps of differential column density across our objects at different temperatures and values of emissivity index.
Using point spread function (PSF) information PPMAP is able to create column density maps without smoothing data to a common resolution. We can therefore use the highest resolution available to disentangle the dust emissivity index $\beta$ and temperature components at any point, and to evaluate their variation across the source.
We apply this technique to study G11.2$-$0.3, G21.5$-$0.9, and G29.7$-$0.3 using the 24\,--\,500\,$\mu$m maps, and compare with the surrounding ISM for which we use only the \textit{Herschel} maps due to potential issues with optical thickness, discussed in more detail later.

PPMAP operates in different dimensions to conventional ($\beta$, T) pixel-to-pixel fits, instead estimating a density distribution of mass throughout the 4D state space $(x, y, T, \beta)$.
The PPMAP procedure is described by \citet{Marsh2015}. In this procedure, astrophysical systems such as SNRs are represented as a collection of primitive objects which each have unit column density.  An object is characterised by its dust temperature (T$_D$), emissivity index ($\beta$), and location on the 2D plane of the sky ($x, y$). 
Assuming that structures are optically thin, flux maps are considered as the total instrumental response to all objects. The distribution of column density is then defined by the number of objects appearing at any point in the 4D state space $(x, y, T_D, \beta)$ . 

PPMAP aims to find the optimal estimate for the distribution of objects by using a procedure based on Bayes' theorem. The expectation number of objects per unit volume in the 4D state space, $\rho(x, y, T_D, \beta)$, can be equated to the differential column density and is found through a stepwise approach. Initially, the measurement noise is artificially increased to the point at which no information can be obtained from the data. At this point the optimal solution is the a priori value of $\rho$. In the absence of further prior information, this function is flat everywhere. 
The noise is then decreased over a series of time-steps until the original signal-to-noise ratio is obtained; $\rho$ is updated at each step, using the previous optimal solution as the new prior. The data at each stage can be described by a model of the form

\begin{equation} \label{eqn:ppmap_data_model}
	\mathbf{d} = \mathbf{A}\mathbf{\Gamma(t)} + \nu(t)
\end{equation}

where \textbf{d} is the data vector at a given time step whose $m^{th}$ component gives the pixel value at $x_m$, $y_m$ for the wavelength $\lambda_m$; the vector $\mathbf{\Gamma(t)}$ describes the actual distribution of objects in state space; $\nu(t)$ gives the measurement noise at each time step, $t$, and is assumed to be Gaussian; and $\mathbf{A}$ is the system response matrix where the $m^{th}$ element gives the $m^{th}$ measurement response to an object in the $n^{th}$ cell of state space i.e. $(x_n, y_n, T_n, \beta_n)$. It is given by

\begin{equation} \label{eqn:ppmap_response}
	A_{mn} = H_{\lambda_m}(X_m - x_n, Y_m - y_m) K_{\lambda_m}(T_n) B_{\lambda_m}(T_n) \kappa_{\lambda_m}(\beta_n) \Delta \Omega_m
\end{equation}

where $H_{\lambda_m}(x, y)$ is the convolution of the PSF at $\lambda_n$ with the profile of a source object; $K_{\lambda_m}(T_n)$ is a colour correction; $B_{\lambda_m}(T_n)$ is the Planck function; $\Delta \Omega_m$ is the solid angle subtended by the $m^{th}$ pixel. By finding an optimal solution for \textbf{A}, we can obtain the most likely distribution of $\beta$ and $T_D$ at each location in $(x, y)$.

The PPMAP procedure is applied to estimate the column density over a grid of 12 temperature bins, centred at temperatures equally spaced in $\mathrm{log}(T_D)$, and up to 7 values of $\beta$ between 0 and 3 \citep{Marsh2015}. As the PWN dust has best-fit modified blackbody temperatures of $>25\,\rm K$, we initially use a temperature grid ranging from 25 to 75\,K. If there is emission from components outside of this temperature range, PPMAP will return higher column density at the bounding temperatures. We therefore repeat the analysis extending our temperature range (20\,--\,90\,K) to search for cooler and warmer components. 
For this study a Gaussian prior is assumed for distribution of material across $\beta$, with a mean of 1.9 \citep{Planck2016beta} and standard deviation of 0.25.
Providing PPMAP with prior information suppresses the temperature--$\beta$ anti-correlation by excluding unrealistic results.

We estimate the cool dust mass of each SNR by summing the column density within the apertures described in Section~\ref{DustMasses} across the entire temperature and $\beta$ grids. By averaging the temperature maps, PPMAP allows the amount of dust column density for each $\beta$ value to be determined.
After summing the differential column density over temperature, we obtain the density-weighted mean $\beta$ along each line of sight.

\subsection{Revised dust masses for our three PWNe} \label{ppmapResults}

\begin{figure*}
	\includegraphics[width=0.41\linewidth]{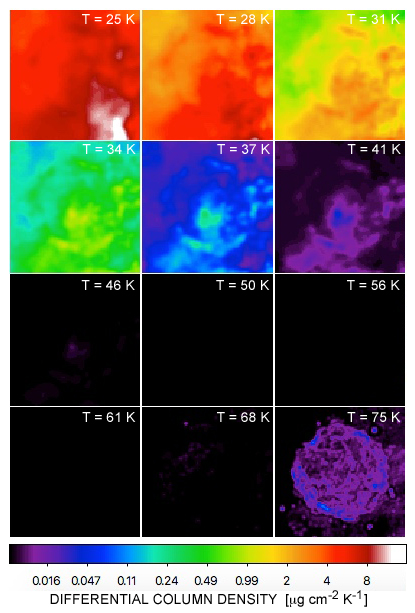}
	\includegraphics[width=0.4\linewidth]{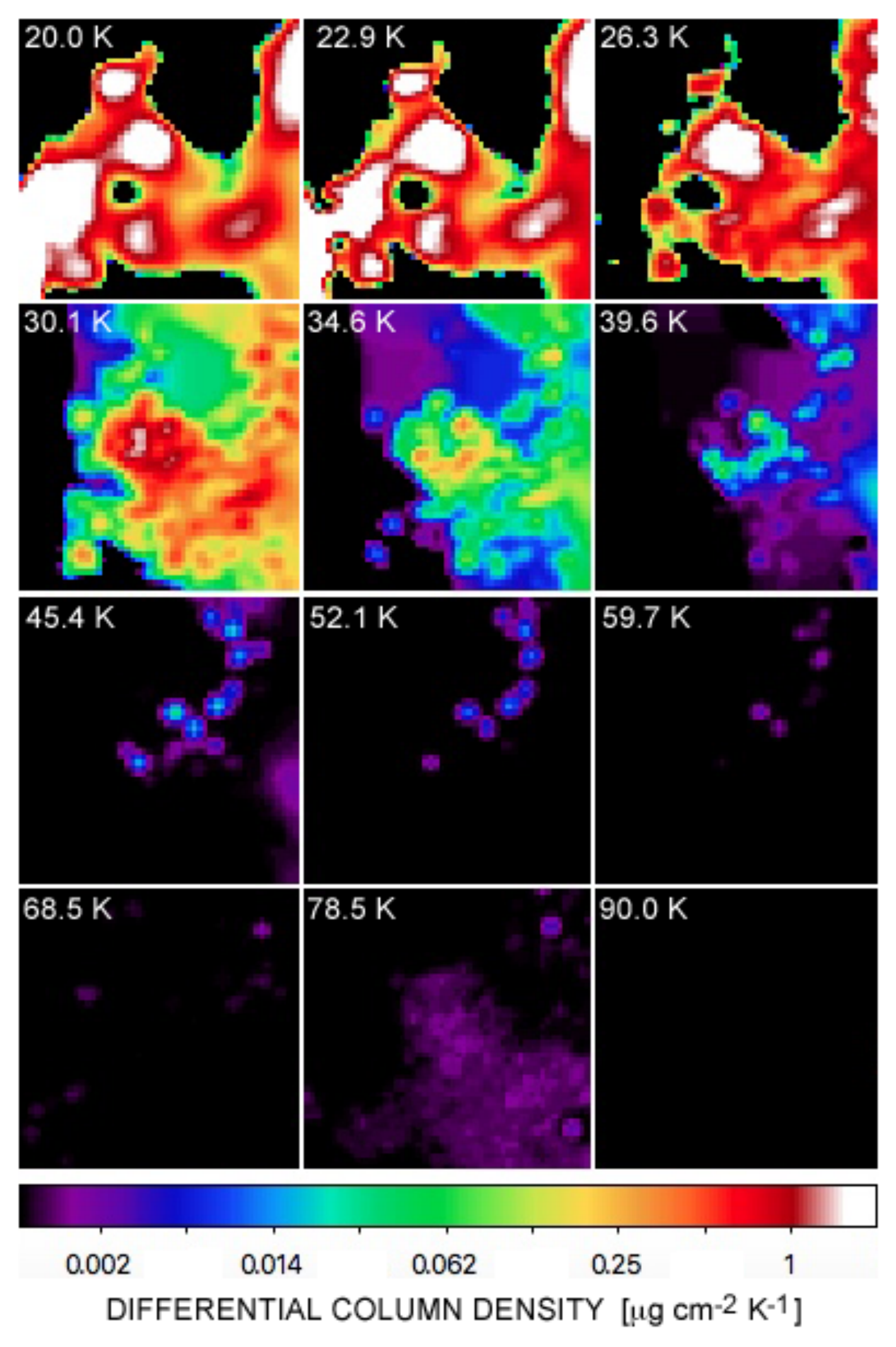}
	\includegraphics[width=0.4\linewidth]{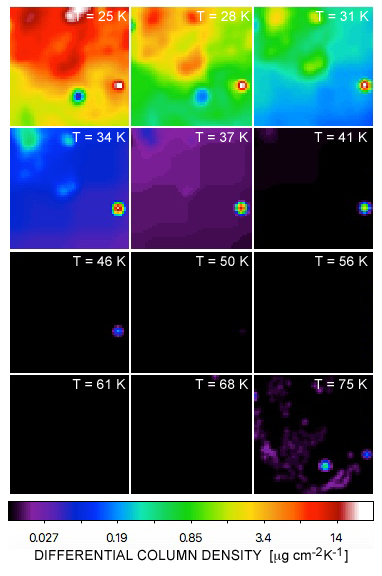}
	\caption{PPMAP generated maps of differential dust column density split in different temperature ranges for \textit{top-left:} G11.2$-$0.3, \textit{top-right:} G21.5$-$0.9, and \textit{bottom:} G29.7$-$0.3. The corresponding dust temperature is indicated in the top-right of each panel. At temperatures less than 25\,K in G11.2$-$0.3 and G29.7-0.3, the column density map begins to be dominated by unrelated interstellar dust along the line of sight and thus temperatures below 25\,K for these sources are not used in the analysis.}
	\label{fig:G11_temps}
\end{figure*}

\begin{figure}
	\includegraphics[width=0.95\linewidth]{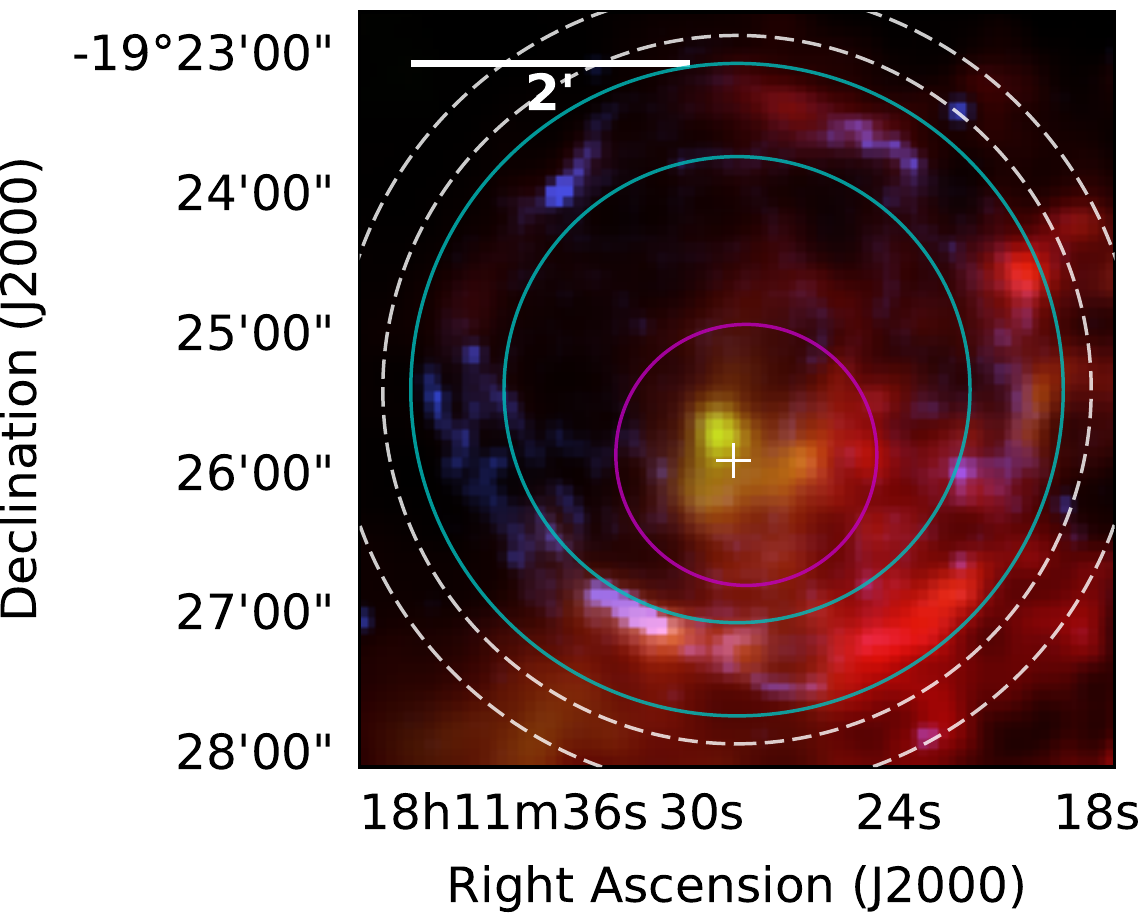}
	\includegraphics[width=0.95\linewidth]{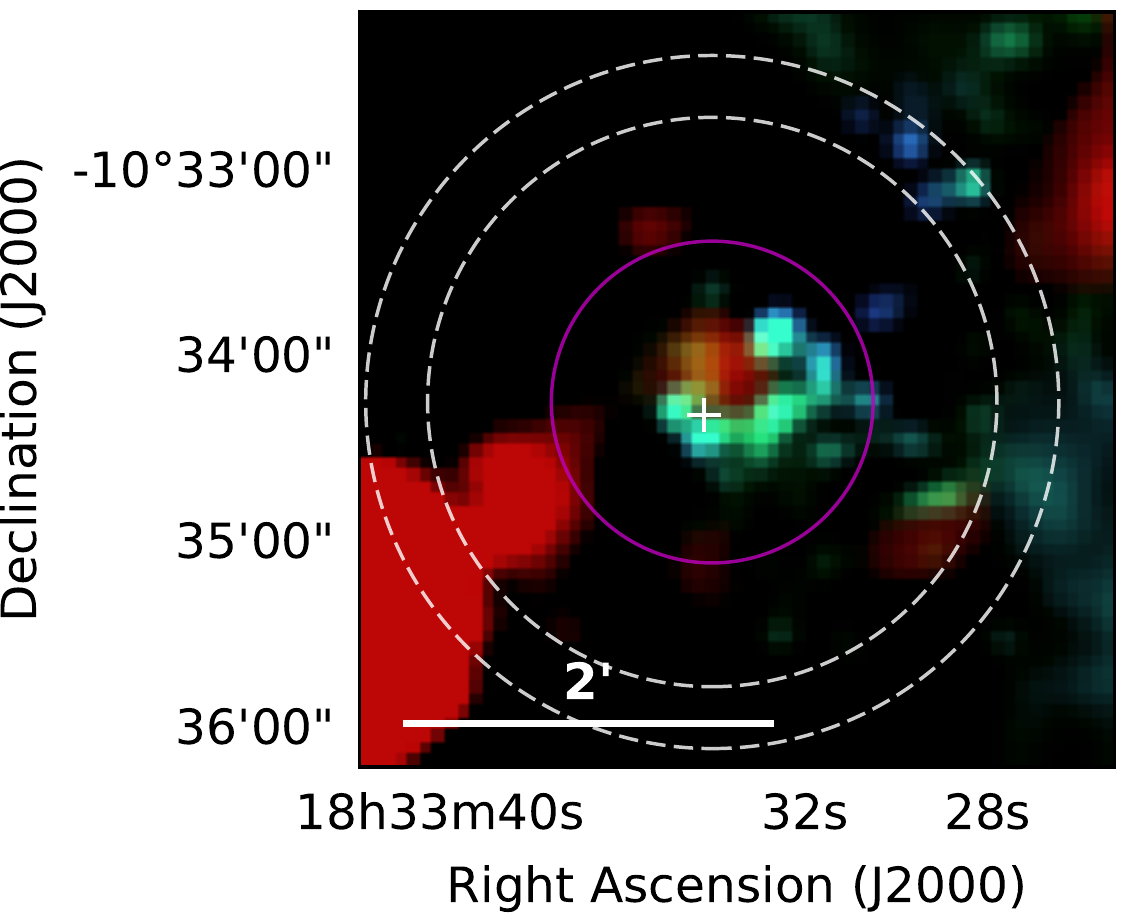}
	\includegraphics[width=0.95\linewidth]{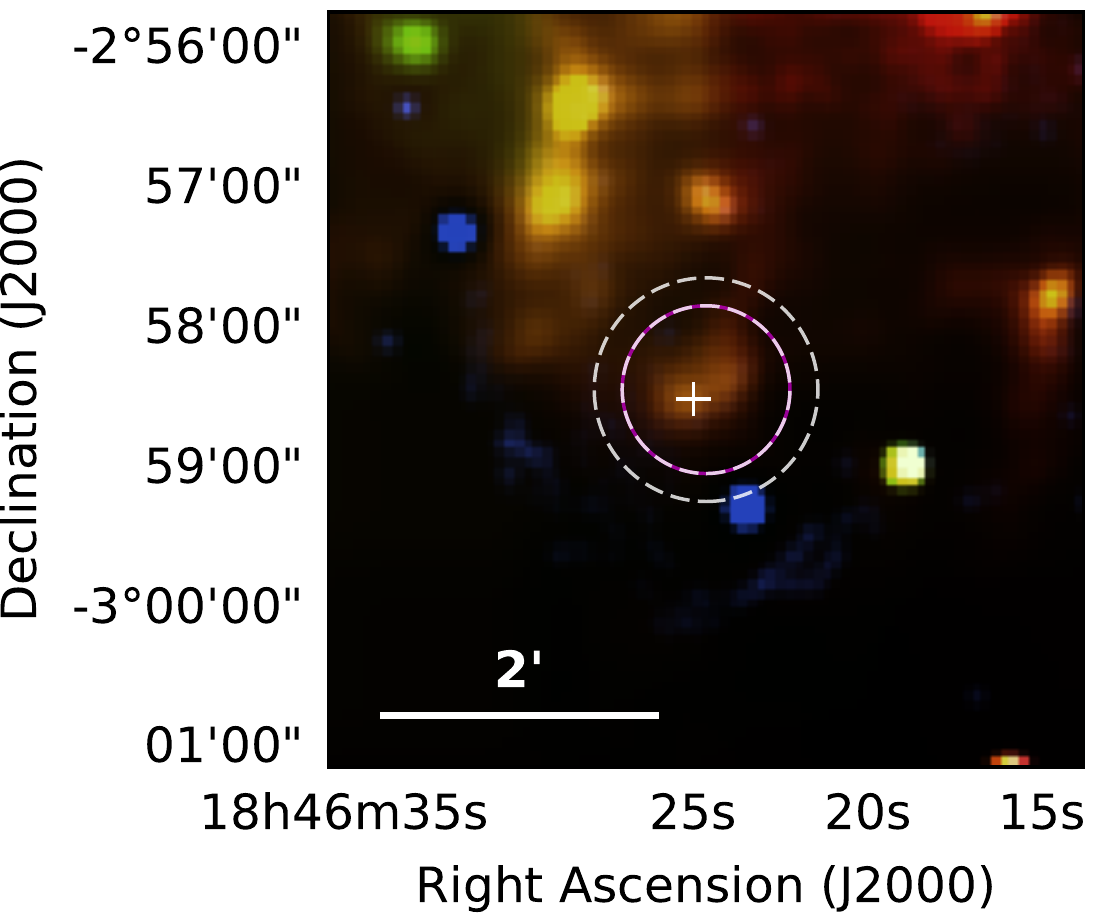}
	\caption{
	PPMAP-generated three colour maps of differential dust column density created using dust temperature slices from Figure~\ref{fig:G11_temps}. Colours show dust in G11.2$-$0.3 (\textit{top}): 31\,K (red), 41\,K (green), and 75\,K (blue). In G21.5$-$0.9 (\textit{middle}): 20\,K (red), 34.6\,K (green), and 39.6\,K (blue). In G29.7$-$0.3 (\textit{bottom}): 28\,K (red), 31\,K (green), and 75\,K (blue). The white crosses indicates the X-ray centres. The magenta circles show the apertures used for PWN dust, the cyan and white circles were used to analyse shell and/or ISM emission respectively.
	}
	\label{fig:SNR_col}
\end{figure}

\begin{figure}
	\includegraphics[width=0.95\linewidth]{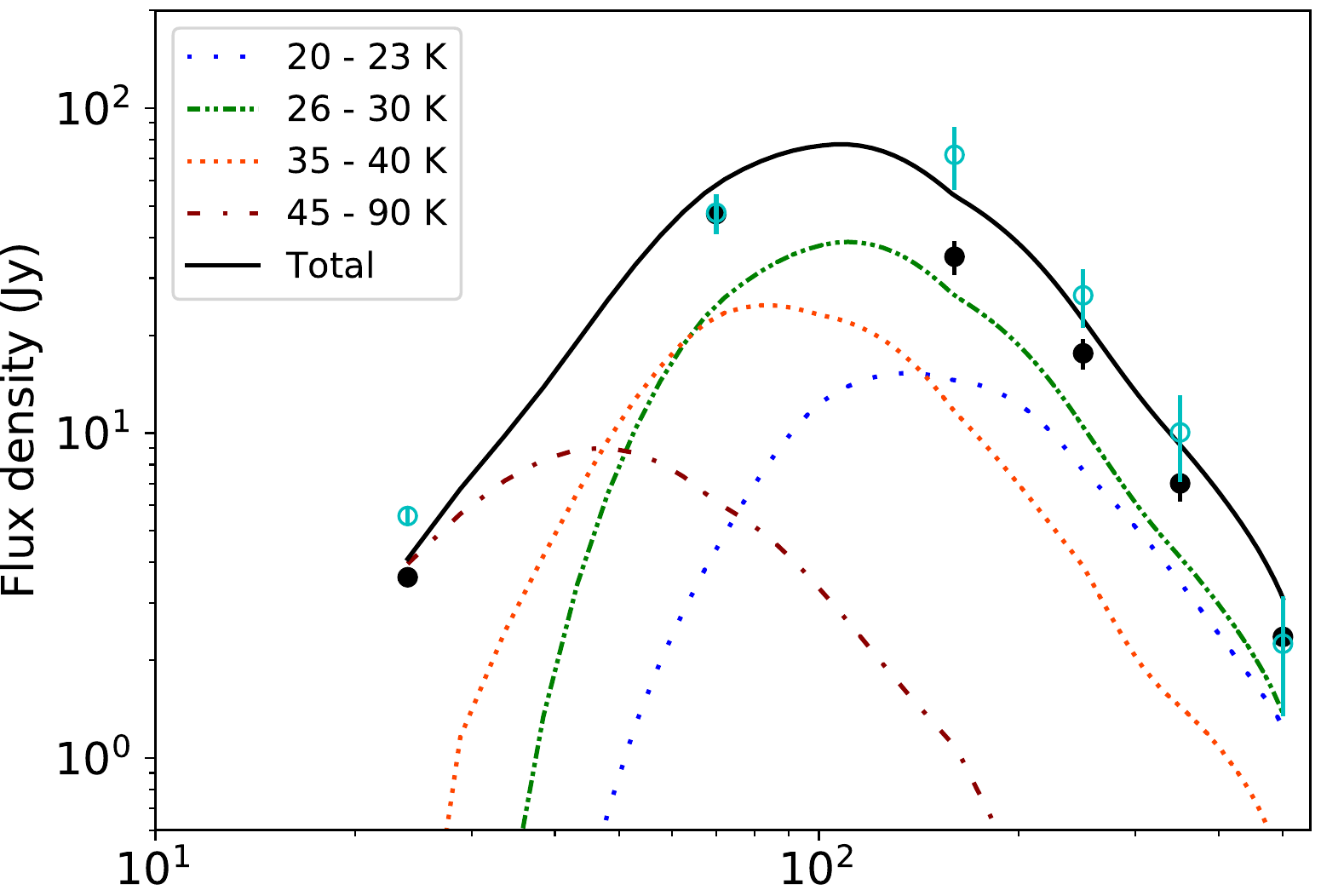}
	\includegraphics[width=0.95\linewidth]{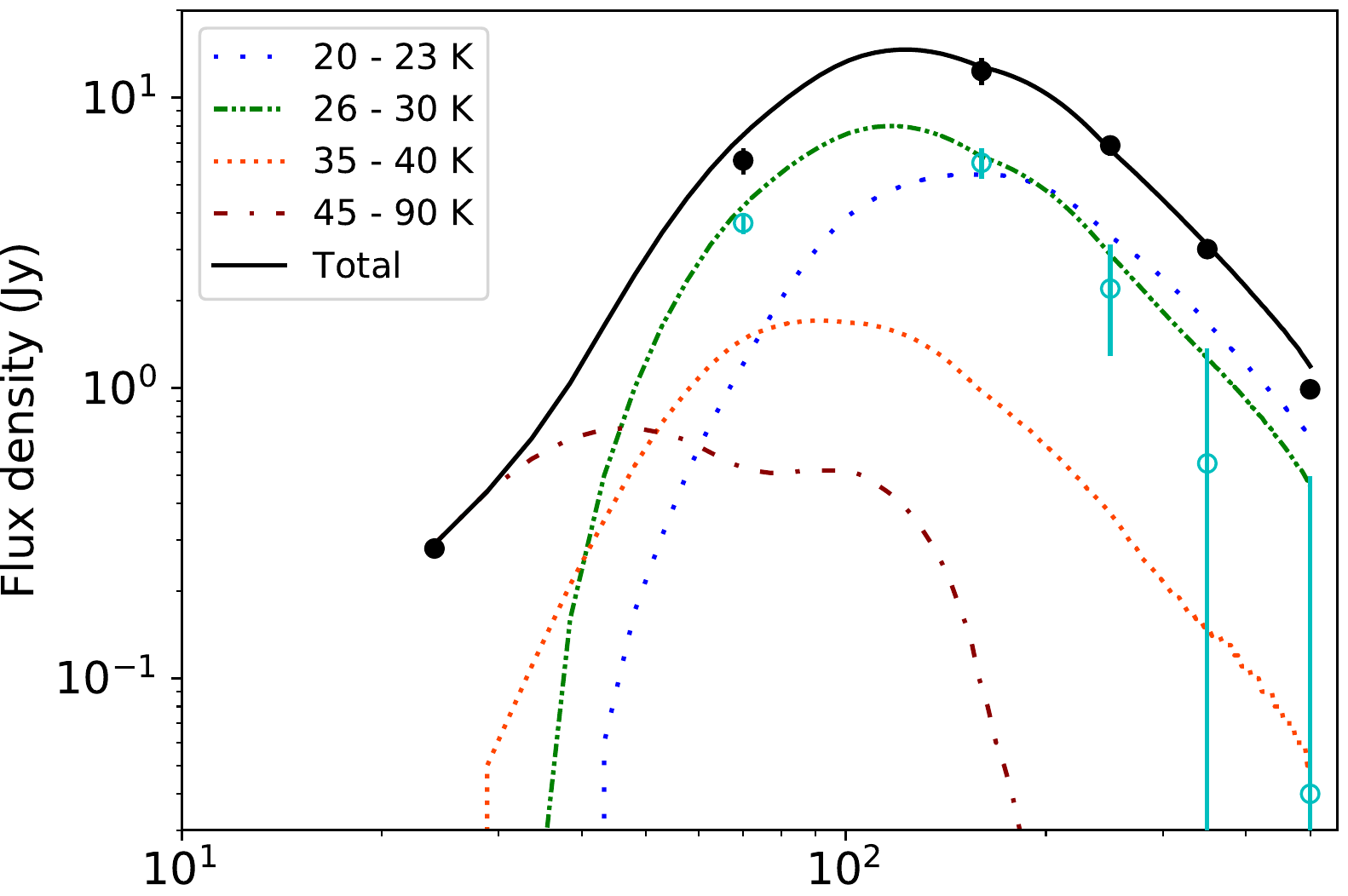}
	\includegraphics[width=0.95\linewidth]{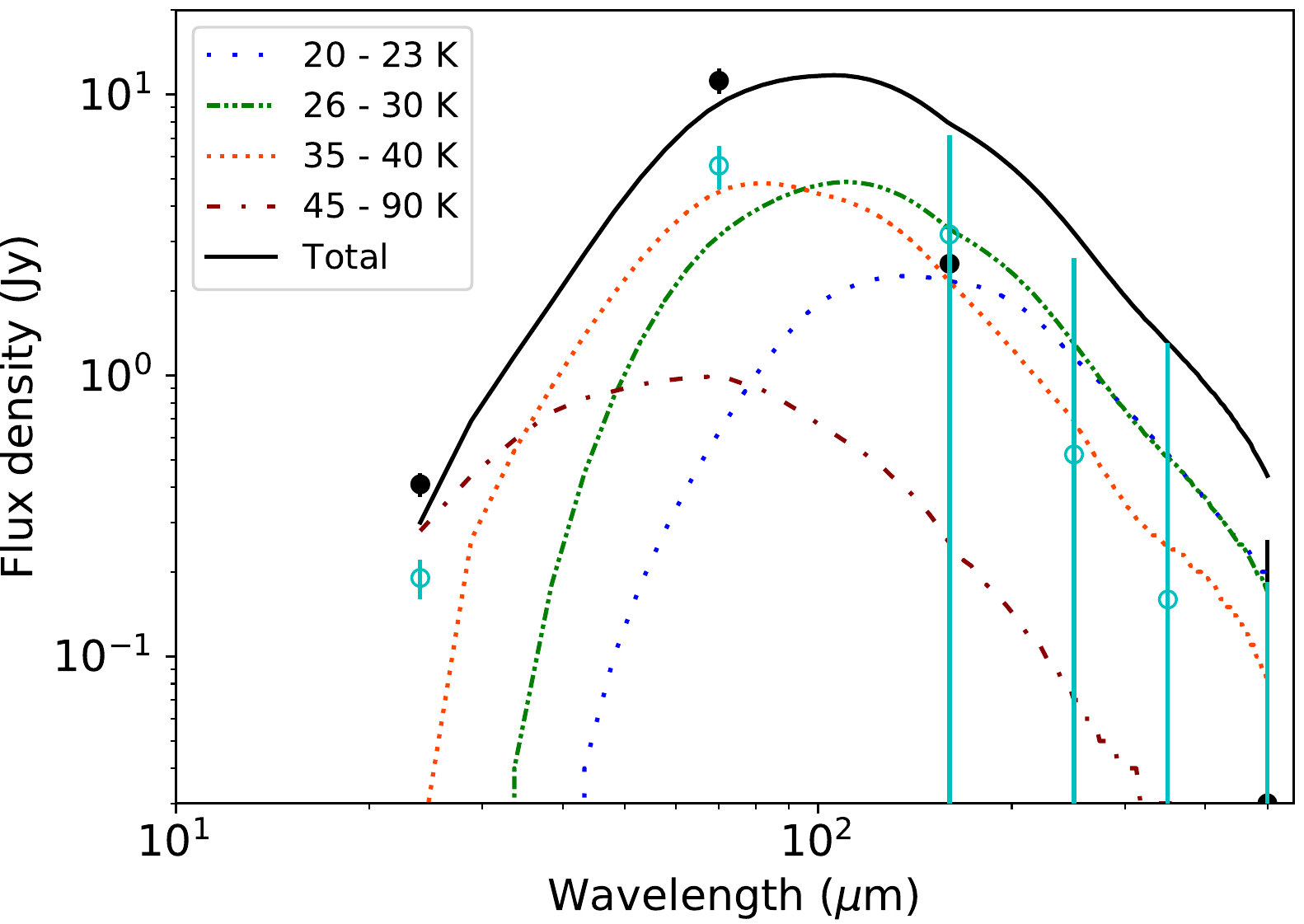}
	\caption{FIR SED as measured with PPMAP, analysing dust at temperatures between 20 and 90\,K, for the PWN in G11.2$-$0.3 (\textit{top}), G21.5$-$0.9 (\textit{middle}), and G29.7$-$0.3 (\textit{bottom}). The open markers indicate the thermal flux estimated by aperture photometry in Section~\ref{DustMasses} and the filled markers indicate the estimated PPMAP flux in Section~\ref{ppmap}, where the difference is due to a variation in the estimate of the ISM level. Both sets of fluxes are background and synchrotron subtracted.
	{There is a considerable contribution to the SEDs of both G11.2$-$0.3 and G29.7$-$0.3 from a cold dust (20\,--\,23\,K) component. By visually inspecting the columns density maps for these sources we find that dust at temperatures below 25\,K are not related to the SNR. These cool dust components are therefore excluded when estimating the dust masses to avoid contamination from the ISM.}
	}
	\label{fig:PPMAP_SED}
\end{figure}

The column density map of dust emission at each of the temperature grid points from PPMAP is shown in Figure~\ref{fig:G11_temps} for the three SNRs, G11.2$-$0.3, G21.5$-$0.9, and G29.7$-$0.3.
Prior to running the PPMAP analysis, an average ISM level is subtracted from the SNRs. This is estimated from annuli around the SNRs, as shown in Figure~\ref{fig:SNR_col}, which are sigma-clipped to remove especially bright objects. The dust mass within each PWN is estimated by aperture photometry on the column density maps. This involves a second background subtraction, again using an average level from the annuli in Figure~\ref{fig:SNR_col}, to ensure that there is minimal ISM contributing to our estimate. Figure~\ref{fig:PPMAP_SED} shows the total SED for each remnant and the contribution from each temperature component, as estimated by PPMAP.


\bigskip

\textbf{G11.2$-$0.3:}
In Figure~\ref{fig:G11_temps} we can see that the SNR is indistinguishable from the surrounding ISM at 25\,K highlighting the difficulty in removing ISM emission where the temperatures are similar to dust in the ISM. Dust in the central region is brightest between 34 and 41\,K and is not detected above 46\,K.

By contrast, there is evidence of two temperature components in the shell as we find that there is dust at 28\,--\,41\,K and at 75\,K. The tightly bound inner shell could indicate that the warmer shell dust may be reverse-shock heated, as seen in Cassiopeia A \citep{Rho2008}. However, it is thought that the reverse shock has already reached the centre of this remnant and expansion into an anisotropic CSM has caused the sharpness of the shell's inner edge \citep{Borkowski2016} so it is unclear what is heating the shell dust.  The dust structures are clearly seen when combining three of the temperature maps from Figure~\ref{fig:G11_temps} (31, 41, and 75\,K) to produce a `super-resolved' colour image in Figure~\ref{fig:SNR_col} (top panel). The dust emission from the PWN towards the centre is very clear in this image compared to the three-colour image derived using the native \textit{Herschel} maps (Figure~\ref{fig:G11.2-0.3Image}) as the PPMAP results match the best available \textit{Herschel} angular resolution i.e. $6.4''$ from the 70$\mu$m images \citep{Traficante2011}.

{The PPMAP generated SED for G11.2$-$0.3 in Figure~\ref{fig:PPMAP_SED} shows how some of the temperature components revealed using PPMAP contribute to the total SED (here we show the full range investigated using PPMAP i.e. 20-90\,K).  We can see that the 20-23\,K component contributes a significant amount of the total FIR flux measured in the apertures even after background subtracting the ISM. Since we have previously discussed that the PPMAP dust structures seen at temperatures of 20-25\,K for this source are likely unrelated to the supernova, we therefore subtract these colder dust components from the final PPMAP derived dust mass. The dust mass is therefore derived by summing the column densities at the range of temperatures where SNR-related emission is seen, providing an estimated dust mass for the central ejecta region of G11.2$-$0.3 of M$_d$\,=\,\PPMAPMassa\,$\pm$\,\PPMAPMassErra\,M$_\odot$ (Table~\ref{tab:TempMass}).\footnote{If we were to include the PPMAP dust components at temperatures below 25\,K, the dust mass for G11.2$-$0.3 would be M$_d$\,=\,\PPMAPMassaWide\,$\pm$\,\PPMAPMassErraWide\,M$_\odot$.
}


{This dust mass estimate is considerably smaller than the the traditional modified blackbody fits in Section~\ref{DustMasses}. This is due to the combination of differences between the two methods in the background subtraction and therefore the final FIR fluxes attributed to the SNR (this discrepancy is larger between 160 and 350\,$\mu$m) and the fact that by inspecting the temperature components visually with PPMAP, we were able to conclude that dust at temperatures below 25\,K were likely not SN-related, hence resulting in an additional subtraction from the final PPMAP dust mass.  Indeed Section~\ref{DustMasses}, suggests that the cool dust in G11.2 is at 26.7\,K whereas PPMAP finds that part of this dust is at warmer temperatures, thus giving a smaller estimate for the total dust mass.
}

\bigskip

\textbf{G21.5$-$0.9:}
As shown in Figure~\ref{fig:G11_temps}, for this source we need to extend the temperature grid down to 20\,K to show all the dust features, as there is a clear detection of PWN dust at temperatures of 20\,--\,25\,K.  The cold dust component is visible in the grid at temperatures from 20\,--\,27\,K across the entire PWN region.  Warmer dust between 30 and 37\,K forms a shell-like structure close around the south-west of the central region in which the pulsar is located. The distribution of dust in this source is similar to seen in G54.1$+$0.3, a bright peak of dust is evident to the north-west of the PWN shell which is significant and is robust to changes in mapping parameters. The PPMAP three-temperature map for G21.5$-$0.9 is shown in Figure~\ref{fig:SNR_col} (middle panel) where these features are clearly visible (with the coldest dust emission highlighted in red).

Using PPMAP, we retrieve a dust mass for G21.5$-$0.9 of M$_d$\,=\,\PPMAPMassb\,$\pm$\,\PPMAPMassErrb\,M$_\odot$ (including all the emission from 20-90\,K). This is $\sim 2.5$ times larger than both the best-fit and median dust masses quoted in Section~\ref{DustMasses}.
{Figure~\ref{fig:PPMAP_SED} indicates that the fluxes used for the PPMAP analysis are greater than those used in Section~\ref{DustMasses}. This is due to a combination of differences in the level of the background emission derived between PPMAP and Section~\ref{DustMasses} (this leads to a factor of 1.5\,--\,2 difference in the FIR fluxes attributed to the SNR) and PPMAP allowing for a wider range of temperatures in each pixel instead of putting all of the dust at a single temperature in the modified blackbody fits. }

Thus G21.5$-$0.9 joins Cas A and G54.1 in the small list of Galactic SNRs for which dust at temperatures $<25$\,K has been detected. PPMAP, with its better resolution and ability to map dust at different temperatures, has revealed significantly more dust in this source than from traditional modified blackbody fits.

\bigskip

\textbf{G29.7$-$0.3:}
In Figure~\ref{fig:G11_temps}, the PWN structure clearly contains dust at 28\,--\,31\,K, although the surrounding medium is also at a similar temperature making the detection more confused than the other sources. There is warm (75\,K) dust in the shell around the south-east of the SNR at the location of 24 and 70\,$\mu$m emission observed in Figure~\ref{fig:G29.7-0.3Image}. This emission likely arises from shocked dust. Similar to G11.2$-$0.3, there is some evidence for warm dust in filaments within the shell, however this is much fainter and more confused in this case.

{Similarly to G11.2$-$0.3, when inspecting the temperature grid of G29.7$-$0.3 between 20 and 90\,K we do not see evidence for any material related to the SNR structure at temperatures below 25\,K.  The PPMAP-derived SED shown in Figure~\ref{fig:PPMAP_SED} shows that dust below these temperatures contributes significantly to the total FIR emission of this source, yet we have no evidence that this is associated with the SNR. Therefore we only use the temperature components revealed by PPMAP at 25\,K and above when determining the dust mass for this source. For the central ejecta region in G29.7$-$0.3, PPMAP therefore estimates a dust mass of M$_d$\,=\,\PPMAPMassc\,$\pm$\,\PPMAPMassErrc\,M$_\odot$\footnote{If we were to include the PPMAP dust components at temperatures below 25\,K, the dust mass for G29.7$-$0.3 would be M$_d$\,=\,\PPMAPMasscWide\,$\pm$\,\PPMAPMassErrcWide\,M$_\odot$. In this case we expect that the estimate suffers from contamination by unrelated ISM and is superficially large.}. Again this is significantly larger than the best-fit values derived in Section~\ref{DustMasses}, although within the uncertainties it is consistent with the median from the Monte Carlo fits. This is not surprising since the single temperature best-fit SEDs can be biased towards allocating a warmer dust temperature ($\sim$46\,K), whereas the Monte Carlo estimate allows for random selection of lower temperature components ($\sim$30\,K) and therefore higher dust masses than the best-fit SED. The temperature grid from PPMAP in Figure~\ref{fig:G11_temps} shows emission from SN-related dust at lower temperatures than the best fit and therefore is more in agreement with the Monte Carlo estimate from Section~\ref{DustMasses}.
}


\subsection{Is the dust emissivity index different in SNe ejecta?}
\label{sec:betadiff}

\begin{figure}
\includegraphics[width=1.0\linewidth]{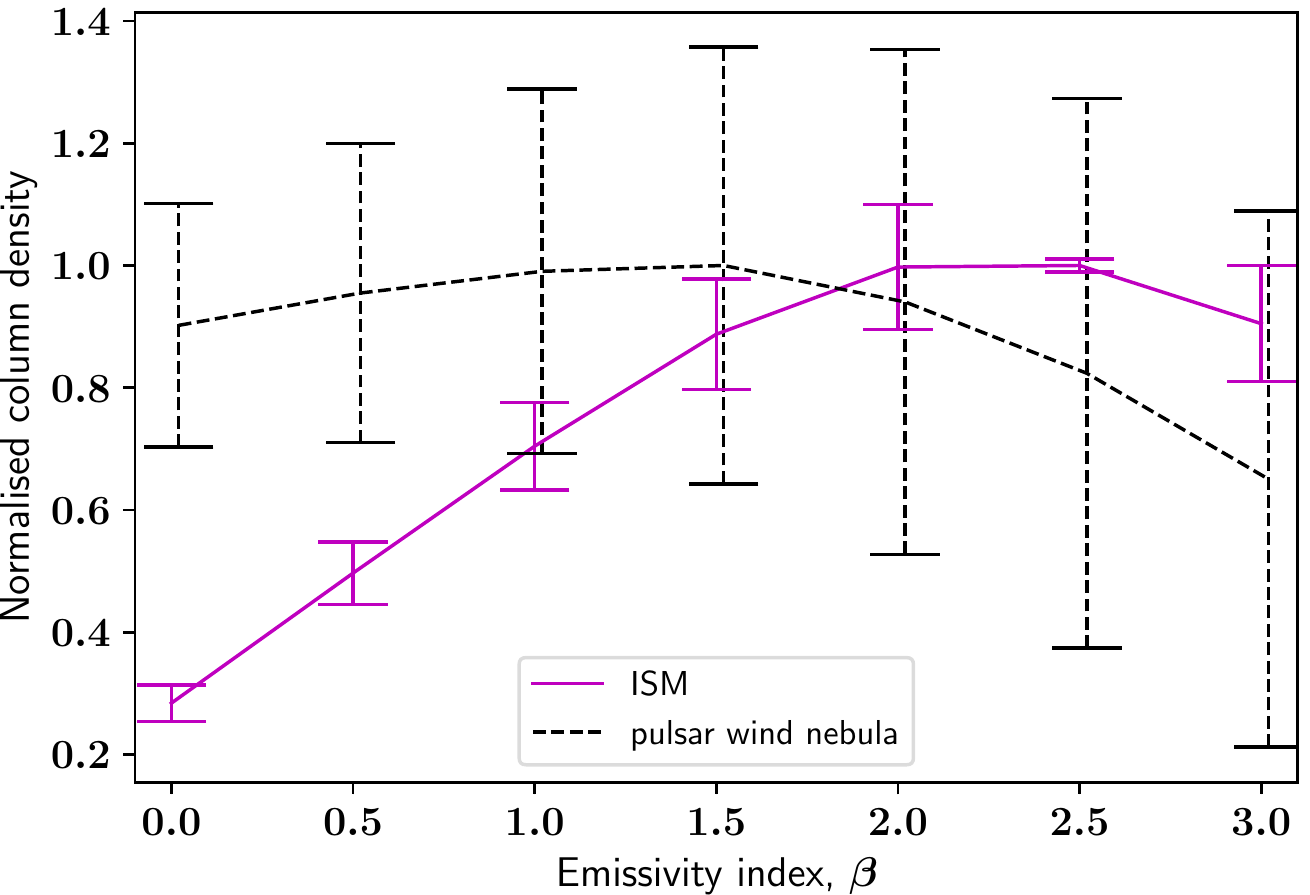}
	\caption{ Estimated column density within each region at a given $\beta$ for  G21.5$-$0.9, measured from temperature averaged PPMAP images. The column density in each region is normalised by dividing by the peak column density of that region for easier comparison of the two curves. This is the resulting dust column density $\beta$ profile assuming a flat $\beta$ prior. (Figure\,A1 shows the same profile from PPMAP assuming a Gaussian prior instead). The broad PWN profile reflects our inability to discriminate different values of $\beta$ with low S/N.
	}
	\label{fig:betadiff}
\end{figure}

PPMAP allows us to evaluate $\beta$ of the SNR and ISM dust as shown in Figure~\ref{fig:betadiff}.
We estimated the density-weighted mean value of $\beta$ within an aperture around the PWN and within in annuli encapsulating the surrounding medium and, for G11.2$-$0.3, the SNR shell (Figure~\ref{fig:SNR_col}). We can therefore compare the emissivity index of material within the SNR components to that of the surrounding ISM using a $\beta$ profile (Figure~\ref{fig:betadiff}).  As the SNR $\beta$ parameter is sensitive to the estimate of the ISM level, a lack of careful analysis can result in spurious variations in $\beta$. We  therefore subtract the ISM from the source flux prior to analysis of the column density using the annuli shown in Figure~\ref{fig:SNR_col}; this process is only applied to the SNRs, not to the ISM. This provides more robust results than subtracting the background of the resulting column density maps.

Additionally, the PPMAP analysis assumes that all media are optically thin, however, we find that the optical depth of the surrounding ISM in the 24\,$\mu$m image is not negligible. We therefore disregard this band when evaluating $\beta$ for the dust emission originating in the ISM, although this does not have a significant impact on any results for the cool dust components. The SNR material is likely optically thin and therefore the PPMAP analysis can be applied to this region using the 24\,$\mu$m image.

We compare the dust column density in the PWN and surrounding (unrelated) interstellar dust found in each $\beta$ `bin' using the standard PPMAP assumption of a Gaussian prior with peak 1.9 and $\sigma =0.25$. However we find inconclusive results for $\beta$ variations between the PWN and ISM in all three of our sources. These profiles are shown in Figure\,A1 for completeness. Although the dust column density profiles with $\beta$ appear to have a `peak' column density for $\beta$ of 1.8\,--\,2.0, we believe this is simply returning the prior distribution of $\beta$ for the dust in the shell, ISM emission and the PWN. To confirm this we vary the prior to a lower mean, in which case PPMAP returns a correspondingly lower peak value of $\beta$. Thus, there is a systematic difference in the $\beta$ of the ISM, shell, and PWN. However, the value of $\beta$ depends on the prior.

Next we test how robust the PPMAP $\beta$ analysis is given changes to the prior assumptions.  Assuming a flat $\beta$ prior instead of the Gaussian assumed earlier, we see no differences in the results for G11.2$-$0.3 and G29.7$-$0.3, but the column density-$\beta$ profile for G21.5$-$0.9 does change (Figure~\ref{fig:betadiff}). The dust column density in the ISM now appears to peak at $\beta_{\rm ISM} \sim 2.5$ whereas the PWN peaks at lower values of $\beta_{\rm PWN}\,=\,1.4 \pm 0.3$, though the error bars are large.
The wide dispersion in $\beta$ of the PWN is a result of the large uncertainties, reflecting our inability to discriminate different values of $\beta$ at the low signal-to-noise ratio.
As we do not apply a $\beta$ prior in this case we are more susceptible to $\beta$-temperature anti-correlation, which is included in this uncertainty. Although there are a range of possible solutions, as indicated by the large uncertainty, there is some indication of a variation.

The density-weighted mean value of $\beta$ along the line of sight for the ISM aperture is $\beta_{\rm ISM}\,=\,1.8 \pm 0.1$ (this takes into account which values of $\beta$ represents more of the mass), whereas for the PWN, this value is still consistent with the peak profile of $\beta\,=\,1.4$.

Finally we re-ran PPMAP using a Gaussian prior with mean of 1.4 (instead of 1.9). Again we found no differences in the results for SNRs G11.2$-$0.3 and G29.7$-$0.3. However, for G21.5$-$0.9, the new $\beta$ profile is very similar to Figure~\ref{fig:betadiff}, but with a reduced $\chi$-squared by $\sim$10\,per\,cent, supporting the (marginal) result that the dust emissivity index in this PWN is smaller than the canonical ISM value.

\subsection{Testing the reliability of our PPMAP results}
\label{sec:reliability}

In order to check whether the results from PPMAP have been affected due to any assumptions, we test for potential biases due to overestimating the flux from hot dust, our chosen grids of dust temperatures and on the prior assumptions for $\beta$.

\begin{itemize}

\item  The warmer dust components detected in the shells of G11.2$-$0.3 and G29.7$-$0.3 may artificially arise from contributions due to line emission at 24\,$\mu$m. We therefore tested the PPMAP process for our SNRs after subtracting 30\,per\,cent of the flux in the 24\,$\mu$m band. This level was chosen as it is similar to the fraction of the Crab Nebula 24\,$\mu$m flux expected to originate from line emission \citep{Temim2012}. We find this does not affect any of the dust mass or $\beta$ results.

\item We next check if PPMAP is, in fact, able to extract values of $\beta$ or is simply returning its prior value. We test this by simulating several 26\,K Gaussian sources with a 1$^\prime$ FWHM, varying $\beta$ between 1.5 and 2.4. The images are convolved to the \textit{Herschel} beam sizes and add Gaussian noise for a signal-to-noise ratio between 3 and 10,000, which is constant across the spectrum. Assuming a flat $\beta$ prior between 0 and 3, we then use PPMAP to estimate $\beta$ for each simulated source in order to verify if PPMAP can return the correct value from the simulations.  We find that the ability of PPMAP to pull out precise values of $\beta$ is critically dependent on the signal-to-noise of the FIR data. At high signal-to-noises ($\sim$100) we find that the returned $\beta$ matches the original value with an error less than 0.15. Figure~\ref{fig:beta_error} shows how the uncertainty in the PPMAP estimated values of $\beta$ varies with the signal-to-noise of the input simulations. The lowest signal-to-noise ratio in this test ($\sim 3$) is typical of our most resolved PWN source, G21.5$-$0.9. The other two SNRs have larger FIR uncertainties (due to the high background levels) and therefore this confirms that we would not be able to derive the value of $\beta$ in G11.2$-$0.3 and G29.7$-$0.3 with any precision.  For G21.5$-$0.9, this suite of simulations therefore suggests our uncertainty in the $\beta$ value derived in Section~\ref{sec:betadiff} is slightly larger, i.e. $\beta_{\rm PWN}\,=\,1.4 \pm$ 0.5.

\end{itemize}

\begin{figure}
	\includegraphics[width=1.0\linewidth]{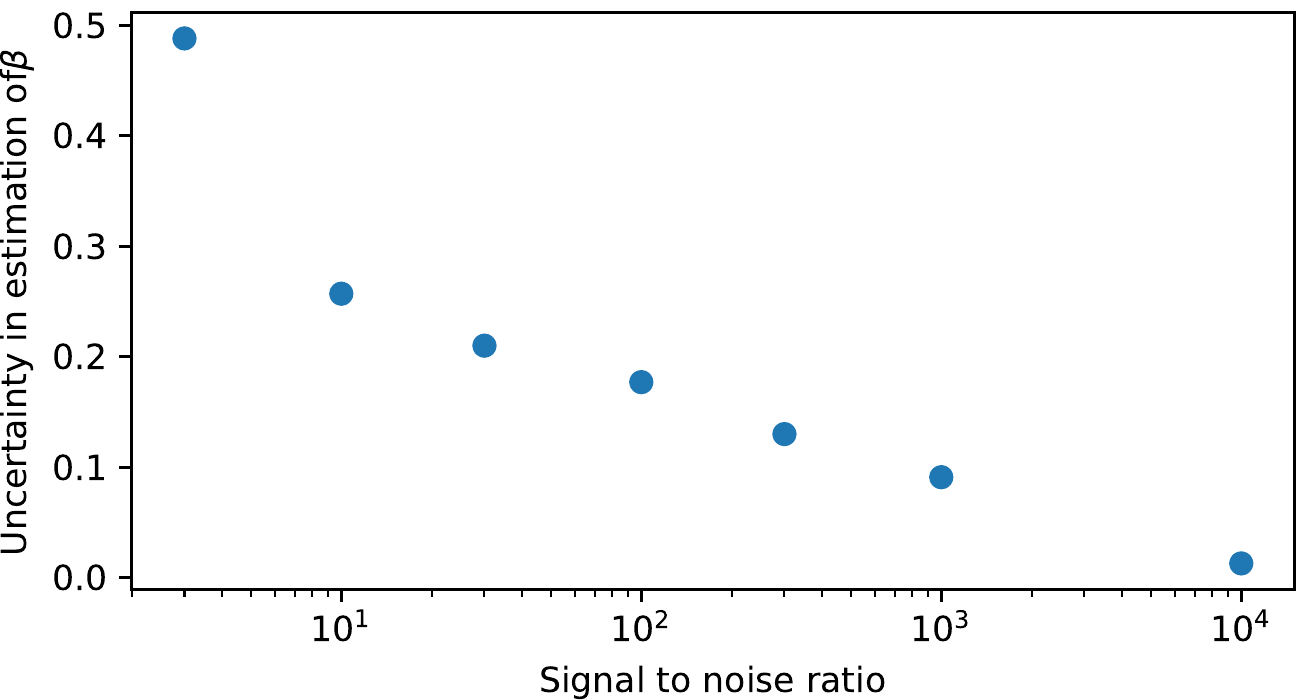}
	\caption{The root mean square uncertainty of the $\beta$ value derived by PPMAP based on simulations for a 26\,K source with Gaussian noise applied.}
	\label{fig:beta_error}
\end{figure}

In our images the signal-to-noise ratio is not constant across the spectrum as we have much more confusion at longer wavelengths. In order to constrain $\beta$, our tests show that higher signal-to-noise at the Rayleigh-Jeans end of the spectrum is required. This is difficult given the current large uncertainties and variation in the ISM flux for most Galactic SNRs observed with \textit{Herschel}. With higher resolution or longer wavelength data we could constrain $\beta$ to within a smaller uncertainty range and potentially draw out any differences in the dust properties between different SNR regions. It may be easier to constrain $\beta$ for SNRs in the regions of the Galactic plane at $\mid l \mid >60^{\circ}$ which, on average, suffer from lower levels of background interstellar dust emission.

\section{Conclusions} \label{conclusions}
We follow R06 and PG11 and classify the FIR structures in Galactic SNRs in levels 1\,--\,4 where 1 = detection (FIR emission which is clearly correlated with radio, MIR, or X-ray structures and can be distinguished from ISM) and level 4 is not detected in the FIR. We add \numNewDetections new SNRs to the current sample of 3 in our Galaxy that contain cool dust (<50\,K) associated with the SNR. This is a lower limit to the number of dusty SNRs in our sample as we suffer from a number of biases.

Dust is detected from the central region of \numCentral sources, 1 of which was recently discussed elsewhere \citealp{Temim2017, Rho2018} and 3 of which (G11.2$-$0.3, G21.5$-$0.9, and G29.7$-$0.3) are \emph{new detections of FIR emission coinciding with the locations of pulsar wind nebulae.} We analyse the dust content of these 3 PWNe by fitting their SED to NIR-radio data using modified blackbody fits. The best fits from this analysis give estimated cold dust temperatures for G11.2$-$0.3, G21.5$-$0.9, and G29.7$-$0.3 respectively of \ColdTempaBest\,K, \ColdTempbBest\,K, and \ColdTempcBest\,K, and dust masses of \ColdMassaBest\,M$_\odot$, \ColdMassbBest\,M$_\odot$, and \ColdMasscBest\,M$_\odot$. We also carry out a Monte Carlo analysis, randomly sampling a set of 1000 flux values at each wavelength with a Gaussian distribution centred at the measured flux. This provides 1000 sets of results for each SNR, with median dust temperatures for G11.2$-$0.3, G21.5$-$0.9, and G29.7$-$0.3 respectively of \ColdTempa\,$^{+\,\ColdTempaP} _{-\,\ColdTempaN}$\,K, \ColdTempb\,$\pm$\,\ColdTempbP\,K, and \ColdTempc\,$^{+\,\ColdTempcP} _{-\,\ColdTempcN}$\,K, and dust masses of \ColdMassa\,$^{+\,\ColdMassaP} _{-\,\ColdMassaN}$\,M$_\odot$, \ColdMassb\,$\pm$\,\ColdMassbP\,M$_\odot$, and \ColdMassc\,$^{+\,\ColdMasscP} _{-\,\ColdMasscN}$\,M$_\odot$. Large uncertainties in our fluxes at the longest FIR wavelengths for G29.7$-$0.3 pull the median estimate to a lower dust temperature and larger dust mass than that of the best fit.

We use PPMAP to more rigorously analyse the material within the three SNRs. This confirms the presence of cold dust (20\,--\,40\,K) within the PWN regions, which is clearly distinguishable from the surrounding ISM, and warm dust in the shells of G11.2$-$0.3 and G29.7$-$0.3. Through this analysis we estimate significant cold dust masses within the PWNe of \PPMAPMassa\,$\pm$\,\PPMAPMassErra\,$\rm M_{\odot}$, \PPMAPMassb\,$\pm$\,\PPMAPMassErrb\,$\rm M_{\odot}$, \PPMAPMassc\,$\pm$\,\PPMAPMassErrc\,$\rm M_{\odot}$, for G11.2$-$0.3, G21.5$-$0.9, and G29.7$-$0.3 respectively. For both G11.2$-$0.3 and G29.7$-$0.3 these estimates are much more constrained than those of the traditional SED-fitting routine. PPMAP allows for a range of dust temperatures across the SNRs making these results more reliable.

Using PPMAP we also analyse the variation in the dust emissivity index, $\beta$, across the SNRs and compare with the values derived for the SN shell and ISM regions. We are unable to constrain $\beta$ within G11.2$-$0.3 and G29.7$-$0.3 due to the low signal-to-noise at long wavelengths.  For G21.5$-$0.9, we estimate $\beta_{\rm PWN}$\,=\,1.4\,$\pm$\,0.5 in the PWN (with uncertainty in $\beta$ derived from simulated data sets), and find $\beta_{\rm ISM}\,=\,1.8 \pm 0.1$ in the ISM dust based on using a flat $\beta$ prior. We show that with higher signal-to-noise in the FIR (or indeed longer wavelength data), our simulations suggest that in future, we could further constrain $\beta$ within the PWNe and potentially draw out any differences in the dust properties with respect to dust in the interstellar medium. This may be easier for Galactic plane SNRs at $\mid l \mid >60^{\circ}$ where there are lower levels of background interstellar dust emission on average.

\section*{Acknowledgements}
We thank the anonymous referee for their thorough and careful reading of this paper and many constructive suggestions.
We acknowledge Nicolas Peretto for help with CO data at the locations of some of our sources.
HC, HLG, and PC acknowledge support from the European Research Council (ERC) in the form of Consolidator Grant
{\sc CosmicDust} (ERC-2014-CoG-647939).  MJB and IDL acknowledge support from the ERC in the form of Advanced Grant SNDUST (ERC-2015-AdG-694520).
MM acknowledges support from an STFC Ernest Rutherford fellowship (ST/L003597/1). \textit{Herschel} is an ESA space observatory with science instruments provided by European-led Principal Investigator consortia and with important participation from NASA.


\bibliographystyle{mnras}
\bibliography{library}


\bsp	
\label{lastpage}
\end{document}